\documentclass[12pt]{article}
\usepackage{epsfig,cite,amsmath,feynarts,axodraw,euscript}

\input paperdef

\graphicspath{{figs/}}

\evensidemargin \oddsidemargin
\makeatletter
\renewcommand\section{\@startsection {section}{1}{\z@}%
                          {-3.5ex \@plus -1ex \@minus -.2ex}%
                          {2.3ex \@plus.2ex}%
                          {\mathversion{bold}\normalfont\Large\bfseries}}
\renewcommand\subsection{\@startsection{subsection}{2}{\z@}%
                           {-3.25ex\@plus -1ex \@minus -.2ex}%
                           {1.5ex \@plus .2ex}%
                           {\mathversion{bold}\normalfont\large\bfseries}}
\renewcommand\subsubsection{\@startsection{subsubsection}{3}{\z@}%
                           {-3.25ex\@plus -1ex \@minus -.2ex}%
                           {1.5ex \@plus .2ex}%
                           {\mathversion{bold}\normalfont\normalsize\bfseries}}
\makeatother

\def\_{\rule{.3em}{.15ex}}

\def\slash#1{\setbox0=\hbox{$#1$}#1\hskip-\wd0\dimen0=5pt\advance
       \dimen0 by-\ht0\advance\dimen0 by\dp0\lower0.5\dimen0\hbox
         to\wd0{\hss\sl/\/\hss}}

\allowdisplaybreaks

\usepackage{mathenv}

\begin{document}
\thispagestyle{empty}

\def\thefootnote{\fnsymbol{footnote}}

\begin{flushright}
DCPT/07/110\\
IPPP/07/55\\
MPP--2007--65\\
arXiv:0710.2972 [hep-ph]
\end{flushright}

\vspace{1cm}

\begin{center}

{\large\sc {\bf  {\boldmath{$Z$}}~Pole Observables in the MSSM}}
 
\vspace{0.4cm}

\vspace{1cm}

{\sc 
S.~Heinemeyer$^{1}$%
\footnote{email: Sven.Heinemeyer@cern.ch}%
, W.~Hollik$^{2}$%
\footnote{email: hollik@mppmu.mpg.de}%
, A.M.~Weber$^{2}$%
\footnote{email: Arne.Weber@mppmu.mpg.de}%
, G.~Weiglein$^{3}$%
\footnote{email: Georg.Weiglein@durham.ac.uk}
}

\vspace*{1cm}

{\sl

$^1$Instituto de Fisica de Cantabria (CSIC-UC), Santander,  Spain 

\vspace*{0.4cm}

$^2$Max-Planck-Institut f\"ur Physik (Werner-Heisenberg-Institut),\\ 
F\"ohringer Ring 6, D--80805 Munich, Germany 

\vspace*{0.4cm}

$^3$IPPP, University of Durham, Durham DH1 3LE, U.K.

}

\end{center}

\vspace*{0.2cm}
\begin{abstract}
We present the currently most accurate prediction of $Z$~pole observables such
as $\sweff$, $\Ga_Z$, $R_b$, $R_l$, and $\si^0_{\rm had}$
in the Minimal Supersymmetric Standard Model (MSSM).  
We take into account the complete one-loop results including the full complex
phase dependence, all available MSSM two-loop corrections as well as the full
SM results. We furthermore include higher-order corrections in the MSSM
Higgs boson sector, entering via virtual Higgs boson contributions.
For $\Ga(Z \to \neu{1} \neu{1})$ we present a full one-loop calculation. 
We analyse the impact of the different sectors of the MSSM with
particular
emphasis on the effects of the complex phases. The predictions for
the $Z$~boson observables and $\MW$ are compared with 
the current experimental
values. Furthermore we provide an estimate of the remaining higher-order
uncertainties in the prediction of $\sweff$.

\end{abstract}

\def\thefootnote{\arabic{footnote}}
\setcounter{page}{0}
\setcounter{footnote}{0}

\newpage


\section{Introduction}

$Z$~boson physics is well established as a cornerstone of the Standard Model
(SM)~\cite{LEPEWWG,LEPEWWG2,TEVEWWG}.  
Many (pseudo-) observables~\cite{Bardin:1997xq} 
have been measured with high accuracy using the
processes (mediated at lowest order by photon and $Z$~boson exchange)
\begin{equation}
e^+e^-\to
f \bar f,\ \ \ \ \  f\neq e,
\label{epemtofermferm}
\end{equation}
at LEP and SLD with a center of mass energy $\sqrt{s} \approx \MZ$. 
In particular these are the effective leptonic weak mixing
angle at the $Z$~boson resonance, $\sweff$, $Z$~boson 
decay widths to SM fermions, 
$\Ga(Z \to f \bar f)$, the invisible width, $\Ga_{\rm inv}$,
the total width,
$\Ga_Z$, the ratios of partial widths, $R_l$ and $R_b$, forward-backward
and left-right asymmetries, $A_{\rm FB}$ and 
$A_{\rm LR}$, and the total hadronic cross section, $\si^0_{\rm had}$.
Together with the measurement of the mass of the $W$~boson, $\MW$, and
the mass of the top quark, $\mt$, the $Z$~pole 
observables have been used 
to constrain indirectly the SM Higgs boson mass,
$\MHSM$, the last free parameter of the model, yielding 
$\MHSM = 76^{+33}_{-24} \gev$ with an upper limit of 
$\MHSM \le 144 \gev$ at the 95\% C.L.~\cite{LEPEWWG,LEPEWWG2}.
The precision observables are also very powerful for testing 
models beyond the SM. In particular the Minimal Supersymmetric Standard Model
(MSSM)~\cite{mssm} has been investigated, see \citere{PomssmRep}
for a review.
Performing fits in constrained SUSY models a certain preference for not
too heavy SUSY particles has been
found~\cite{ehow3,ehow4,ehoww,AllanachFit,Rotze,mastercode}. 
The 
prospective improvements in the experimental accuracies, in particular
at the ILC with GigaZ option, will provide a high sensitivity to deviations
both from the SM and the MSSM.

In order to fully exploit the high-precision measurements, the
theoretical uncertainty in the predictions of the (pseudo-) observables
should be sufficiently smaller than the experimental errors.
Within the SM the complete one-loop and two-loop
results~\cite{BayernFermionic,UliMeierFussballgottFermionic,UliMeierFussballgottBosonic,dkappaSMbos2L}
as well as leading higher-order  
contributions~\cite{drSMgfals2,MWSMQCD3LII,drSMgf3mh0,drSMgf3,derhoalalscube}
are available for $\sweff$.
For the leptonic and hadronic $Z$~widths partial results for
process-specific two-loop corrections are
known~\cite{Bardin:1997xq,Czarnecki:1996ei,Fleischer:1999iq,GammaZ2L}.

The theoretical evaluation of the $Z$~pole observables within the MSSM is 
not as advanced as 
in the SM. So far, the one-loop contributions have been
evaluated 
completely, restricted however to the special case of
vanishing complex phases (contributions 
to the $\rho$ parameter with non-vanishing complex phases in the scalar
top and bottom mass matrices have been
considered in \citere{kangkim}). At the \twol\ 
level, the leading $\oaas$ corrections~\cite{dr2lA} and
the leading electroweak corrections of \order{\alt^2}, 
\order{\alt \alb}, \order{\alb^2}  to $\De\rho$ have been
obtained~\cite{drMSSMal2A,drMSSMal2B}~($\alt$ and $\alb$ are defined in
terms of the Yukawa couplings $y_f$ as $\al_f=\frac{y_f^2}{4\pi}$).  
Going beyond the minimal SUSY model and allowing for non-minimal
flavor violation the leading one-loop contributions to $\De\rho$ are
known~\cite{delrhoNMFV}. 

In order to confront the predictions of
supersymmetry (SUSY) with the electroweak precision data
and to derive constraints on the supersymmetric parameters,
it is desirable to achieve the same level of accuracy for the
SUSY predictions as for the SM.
In this paper we present 
complete one-loop results for the $Z$~boson observables in the MSSM
with complex parameters, 
taking into account the full
complex phase dependence. Besides the decays of the $Z$~boson into
quarks and leptons, we also provide a complete one-loop result for
the partial width of the $Z$~boson decay into the lightest neutralino,
$\Ga(Z \to \neu{1}\neu{1})$, which is the first one-loop result for this
decay in the full MSSM. If the $Z$~boson decay into the lightest
neutralino is kinematically possible it contributes to the invisible width of
the $Z$~boson.
We combine our new one-loop results with the full set of available 
higher-order corrections in the MSSM. In order to recover the
state-of-the-art SM results in the decoupling limit where all
supersymmetric particles are heavy, we consistently incorporate also
those SM-type higher-order corrections which go beyond the
results obtained within the MSSM so far. 
In this way we
provide the currently most complete results for the $Z$~boson
observables in the MSSM (for the corresponding results for the $W$~boson
mass, $\MW$, see
\citere{MWweber}). A public computer code based on our result for
the electroweak precision observables (EWPO), i.e.\ the $Z$~pole
observables and $\MW$, is in preparation~\cite{pope}.  
It provides predictions for the EWPO in terms of the low-energy
parameters of the MSSM, which can be freely chosen as independent
inputs.

We analyse the numerical results for the EWPO for various MSSM
scenarios, such as SPS benchmark scenarios~\cite{sps}, scenarios with
heavy scalar masses~\cite{split,focus} and the CPX~scenario~\cite{cpx}.
The dependence of the results for the EWPO 
on the complex phases is investigated. For 
$\sweff$, showing the largest sensitivity of the $Z$~pole
observables to the SUSY loop corrections, we provide an estimate of the 
remaining theoretical uncertainties
from unknown higher-order corrections.

The rest of the paper is organised as follows: In
\refse{sec:conventions} we introduce our notations and
conventions. The $Z$~pole observables are discussed in
\refse{sec:basics}. Details about the evaluation of the
higher-order corrections are given in \refse{sec:calceffcoup}. In
\refse{sec:numanal} we present our numerical analysis. The estimate
of the theory uncertainties in the prediction for $\sweff$
from unknown higher-order
corrections is given in \refse{sec:theounc}. We conclude with
\refse{sec:conclusions}.


\section{Notations and conventions}
\label{sec:conventions}

In the complex MSSM, the $Z$~pole observables depend on all the
free parameters of the model, such as SUSY particle 
masses, mixing angles and couplings. 
In order to be self-contained we list in this section the notations and
conventions used in our calculations. We briefly describe the relevant
quantities in 
the sfermion, the chargino/neutralino, and the Higgs boson sector of the
MSSM. 

\subsection{Sfermions}
\label{sec:sfermions}

The mass matrix for the two sfermions of a given 
flavour, in the $\sfl, \sfr$ basis, 
is given by 
\begin{align}
{\matr{ M}}_{\tilde f} =
\begin{pmatrix}
        M_L^2 + \mf^2 & \mf \; \Xf^* \\
        \mf \; \Xf    & M_R^2 + \mf^2
\end{pmatrix} ,
\label{squarkmassmatrix}
\end{align}
with
\begin{eqnarray}
M_L^2 &=& M_{\tilde F}^2 + \MZ^2 \CZb (I^f_3 - Q_f \sw^2), \non \\
M_R^2 &=& M_{\tilde F'}^2 + \MZ^2 \CZb Q_f \sw^2, \\ \non
\Xf &=& A_f - \mu^* \{\CTb, \tb\} ,
\label{squarksoftSUSYbreaking}
\end{eqnarray}
where $\{\CTb, \tb\}$ applies for up- and down-type sfermions,
respectively, and $\tb$ is the ratio of the two vacuum expectation values of
the two Higgs doublets (see \refse{sec:higgs}), $\tb \equiv v_2/v_1$. 
We have furthermore used $\sw^2 \equiv \sin^2\theta_{\rm W} = 1 - \MW^2/\MZ^2$.
In the Higgs and scalar fermion sector of the complex MSSM, $N_f + 1$
phases are present, one for each $A_f$ and one
for $\mu$, i.e.\ $N_f + 1$ new parameters appear
in comparison to the MSSM with only real parameters.
As an abbreviation,
\begin{equation}
\phi_{\Af} \equiv \arg(\Af) \, 
\end{equation}
will be used.
As an independent parameter one can trade $\phi_{\Af}$ for 
$\phi_{\Xf} \equiv \arg(\Xf)$.
The sfermion mass eigenstates are obtained by the transformation
\begin{equation}
\VL \sfe \\ \sfz \VR = {\matr{U}_{\tilde{f}}}_{\sf} \VL \sfl \\ \sfr \VR,
\end{equation}
with a unitary matrix $\matr{U}_{\tilde{f}}$.  
The mass eigenvalues are given by
\begin{equation}
m_{\tilde f_{1,2}}^2 = \mf^2
  + \edz \KKL M_L^2 + M_R^2
           \mp \sqrt{( M_L^2 - M_R^2)^2 + 4 \mf^2 |\Xf|^2}~\KKR ,
\label{sfermmasses}
\end{equation}
and are independent of the phase of $\Xf$.


\subsection{Gaugino sector}
\label{sec:charneu}

The physical masses of the charginos are determined by the matrix
\begin{align}
  \matr{X} =
  \begin{pmatrix}
    M_2 & \sqrt{2} \sinb \MW \\
    \sqrt{2} \cosb \MW & \mu
  \end{pmatrix},
\label{Cmatrix}
\end{align}
which contains the soft breaking term $M_2$ and the Higgsino mass term $\mu$,
both of which may have complex values in the complex MSSM. Their complex
phases are denoted by 
\begin{equation}
\phi_{M_2} \equiv  {\rm arg}\KL M_2 \KR  \ \ \ \ \  \textup{and} \ \ \ \ \ 
\phi_{\mu} \equiv {\rm arg}\KL \mu \KR . 
\end{equation}
The physical masses are denoted as $m_{\tilde{\chi}^\pm_{1,2}}$ and
are obtained by applying the diagonalisation matrices 
   ${{\matr U}_{{\tilde{\chi}}^\pm}}$ and 
   ${{\matr V}_{{\tilde{\chi}}^\pm}}$
\begin{equation}
{\bf U^\ast_{{\tilde{\chi}}^\pm}}  \matr{X} {{\matr
    {V}}^\dag_{{\tilde{\chi}}^\pm}} = 
\textup{diag}\left(m_{\tilde{\chi}^\pm_1} ,m_{\tilde{\chi}^\pm_2} \right).
\end{equation}
The situation is similar for the neutralino masses, 
which can be calculated from the mass matrix
($\sw = \sin\theta_{\rm w}$, $\cw = \cos\theta_{\rm w}$)
\begin{align}
  \matr{Y} =
  \begin{pmatrix}
    M_1                    & 0                & -\MZ \, \sw \cosb
    & \MZ \, \sw \sinb \\ 
    0                      & M_2              & \quad \MZ \, \cw \cosb
    & -\MZ \, \cw \sinb \\ 
    -\MZ \, \sw \cosb      & \MZ \, \cw \cosb & 0
    & -\mu             \\ 
    \quad \MZ \, \sw \sinb & -\MZ \, \cw \sinb & -\mu                   & 0
  \end{pmatrix}.
\label{Nmatrix}
\end{align}
This symmetric matrix contains the additional complex soft-breaking
parameter $M_1$, where the complex phase of $M_1$ is given by
\begin{equation}
\phi_{M_1} \equiv  {\rm arg}\KL M_1 \KR.
\end{equation}

The physical masses are denoted as $m_{\tilde{\chi}^0_{1,2,3,4}}$ and
are obtained in a diagonalisation procedure using the matrix
$\matr{N}_{\tilde{\chi}^0}$ 
\begin{align}
\matr{N}_{\tilde{\chi}^0}^\ast \matr{Y} \matr{N}_{\tilde{\chi}^0}^\dag=
\textup{diag}\left(m_{\tilde{\chi}^0_{1}},m_{\tilde{\chi}^0_{2}},
                   m_{\tilde{\chi}^0_{3}},m_{\tilde{\chi}^0_{4}}\right). 
\label{NNmatrix}
\end{align}

The gluino enters the predictions for the hadronic decays of the $Z$~boson 
(and accordingly the total $Z$~width) at the one-loop level, while for the
EWPO with leptons in the final state it enters only at \order{\al\als}.
The soft-breaking gluino mass parameter $M_3$ is in general
complex,
\begin{equation}
M_3 = |M_3| e^{i \phigl} , 
\label{eq:gluinophase}
\end{equation}
and the gluino mass is given by $\mgl = |M_3|$.
The phase can be absorbed by a redefinition of the gluino Majorana
spinor such that it appears only in the gluino couplings but not in
the mass term.
In our calculation of the EWPO
below we will incorporate the full phase
dependence of the complex parameters at the one-loop level, while we
neglect the explicit dependence on the complex phases beyond the
one-loop order. 
Accordingly, we incorporate the gluino phase appearing in
\refeq{eq:gluinophase} into our predictions for the hadronic
Z~observables (and also in Higgs-sector corrections associated with the
bottom Yukawa coupling, see below), while we treat the two-loop
corrections to the EWPO in the approximation of vanishing gluino phase.


\subsection{Higgs bosons}
\label{sec:higgs}

$\cp$-violating phases can have an important impact on the Higgs sector
of the 
MSSM with complex 
parameters~\cite{mhiggsCPXgen,mhiggsCPXRG1,mhiggsCPXRG2,mhiggsCPXFD1,mhcMSSMlong}.  
In our (one-loop) analysis of the complex phase dependence 
of the predictions for the $Z$~pole observables we take into account
also the potentially large effects of $\cp$-mixing in the complex MSSM Higgs 
sector~\cite{mhcMSSMlong,mhcMSSM2L}, although formally
$\cp$-violating contributions in the complex MSSM Higgs sector enter the
$Z$~pole observables only at the two-loop level. 
We 
consistently
incorporate the higher-order corrected Higgs states in our analytical
results for the $Z$~pole observables and the $W$~boson
mass.

Once higher-order terms are included in the complex MSSM~\cite{mhiggsCPXgen}, 
Higgs fields are mixtures of the
$\cp$-even, $h$ and $H$, and the $\cp$-odd states, $A$ and $G$.
The Higgs boson propagator matrix receives contributions from the Higgs boson
self energies. In the propagator matrix the 
mixing between the Higgs fields $h,H,A$ and the
Goldstone~boson $G$ is of sub-leading two-loop order and can therefore 
safely be neglected~\cite{Frankphd,mhcMSSMlong}.
The three mass eigenvalues of the remaining $(3 \times 3)$ matrix,
\begin{equation}
\MHe \le \MHz \le \MHd ,
\end{equation} 
corresponding to the mass eigenstates $\He$, $\Hz$, $\Hd$, are then
determined by the propagator poles~\cite{mhcMSSMlong}.

It is sometimes convenient for phenomenological analyses to introduce 
effective couplings that incorporate leading higher-order effects. 
There is no unique procedure how to define these effective couplings,
and care has to be taken not to spoil gauge or unitarity cancellations by
a partial inclusion of higher-order contributions. One possibility is to
use the ``$p^2 = 0$'' approximation (see \citere{mhcMSSMlong} for
details), i.e.\ to evaluate all Higgs boson self energies at zero
external momentum, 
which is equivalent to the effective potential approach.
In this way a unitary effective mixing matrix 
$\matr{U}_{\rm eff}$ can be defined (ensuring decoupling to the SM for
heavy SUSY particles), transforming in this approximation 
the lowest-order states $h, H, A$ into the mass eigenstates $h_1$,
$h_2$, $h_3$,
\begin{equation}
\left(\begin{array}{c} \He \\ \Hz \\ \Hd \end{array}\right)_{p^2 = 0}
= \matr{U}_{\rm eff}
\left(\begin{array}{c} h \\ H \\ A \end{array}\right). 
\end{equation}
The elements of the effective mixing matrix $\matr{U}_{\rm eff}$ 
can be interpreted as effective couplings of the Higgs bosons,
incorporating leading higher-order corrections from Higgs boson
self energies into the Higgs boson couplings~\cite{mhcMSSMlong}.
In our numerical calculation the Higgs sector parameters are 
evaluated with the help of the program
\fhtt~\cite{feynhiggs,mhiggsAEC,mhcMSSMlong}. 

\medskip
Another numerically important correction appears in the relation
between the bottom-quark mass and the bottom Yukawa coupling, $y_b$. 
The leading $\tb$-enhanced contributions to the relation arise from
one-loop contributions with gluino--sbottom and chargino--stop loops. 
We include the leading effects via the quantity $\db$~\cite{deltamb2}
(see also \citeres{deltamb1,deltamb2b,deltamb3}). 
The $\db$ corrections can affect for instance the prediction for $\Ga(Z
\to b \bar b)$, where $y_b$ enters at the one-loop level. Thus the
$\db$ corrections are a two-loop effect, which, however, can in
principle have a noticable impact.

Numerically the correction expressed by $\db$ to the relation between
the bottom-quark 
mass and the bottom Yukawa coupling is usually by far the dominant 
part of the contributions from the sbottom sector (see also
\citeres{mhiggsEP4,mhiggsFD2}). 
In the limit of $M_{\tilde F}, M_{\tilde F'} \gg \mt$ and $\tb \gg 1$,
$\db$ is given 
by~\cite{deltamb2} 
\begin{equation}
\db = \frac{2\als}{3\,\pi} \, M_3^* \, \mu^* \, \tb \,
                    \times \, I(\msbe, \msbz, \mgl) +
      \frac{\alt}{4\,\pi} \, \At^* \, \mu^* \, \tb \,
                    \times \, I(\mste, \mstz, |\mu|)~.
\label{def:dmb}
\end{equation}
The function $I$ is defined as
\BEA
I(a, b, c) &=& \ed{(a^2 - b^2)(b^2 - c^2)(a^2 - c^2)} \,
               \KL a^2 b^2 \log\frac{a^2}{b^2} +
                   b^2 c^2 \log\frac{b^2}{c^2} +
                   c^2 a^2 \log\frac{c^2}{a^2} \KR~. 
\EEA
For the numerical evaluation we use the implementation of $\db$ and the
corresponding Higgs couplings in \fh.


\section{Description of the $Z$~boson resonance}
\label{sec:basics}

\subsection{Effective coupling approach}
\begin{figure}[htb!]
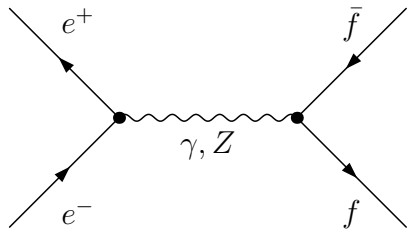

\centerline{
\unitlength=1.cm%
\begin{feynartspicture}(4.,4)(1,1)
\FADiagram{}
\FAProp(-8.,2)(0,10)(0.,){/Straight}{1}
\FALabel(-3,4)[t]{$e^-$}
\FAProp(-8.,18)(0,10)(0.,){/Straight}{-1}
\FALabel(-3,18)[t]{$e^+$}
\FAVert(0,10){0}
\FAProp(0,10)(13,10)(0.,){/Sine}{0}
\FALabel(6.5,9)[t]{$\ga,Z$}
\FAVert(13,10){0}
\FAProp(13,10)(21,18)(0.,){/Straight}{-1}
\FALabel(17,18)[t]{$\bar f$}
\FAProp(13,10)(21,2)(0.,){/Straight}{1}
\FALabel(17,4)[t]{$f$}
\end{feynartspicture}
}
\vspace{-1em}
\caption[$e^+e^-\to f\bar f$ at Born level]
        {$e^+e^-\to f \bar f$ at Born level (neglecting Higgs boson 
         exchange).}  
\label{fig:epemfbf}
\end{figure}

$e^+e^-$ collisions at the $Z$~boson resonance are commonly described in an
effective coupling approach. 
These effective couplings are subsequently used to define the so-called
$Z$~resonance pseudo observables. 
In the following we will mostly use the conventions given in
\citere{Bardin:1997xq}.
At Born level the matrix element of the process in
\refeq{epemtofermferm}, depicted in \reffi{fig:epemfbf}, is given by 
\begin{eqnarray}
\mathcal{M_{\textup{Born}}}\propto\frac{1}{s}\bigg\{Q_e Q_f
\left(\ga_\al \otimes \ga^\al\right)+\chi\Big[
g_{v,(0)}^{e} 
g_{v,(0)}^{f}
\left(\ga_\al \otimes \ga^\al\right) - 
g_{v,(0)}^{e} 
g_{a,(0)}^{f}
\left(\ga_\al\otimes \ga^\al \ga_5\right)\nn\\ 
-
g_{a,(0)}^{e} 
g_{v,(0)}^{f} 
\left(\ga_\al\ga_5 \otimes \ga^\al \right)+ 
g_{a,(0)}^{e} 
g_{a,(0)}^{f} 
\left(\ga_\al \ga_5 \otimes \ga^\al \ga_5\right)\Big] 
\bigg\},
\label{mateetoff}
\end{eqnarray}
with the propagator $\chi$ defined as%
\footnote{The form of the propagator in \refeq{eq:zprop} corresponds to a
Breit-Wigner function with a running width, which is the convention
normally adopted in the experimental determination of the gauge-boson
masses. It should be noted, however, that from two-loop order on the
$Z$~boson propagator has a complex pole. Expanding around the complex
pole and defining the mass according to the (gauge-invariant) real part
of the complex pole leads to a Breit-Wigner parametrisation of the
resonance line shape with a constant decay width (see
\citere{MWSMferm2L} for details). The (numerically sizable) difference
between the two mass definitions needs to be properly taken into account
when incorporating higher-order corrections obtained with the fixed-width
parametrisation.}
\begin{equation}
\chi=\frac{s}{s-\MZ^2+is \, \Gamma_Z/\MZ} .
\label{eq:zprop}
\end{equation}
In \refeq{eq:zprop} $s$ denotes the center-of-mass energy, 
$\MZ$ is the $Z$~boson mass and
$\Ga_Z$ its width. We use the shorthand notation
\begin{equation}
\left(A_\al \otimes B^\al\right)=\left[\bar v_e A_\al
  u_e\right]\times\left[\bar u_f B^\al v_f\right]. 
\end{equation}
Furthermore the lowest-order
vector and axial vector couplings of a fermion $f$ 
\begin{eqnarray}
&&
g_{v,(0)}^{f}
=\frac{I^f_3-2 Q_f \sw^2}{2 \cw \sw},\nn \\
&&
g_{a,(0)}^{f}
=\frac{I^f_3}{2 \cw \sw},
\label{borncouplvecax}
\end{eqnarray}
were introduced in \refeq{mateetoff}. $Q_f$ denotes the charge of the
fermion $f$ as
fraction of the elementary charge $e$, $I^f_3$ is its third weak isospin
component. 
The simple structure of the matrix element in \refeq{mateetoff} easily
allows the identification of QED contributions due to photon exchange
graphs (first term) and electroweak $Z$~boson exchange graphs (terms in
square brackets).  
%
%

The inclusion of higher-order corrections in general leads to a 
modification of this simple structure.
At the one-loop level one can easily classify the corrections into
QED, QCD and electroweak corrections.  
Supplementing the Born-level diagrams (see \reffi{fig:epemfbf})
with an additional photon line gives the one-loop QED corrections.
They form a gauge-invariant, UV-finite, but IR-divergent subset of loop
corrections to the process $e^+e^-\to f \bar f$.
IR finiteness is obtained by taking real Bremsstrahlung contributions
into account $(e^+e^-\to f \bar f+\ga)$. 
This implies that the QED
corrections are dependent on the experimental setup. 
The situation is similar for the QCD corrections. At one-loop order
these corrections occur for quark pair production $(e^+e^-\to q\bar q)$
and contain corrections due to virtual gluon and
gluino exchange, as well as real gluon emission, the latter again
leading to a dependence on the experimental setup.
Beyond one-loop order a
clean distinction between QCD and electroweak corrections 
is no longer possible. Mixed \order{\al\als} corrections appear for
instance as gluon- and gluino-exchange contributions in virtual quark
loops of the $W$~and $Z$~boson propagators. In the following we will
explicitly indicate final-state QCD corrections and initial- and
final-state QED corrections where relevant, while we use a common notation
for all other contributions. The self energy, vertex and box
contributions can be expressed in terms of complex form factors
$\mathcal{F}^{ef}_{ij}$. These form factors depend on the Mandelstam
variables $s$ and $t$, where the $t$~dependence enters only via box
contributions.

At the $Z$~boson resonance those contributions that are not enhanced by
a resonant $Z$~propagator are relatively small. In particular, box
diagrams contribute only a fraction of less than $10^{-4}$ at the
one-loop level. It is therefore convenient for describing physics at the 
$Z$~boson resonance to treat the non-resonant higher-order corrections
separately as part of a ``deconvolution'' procedure (see the discussion
in \refse{ZPoleObs} below) and to express the dominant contributions 
in terms of a Born-type matrix element. 
The effective couplings in this matrix element are given by
the form factors $\mathcal{F}^{ef}_{ij}$ ($i,j = V,A$)
in the approximation where higher-order  
non-resonant contributions are neglected, so that they only depend on
the Mandelstam variable~$s$,
\begin{eqnarray}
\mathcal{M^{\textup{eff}}}\propto \frac{1}{s}\bigg\{\al(s)
\left(\ga_\al \otimes
  \ga^\al\right)+\chi\Big[\mathcal{F}_{VV}^{ef}(s) \left(\ga_\al
  \otimes \ga^\al\right) - \mathcal{F}_{VA}^{ef}(s)
\left(\ga_\al\otimes \ga^\al \ga_5\right)\nn\\ 
-\mathcal{F}_{AV}^{ef} (s)\left(\ga_\al\ga_5 \otimes \ga^\al
\right)+ \mathcal{F}_{AA}^{ef}(s)  \left(\ga_\al \ga_5 \otimes
  \ga^\al \ga_5\right)\Big] 
\bigg\}.
\label{eq:effmatrel}
\end{eqnarray}
As the only relevant contribution to the form factor of the photon-exchange
part,  the running QED coupling $\al(s)$ is kept.

The factorization is the result of a variety of approximations
that is valid at the $Z$ resonance to the accuracy needed 
(see~\citere{Bardin:1997xq} for a more detailed discussion).
At this level, 
the form factors $\mathcal{F}^{ef}_{ij}$in~\refeq{eq:effmatrel} 
factorise into contributions from the production and the
decay. 
The final step in the ``$Z$~pole approximation''
is to set $s=\MZ^2$ in the effective matrix element of
\refeq{eq:effmatrel}. This yields for the form factors
\begin{equation}
\mathcal{F}^{ef}_{ij}(s = \MZ^2)= g^e_{i} g^f_{j}, 
\label{eq:factorisation}
\end{equation}
where the couplings $g^f_{\{V,A\}}$ have the loop expansion
\begin{eqnarray}
&&g^f_{V}= g^f_{v,(0)}\left[1 + g^f_{v,(1)}+ g^f_{v,(2)}+\dots\right],\nn\\
&&g^f_{A}= g^f_{a,(0)}\left[1 + g^{f}_{a,(1)}+ g^f_{a,(2)}+\dots\right] .
\label{effcouplingsexpanded}
\end{eqnarray}
We use indices $\{V,A\}$ in capitals to label generic vector and axial
vector couplings which contain all higher order terms. 
In contrast, lower case $\{v,a\}$ is used for couplings of a specific
loop order. The couplings $g^f_{\{v,a\},(0)}$ in the above equation 
are thus the lowest-order couplings from
\refeq{borncouplvecax}. The terms $g^f_{\{v,a\},\{(1),(2),\dots\}}$
represent loop corrections of order~$\{(1),(2),\dots\}$.  
Note that, working in our conventions, the Born level couplings
$g_{f,\{v,a\}}^{(0)}$ are factored out in \refeq{effcouplingsexpanded}
and are hence not contained in the higher order terms
$g^f_{\{v,a\},\{(1),(2),\dots\}}$.

\begin{figure}[htb!]
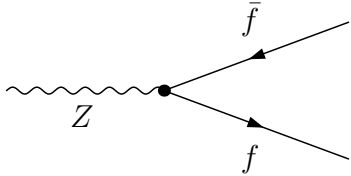

\centerline{
\unitlength=1.cm%
%
%
\begin{feynartspicture}(4,4)(1,1)
\FADiagram{}
\FAProp(-5.,10.)(6.5,10.)(0.,){/Sine}{0}
\FALabel(0.5,8.93)[t]{$Z$}
\FAProp(6.5,10.)(20.,15.)(0.,){/Straight}{-1}
\FALabel(13.5,14)[br]{$\bar f$}
\FAProp(6.5,10.)(20.,5.)(0.,){/Straight}{1}
\FALabel(13.5,6)[tr]{$f$}
\FAVert(6.5,10.){0}
\end{feynartspicture}
}
\vspace{-1em}
\caption{$Z$~boson decay to fermions at Born level. 
} 
\label{fig:ZDecayBorn}
\end{figure}

The $Z$~boson decay, which almost entirely proceeds through the decay into two
fermions (the Born level graph is depicted in \reffi{fig:ZDecayBorn}) 
\begin{equation}
Z\to f \bar f,
\label{ZDecayGeneric}
\end{equation}
is with these prerequisites of the form
\begin{equation}
\mathcal{M}^{\textup{eff}}_{Z\bar f f}=\bar u_f \ga_\al \left[ g^f_V
  - g^f_A \ga_5 \right] v_f \epsilon_Z^\al,
\end{equation}
where $\bar u_f, v_f$  are the Dirac spinors of 
the fermion anti-fermion pair 
and $\epsilon_Z^\al$ is the polarisation vector of the $Z$~boson.


\subsection{Observables at the $Z$~boson resonance}
\label{ZPoleObs}

The $Z$~pole (pseudo-) observables determined from the measurements at
the $Z$~boson resonance in $e^+e^-$ annihilation allow to perform
precision tests of the SM~\cite{Bardin:1997xq} and the
MSSM~\cite{PomssmRep}.  
The most prominent $Z$~pole observables, some of which are interrelated
(see the discussion in the subsections below), 
can be classified as follows:
\begin{itemize}
\item[(a)] inclusive quantities:
  \begin{itemize}
   \item[$-$] the partial leptonic and hadronic decay widths
                    $\Ga_f$,
   \item[$-$] the total decay width $\Ga_Z$,
   \item[$-$] the hadronic peak cross section $\si^0_{\rm had}$,
   \item[$-$] the ratio of the hadronic to the electronic decay
                  width of the $Z$~boson, $R_l$,
   \item[$-$] the ratio of the partial decay width for 
                    $Z\to c\bar{c} \, ( b \bar{b} )$ 
                   to the hadronic width, $R_{c(b)}$.
  \end{itemize}
\item[(b)] asymmetries and effective  fermionic weak mixing angles:
  \begin{itemize}
   \item[$-$] the \it forward-backward  \rm pole asymmetries 
                    $A_{\rm FB}^{0,f}$,
   \item[$-$] the \it left-right  \rm pole  asymmetries 
                    $A_{\rm LR}^{0,f}$,
   \item[$-$] the effective fermionic weak mixing angles $\sweff^{f}$.
  \end{itemize}
\end{itemize}
The quantities that can be measured directly in collider experiments are
energy dependent cross sections~$\sigma^f(s)$, 
left-right~$A^f_{\textup{LR}}(s)$ and forward-backward
asymmetries~$A^f_{\textup{FB}}(s)$ 
of the processes 
$e^+e^-\to (\gamma,Z)\to f\bar f (n\gamma)\ / \ f \bar f(ng)$ 
and the corresponding Bhabha reactions for ${f=e}$. 
Applying deconvolution (unfolding) procedures to these quantities,
pseudo-observables at the $Z$~resonance
such as masses, partial widths, pole asymmetries and effective mixing
angles can be obtained.
The deconvolution procedure is performed under certain assumptions, in
the present case the $Z$~pole approximation, with the support of
sophisticated programs such as {\tt ZFitter}~\cite{ZFitter} or
{\tt TopaZ0}~\cite{TopaZ0}. 
This means in particular that by applying the $Z$~pole approximations
none of the information from the theory side is lost, but is already
accounted for when the pseudo-observables are calculated from the raw
data.  


\subsubsection{Effective mixing angles}
\label{subsec:effma}
The effective electroweak mixing angles of fermions $f$, in particular
leptons, are of great relevance for testing the SM
and its extensions. They have been measured with high accuracy,
while the theory predictions  sensitively depend on the model under
consideration. 
The effective electroweak mixing angle of a
fermion~$f$ is defined as 
\begin{equation}
4 |Q_f| \sweff^f := 1 - \textup{Re}\left[\frac{g^f_V}{g^f_{A}}\right] .
\label{effmixangledef}
\end{equation}
This definition allows to absorb all higher-order corrections into a
quantity $\De \kappa_f$
(similar to the case of the quantity $\De r$ in the
context of the muon decay, cf.\ \citere{MWweber}),  
\begin{eqnarray}
 \sweff^f = \frac{1}{4 |Q_f| }\Bigg(1 -
 \textup{Re}\left[\frac{g^f_{V}}{g^f_{A}}\right]\Bigg)=\sw^2
 \textup{Re}\left[\kappa_f\right]=\sw^2(1+ \textup{Re}\left[\De
   \kappa_f\right]), 
\end{eqnarray}
which expanded up to two-loop order reads
\begin{equation}
\De \kappa_f = \left(g^f_{v,(1)}-g^f_{a,(1)}\right) +
\left(g^f_{v,(2)}-g^f_{a,(2)}+g^f_{a,(1)}\left(g^f_{a,(1)}-g^f_{v,(1)}\right)\right)+\dots \ . 
\end{equation}
The theoretical prediction for the quantity $\De\kappa_f$ can be 
obtained in different models, depending on
all free parameters of the model through virtual corrections,
\begin{equation}
\De \kappa_f = \De \kappa_f (\MW,\MZ,\mt,\alpha,\alpha_s,\ldots,X),
\label{MWit}
\end{equation}
where
\begin{eqnarray}
                  && X = \MHSM ~(\SM) , \nn\\ 
                  && X= \Mh, \MH, \MA,  \MHp, \tb, \msusy, A_f,
    m_{\tilde\chi^{0,\pm}}, \dots~\textup{(MSSM)} . \nn
\end{eqnarray}
Furthermore the  prefactor 
\begin{equation}
 \sw^2=1-\frac{\MW^2}{\MZ^2}=1-
\frac{\MW^2(\MZ,\mt,\alpha,\alpha_s,\ldots,X)}{\MZ^2},
\end{equation}
itself depends on the experimental input $\MZ,\mt,\alpha,\alpha_s,\ldots$ 
and the unknown model parameters $X$. This is due to the fact that the 
theory prediction for $\MW$ is employed,
which is  obtained from muon decay by trading the
precisely measured Fermi constant $\gf$ as input for $\MW$ (see
\citere{MWweber} for the most up-to-date prediction for $\MW$ in the
MSSM).


\subsubsection{Asymmetries and asymmetry parameters}
Though directly related to the effective mixing angles, there
are further observables which are often referred to in the
literature. For completeness we include them in our discussion.

We start with the asymmetry parameters 
${\mathcal A}_f$.   
They are by definition given in terms of the real parts of the ratios of
the effective vector and axial vector couplings, in contrast to the
partial widths that include the full complex effective
couplings,
\begin{equation}
{\mathcal A}_f = 2 \frac{\re\frac{g^f_{V}}{g^f_{A}}}{1
  +\big[\re\frac{g^f_{V}}{g^f_{A}}\big]^2 }. 
\label{assypara}
\end{equation}
The asymmetry parameters ${\mathcal A}_f$ are obviously directly
correlated with the effective mixing angles defined in
\refeq{effmixangledef}. 
They  can be used to calculate the forward-backward pole asymmetries
$A^{0,f}_{\textup{FB}}$ and the left-right pole asymmetries
$A^{0,f}_{\textup{LR}}$,
\begin{eqnarray}
A^{0,f}_{\textup{FB}} = \frac{3}{4}{\mathcal A}_e{\mathcal A}_f,\\
A^{0,f}_{\textup{LR}} = {\mathcal A}_e.
\end{eqnarray}
Note that the $A^{0,f}_{\rm FB,LR}$ are theoretical, 
to a certain level artificial,
quantities (pseudo-observables in the terminology
of~\citere{Bardin:1997xq}). 
Besides the unfolding from the QED corrections,
also small effects from $\gamma$-$Z$ interference and pure 
$\gamma$ exchange have to be taken into account, as mentioned 
in the beginning of \refse{ZPoleObs}. Their use is justified
by their dominance at the $Z$ peak, by their
simple parametrizations in terms of the effective couplings
resp.\ mixing angles, and by the practically
model-independent correction terms relating them to realistic observables.  
For a detailed discussion, see also~\citere{Hollik:2000ap}.


\subsubsection{Decay widths}
\label{subsec:dcw}
The partial decay width for the decay into fermions $f$ is given by
(cf.\ \citere{Bardin:1997xq})
\begin{eqnarray}
\Ga_f = N_c^f \frac{\alpha}{3} M_Z \left(\left|g^f_{V}\right|^2   R_V^f +
  \left|g^f_{A}\right|^2  R_A^f\right), 
\label{eq:Width}
\end{eqnarray}
where the radiation factors $R^f_{V,A}$ describe the final state QED
and QCD interactions and account for the fermion masses $m_f$, see
\refeq{leptradfac} below. The latter contributions are of particular 
relevance for
the decay into bottom-quarks~\cite{KuhnQCD}. 

Equivalently the partial decay width can also be expressed as
(cf.\ \citere{Hollik:2000ap}) 
\begin{equation}
\Ga_f= N_c^f \bar\Ga_0\left|\rho_f\right|
  \left(4(I^f_3-2Q_f\sw^2\left|\kappa_f\right|)^2 R_V^f +R_A^f\right), 
\label{eq:WidthalaHollik}
\end{equation}
with
\begin{equation}
\bar\Ga_0=\frac{\gf M_Z^3}{24\sqrt{2}\pi}.
\end{equation}
The factor $\sqrt{\rho_f}$ normalises the overall matrix element to 
$Z$~boson decay. The quantity $\kappa_f$ is a measure for the relative 
coupling strength of the effective vector and axial vector couplings, see
\refse{subsec:effma}.  
The effective couplings $g^f_{V,A}$ used in \refeq{eq:Width} and the
quantities $\rho_f$ and $\kappa_f$ from \refeq{eq:WidthalaHollik} are
related via 
\begin{eqnarray}
&&\rho_f=\frac{1}{1+\De
  r}\left((g_{a,(0)}^{f})^{-1}g^f_{A}\right)^2=\frac{1}{1+\De
  r}(1+g^f_{a,(1)}+g^f_{a,(2)}+\dots)^2,\nn\\ 
&&1-4 |Q_f|\sw^2\kappa_f =\frac{g^f_{V}}{g^f_{A}}.
\label{eq:kapparhogagv}
\end{eqnarray}
It is convenient to express
$\rho_f$ and $\kappa_f$ in terms of universal
corrections (that are independent of the fermion species) and a 
non-universal part
(depending on the fermion species),
\begin{eqnarray}
&&\rho_f=1+\De \rho_f=1+\De \rho_{\textup{univ}}+\De
\rho_{f,\textup{non-univ}},\nn\\
&&\kappa_f=1+\De \kappa_f=1+\De \kappa_{\textup{univ}}+\De
\kappa_{f,\textup{non-univ}} .
\end{eqnarray}
The universal part arises from vector-boson self energies and
fermion-independent counterterms, and the non-universal part 
from vertex corrections and fermion-dependent counterterms 
(see e.g.\ eq.~(\ref{effcoupl1L})). The leading contribution to the
universal corrections arises from mass splitting in isospin
doublets, like an isospin
doublet of a given family of SM fermions or MSSM sfermions,
entering via the quantity $\De \rho$~\cite{rho} according to
\begin{eqnarray}
&&
\De \rho_{\textup{univ}}=\De\rho+\dots \ ,\nn\\
&&
\De \kappa_{\textup{univ}}=\frac{\cw^2}{\sw^2}\De\rho+\dots \ ,
\label{leadingirredintozobs}
\end{eqnarray}
where
\begin{equation}
\De \rho = \frac{\Si^{ZZ}(0)}{\MZ^2} - \frac{\Si^{WW}(0)}{\MW^2} 
\label{eq:delrho}
\end{equation}
contains the transverse parts of the $Z$ and $W$~boson self energies
evaluated at vanishing momentum-squared.
In the approximation where all fermion masses except the top-quark mass
are neglected ($\mt\gg\mb,m_\tau,\dots$), $\De\rho$ in the SM, 
at the on-loop level, is given by the simple expression
\begin{equation}
\De\rho=N_c \frac{\al}{16\pi\sw^2\cw^2}\frac{\mt^2}{\MZ^2}.
\label{DeRhoSM1LAl}
\end{equation}
Expressing the weak coupling in terms of the Fermi constant $\gf$
rather than the fine-structure constant $\al$ yields
\begin{equation}
\De\bar\rho=N_c\frac{\gf\mt^2}{8\pi^2\sqrt{2}},
\label{DeRhoSM1LGf}
\end{equation}
where the bar is used to distinguish the two parametrisations.
Leading reducible two-loop terms can be obtained by performing
the substitutions~\cite{CHJ} 
\begin{eqnarray}
&&\rho_f=1+\De
\rho+\dots\to\frac{1}{1-\De\bar\rho}+\dots\ ,\nn\\  
&&\kappa_f=1+\frac{\cw^2}{\sw^2}\De\rho\dots\to
1+\frac{\cw^2}{\sw^2}\De\bar\rho+\dots~.                               
\label{eq:rhokappaexp}
\end{eqnarray}
It should be noted that the substitutions in \refeq{eq:rhokappaexp}
replace $\De\rho$ (parametrised in terms of $\al$) by $\De\bar\rho$ 
(parametrised in terms of $\gf$). 
This introduces additional terms at higher orders according to
\begin{equation}
\De\rho(\al) \to \De \bar{\rho} (\gf) = (1+\De r)\De \rho(\al),
\label{eq:Rhoaltorhogf}
\end{equation}
which follows from the explicit expressions given in
\refeq{DeRhoSM1LAl} and \refeq{DeRhoSM1LGf}. 

It was already mentioned that the $Z$ predominantly decays into fermion
pairs. Contributions from other decay modes such as $Z\to
\gamma\gamma\gamma,ggg,b\bar b b \bar b$ are
insignificant~\cite{pdg}. 
In the SM the total width $\Ga_{Z}$ is thus given as the sum of the
leptonic width $\Ga_l$ and the hadronic width $\Ga_h$,
\begin{equation}
\Ga_{Z}^{\textup{SM}}= \Ga_l + \Ga_h,
\end{equation}
where
\begin{equation}
\Ga_l = \Ga_{\textup{inv}} + \Ga_e + \Ga_{\mu} + \Ga_{\tau} ,
\label{gammal}
\end{equation}
and
\begin{equation}
\Ga_h = \Ga_{u} + \Ga_{d} +  \Ga_{c} +  \Ga_{s} +  \Ga_{b}
\label{gammah}
\end{equation}

The invisible width $\Ga_{\textup{inv}}$ in the SM used
in \refeq{gammal} arises from the partial widths of $Z$~boson decays
into neutrinos,
\begin{equation}
\Ga_{\textup{inv}} = \Ga_{\nu_e} + \Ga_{\nu_\mu}
+ \Ga_{\nu_\tau}. 
\end{equation}
In the MSSM further contributions to the invisible width 
can arise from the decay into the 
lightest neutralino (if this particle does not decay further within the
detector).  
We therefore include the process $Z\to \neu{1}\neu{1}$
into our computations, where $\neu{1}$ is the lightest
neutralino.
This decay channel of the $Z$~boson is open if
$\mneu{1}<\MZ/2$. 
The total $Z$~boson width in the MSSM thus reads 
\begin{equation}
\Ga_Z=\Ga_l+\Ga_h+\Gamma_{\neu{1}},
\end{equation}
where the partial width $\Gamma_{\neu{1}}$ is in analogy to
\refeqs{eq:Width},~(\ref{eq:WidthalaHollik}) given by 
\begin{eqnarray}
\Ga_{\neu{1}} = \frac{\alpha}{3} 
               M_Z \left|g^{\neu{1}}_A\right|^2  R_A^{\neu{1}}, \ \ \ \ \ 
\Ga_{\neu{1}}=  \bar\Ga_0|\rho_{\neu{1}}|R_A^{\neu{1}},
\label{eq:WidthNeu1}
\end{eqnarray}
with the radiation factor $R_A^{\neu{1}}$ and the effective axial vector
coupling $g^{\neu{1}}_A$ specified below. 
$g^{\neu{1}}_A$~is related to $\rho_{\neu{1}}$ via
\refeq{eq:kapparhogagv}, with $f\to\neu{1}$. 
As neutralinos are described in terms of Majorana fermions, the
coupling of $\neu{1}$ to the  $Z$~boson does not contain a vector part. 


\subsubsection{Peak cross-sections and ratios of partial widths}
\label{subsec:peakcross}
The cross-section for the process
\refeq{epemtofermferm} is given by  
\begin{equation}
\sigma^0_f = 12 \pi \frac{\Ga_e \Ga_f}{\MZ^2 \Ga_Z^2},
\label{peakcross}
\end{equation}
at the $Z$~boson resonance, i.e.~$s=\MZ^2$.
The hadronic peak cross section is obtained in complete analogy as
\begin{equation}
\sigma^0_{\textup{had}} = 12 \pi \frac{\Ga_e \Ga_h}{\MZ^2 \Ga_Z^2},
\label{peakcrosshad}
\end{equation}
 the only difference being that all decay modes into hadronic final-state 
particles are considered. 
Furthermore certain ratios of partial widths are often analysed,
\begin{eqnarray}
R_l &=& \frac{\Ga_h}{\Ga_e},\\
R_{b,c} &=& \frac{\Ga_{b,c}}{\Ga_h},
\end{eqnarray}
with the partial leptonic (electronic) $\Ga_e$ and 
hadronic $\Ga_b, \Ga_c$ widths from
\refeq{gammal} and \refeq{gammah}, as well as the partial decay widths
to charm $\Ga_c$ and bottom-quarks $\Ga_b$.


\section{Calculation of effective couplings and pseudo 
    observables} 
\label{sec:calceffcoup}

\subsection{Complete one-loop result in the complex MSSM}
\label{subsubsec:oneloop}
In order to calculate the $Z$~pole observables, one of the main tasks is to
evaluate the effective couplings $g^f_{\{\textup V,A\}}$.  
This requires the computation of the $Zf\bar f$ vertex graphs (see
\reffis{fig:vertexoneloop}, \ref{fig:vertexSUSYQCDoneloop}, 
\ref{fig:vertexoneloopNeu1} given in the Appendix) 
and  the $Z\ga$~self energy,
as well as the evaluation of the respective counterterms. 
Expressed in terms of unrenormalised vertex graphs and on-shell counterterms 
the effective  couplings expanded up to one-loop order are found to be
\begin{eqnarray}
\left(
g_A^{f}
\right)^{(\al)} 
&=& g^f_{a,(0)}\left[ 1+ g^f_{a,(1)}\right] \nn\\
&=&  g^f_{a,(0)}
\left[
1+
\frac{\textup{(vertex)}^f_{a}}{g^f_{a,(0)}}+ \frac{\de e}{e} + \frac{\de
  \sw^2}{\sw^2} 
\frac{\sw^2-\cw^2}{2\cw^2}
\right.
\nn\\ 
&&\left.
+\frac{1}{2}\de Z^{ZZ} +
\frac{\frac{\sw}{\cw}Q_f\frac{1}{2}\left(\de Z^{f}_R+\de
  {Z^{f}_R}^\dag\right)+\frac{I^f_3-\sw^2 Q_f}{\sw\cw}\frac{1}{2}\left(\de
  Z^{f}_L+\de {Z^{f}_L}^\dag\right)}{2 g^f_{a,(0)}}
\right],\nn\\ 
\left(
g_V^{f}
\right)^{(\al)} 
&=& g^f_{v,(0)}\left[ 1+ g^f_{v,(1)}\right] \nn\\
&=&
g^f_{v,(0)}
\left[
1+
\frac{\textup{(vertex)}^f_{v}}{g^f_{v,(0)}} 
+ \frac{\de e}{e} - \frac{\de
  \sw^2}{\sw^2} \frac{-I^f_3 \sw^2 + 2 Q_f \sw^4 + \cw^2 \left(I^f_3 + 2 Q_f
  \sw^2\right)}{2\cw^2\left(I^f_3 - 2 Q_f \sw^2\right)} 
\right.
\nn\\ 
&& 
\left.
+\frac{1}{2}\de Z^{ZZ} +
\frac{-\frac{\sw}{\cw}Q_f\frac{1}{2}\left(\de Z^{f}_R+\de
  {Z^{f}_R}^\dag\right)+\frac{I^f_3-\sw^2 Q_f}{\cw\sw}\frac{1}{2}\left(\de
  Z^{f}_L+\de {Z^{f}_L}^\dag\right)}{2 g^f_{v,(0)}}
\right.
\nn\\ 
&&
\left.
-\frac{Q_f}{g^f_{v,{(0)}}}\left( \frac{\Sigma^{\ga
    Z}(\MZ^2)}{\MZ}+\frac{1}{2}\de Z_{\ga Z}\right)\right],  
\label{effcoupl1L}
\end{eqnarray}
where (vertex)$^f_{a/v}$ stands for the axial/vector part of the
unrenormalised vertex graphs, schematically depicted in
\reffis{fig:vertexoneloop} and \ref{fig:vertexSUSYQCDoneloop}. 
The explicit form of the on-shell counterterms in~\refeq{effcoupl1L},  
expressed in terms of self-energies,
can be found in \citere{DennerHabil}
(we use the conventions given therein). 
They have the same structure as in the SM, 
but the self-energies have to augmented by the non-standard
contributions, accordingly.
The charge renormalisation counterterm, $\de e$, receives contributions 
of large logarithms from light fermions, giving rise to
a numerically sizable shift in the fine-structure constant,
$\De\al=\De\al^{(5)}_{\textup{had}}+\De\al_{\textup{lept}}$.
The  correction $\De\rho$, see \refeq{eq:delrho}, enters via the
counterterm to the weak mixing angle, $\de \sw^2$.

Computing the occurring Feynman graphs 
we include the complete set of MSSM one-loop
diagrams, keeping the full complex phase dependence. 
All relevant Feynman graphs are calculated making use of the packages
{\code FeynArts}~\cite{feynarts} and {\code FormCalc}~\cite{formcalc}.  
As regularisation scheme dimensional reduction~\cite{dred} is used,
which allows a mathematically consistent treatment of
UV~divergences in supersymmetric theories at the one-loop level.
Our results are presented in the same conventions as in
\citere{MWweber}. The higher-order corrections in the Higgs
sector were implemented into our predictions for the $Z$~pole
observables and $\MW$ as described in \refse{sec:higgs}.  
Concerning the light quarks and leptons, in our calculations we take
into account the masses of the bottom-quark and the tau-lepton, while the
masses of the other leptons and light quarks are neglected (except for
the contributions giving rise to the shift $\De\al$ in the fine
structure constant, as discussed above). 
As a consequence of keeping a non-zero bottom-quark and tau-lepton mass,
in addition to the terms given in
\refeq{eq:effmatrel} the matrix elements in our calculations 
contain also contributions 
$\propto \bar u_{\tau,b}(k_1) u_{\tau,b}(k_2)(k_{1,2}\epsilon_Z)$
and
$\propto \bar u_{\tau,b}(k_1) \gamma_5 u_{\tau,b}(k_2)(k_{1,2}\epsilon_Z)$.
We checked by explicit computation that these terms are numerically
negligible. 

\medskip
As mentioned above, the decay channel $Z \to \neu{1} \neu{1}$
can provide additional contributions to the
invisible width of the $Z$~boson and thus has to be taken into account.
In order to obtain results at the same level of precision as for the
leptonic and hadronic observables,
we have performed for the first time
a full MSSM one-loop calculation for the decay 
$Z \to \neu{1} \neu{1}$.
At Born level the axial vector couplings are given by (there are no further
contributions from vector couplings owing to the Majorana nature of
the neutralino)
\begin{eqnarray}
g^{\neu{1}}_{a,(0)}=\frac{2N_{1, 4}N_{1, 4}^\ast
                         -2N_{1, 3}N_{1,3}^\ast}{4\cw\sw}. 
\end{eqnarray}
Expanded up to \order{\al} this turns into the effective axial vector coupling
\begin{eqnarray}
\left(g^{\neu{1}}_{A}\right)^{(\al)}=&& \frac{1}{4 \cw\sw}
\left[
{\left(2N_{1, 4}N_{1, 4}^\ast-2N_{1, 3}N_{1, 3}^\ast\right)} 
\right.
\nn\\
&&
\left.
+ 4 \cw\sw \textup{(vertex)}^{\neu{1}}_a+\frac{\de\sw^2}{\sw^2}
\frac{\cw^2 - \sw^2}{\cw^2} ( N_{1, 3} N_{1, 3}^\ast - N_{1, 4}
N^\ast_{1, 4})
\right.
\nn\\ 
&& 
\left.
+ \frac{\de e}{e}    ( 2 N_{1, 4} N_{1, 4}^\ast - 2 N_{1, 3} N_{1,
  3}^\ast) 
 + \de Z_{ZZ} (  N_{1, 4} N_{1, 4}^\ast - N_{1, 3} N_{1, 3}^\ast
 )
\right.
\nn\\  
&& 
\left.
+ \de  Z^{\neu{1,1}}_L         ( N_{1, 4} N_{1, 4}^\ast -
N_{1, 3} N_{1, 3}^\ast ) 
+ {\de Z^{\neu{1,1}}_L}^\dag      ( N_{1, 4} N_{1, 4}^\ast -
N_{1, 3} N_{1, 3}^\ast ) 
\right.
\nn\\ 
&& 
\left.
+ {\de Z^{\neu{2,1}}_L}        ( N_{1, 4} N_{2, 4}^\ast -
N_{1, 3} N_{2, 3}^\ast ) 
+ {\de Z^{\neu{2,1}}_L}^\dag      ( N_{2, 4} N_{1, 4}^\ast -
N_{2, 3} N_{1, 3}^\ast ) 
\right.
\nn\\ 
&& 
\left.
+ {\de Z^{\neu{3,1}}_L}        ( N_{1, 4} N_{3, 4}^\ast -
N_{1, 3} N_{3, 3}^\ast ) 
+ {\de Z^{\neu{3,1}}_L}^\dag      ( N_{3, 4} N_{1, 4}^\ast -
N_{3, 3} N_{1, 3}^\ast ) 
\right.
\nn\\ 
&& 
\left.
+ {\de Z^{\neu{4,1}}_L}^\dag   ( N_{4, 4} N_{1, 4}^\ast -
N_{4, 3} N_{1, 3}^\ast )  
+ {\de Z^{\neu{4,1}}_L}           ( N_{1, 4} N_{4, 4}^\ast -
N_{1, 3} N_{4, 3}^\ast ) 
\right]
. 
\label{eq:gchi}
\end{eqnarray}
Here $N_{i,j} (i,j=1\dots4)$ represent the in general complex entries of the
neutralino diagonalisation matrix $\matr{N}_{\tilde \chi^0}$, see \refeq{NNmatrix}.
We use the notation $\textup{(vertex)}^{\neu{1}}_{a}$ to denote the axial part
of the vertex graphs displayed in \reffi{fig:vertexoneloopNeu1}. The
counterterms of charge, electroweak mixing angle, and $Z$~field 
renormalisation are again evaluated in the on-shell scheme. 
For the neutralino field renormalisation constants  ${\de Z^{\neu{i,1}}_L}$ 
we employ the on-shell renormalisation detailed in \citeres{Fritzsche:2002bi,tfphd}.
The parameter $\tb$ is renormalised by imposing a \drbar~condition as specified in~\citeres{tanbetaren0,tanbetaren}.
Analytic expressions for the neutralino field renormalisation constants in \refeq{eq:gchi} can be found in \citere{Fritzsche:2002bi}.  


\subsection{Incorporation of higher-order contributions} 
\label{subsec:higherorders}
Having computed the results for the effective couplings at the one-loop
level (for the first time under consideration of the full complex
parameter dependence, $\cp$ mixing in the Higgs sector, and resummed $\tb$
enhanced Yukawa couplings) we now include the available
higher-order contributions in the SM and the MSSM. As a result, we
obtain the currently most accurate prediction of the $Z$~pole
observables in the MSSM.


\subsubsection{Combining SM and MSSM contributions} 
\label{subsec:higherordersSM}

As mentioned before, the theoretical evaluation of the $Z$~pole observables in
the SM is significantly more advanced than in the MSSM. In
order to obtain the most accurate predictions within the MSSM it is
therefore useful to take all known  SM corrections into
account. This can be done by writing 
the MSSM prediction for a quantity $x=g^f_{V,A},\rho_f,\kappa_f,\dots$ as
\begin{equation}
x^{\rm MSSM} = x^{\rm SM}\big|_{\MHSM = \MHe} + x^{{\rm MSSM} 
 -{\rm SM}}\equiv x^{\rm SM}\big|_{\MHSM = \MHe}+ x^{\textup{SUSY}}, 
\label{eq:obsSMSUSY}
\end{equation}
where $x^{\rm SM}$ is the prediction in the SM with the SM Higgs boson
mass set to the lightest MSSM Higgs boson mass, $\MHe$,
and $x^{{\rm MSSM}-{\rm SM}} \equiv x^{\textup{SUSY}}$  denotes the difference
between the MSSM and the SM prediction.

In order to obtain $x^{\rm MSSM}$ according to
\refeq{eq:obsSMSUSY} we
evaluate $x^{{\rm MSSM}-{\rm SM}}$ at the level of 
precision of the known MSSM corrections, while for $x^{\rm SM}$
we use the currently most advanced result in the SM including all known
higher-order corrections. As a consequence, $x^{\rm SM}$ 
takes into account higher-order contributions which are only
known for SM particles in the loop, but not for their superpartners
(e.g.\ two-loop electroweak corrections to $\De\kappa$ beyond the
leading Yukawa contributions). 

It is obvious that the incorporation of all known SM contributions
according to \refeq{eq:obsSMSUSY} is advantageous in the decoupling
limit, where all superpartners 
are heavy and the Higgs sector becomes SM-like. In this case the 
second term in \refeq{eq:obsSMSUSY} goes to zero, so that the MSSM
result approaches the SM result with $\MHSM = \MHe$.
For lower values of the scale of supersymmetry the contribution from 
supersymmetric particles in the loop can be of comparable size as
the known SM corrections. In view of the experimental bounds on the
masses of the supersymmetric particles (and the fact that supersymmetry
has to be broken), however, a complete cancellation between
the SM and supersymmetric contributions is not expected. 
Furthermore, the leading Yukawa enhanced corrections with MSSM
Higgs boson exchange were found to be very well approximated by the
corresponding Yukawa terms in the SM~\cite{drMSSMal2A,drMSSMal2B}. 
Therefore it seems appropriate to apply \refeq{eq:obsSMSUSY} also for
rather light SUSY particle spectra.


\subsubsection{Universal SM contributions beyond one-loop}
\label{subsubsec:twoloopSM}

It is convenient to parametrise higher-order SM corrections to the 
$Z$~pole observables in terms of the quantities $\rho_f$ and $\kappa_f$,
defined in \refeq{eq:kapparhogagv}. For the universal contribution 
$\De r$, 
entering via $\MW$,
the complete two-loop result is available in the 
SM~\cite{MWSMferm2L,MWSMferm2Lczak,MWSMbos2L,MWSM}. Further reducible 
SM higher-order contributions, appearing for instance as products of
one-loop contributions in two-loop counterterms, can be incorporated 
with the help of substitutions like  \refeq{eq:Rhoaltorhogf}.
Beyond the two-loop order, irreducible higher-order corrections 
have been obtained for $\De \rho$, the leading universal contribution 
from the mass splitting in an isospin doublet (for two-loop
contributions to $\De\rho$, see
\citeres{drSMgfals,deltarSMgfals,drSMgf2mt4,drSMgf2mt4B}). 
Higher-order QCD contributions to $\De\rho$ in the SM are known up to
\order{\al\als^2}~\cite{drSMgfals2}. The \order{\al\als^2} corrections
to $\De r$ are also known~\cite{MWSMQCD3LII}. 
Electroweak three-loop contributions to $\De\rho$
of \order{\mt^6\gf^3} and mixed
electroweak and QCD corrections of \order{\mt^4 \gf^2 \als} have 
been obtained in \citeres{drSMgf3mh0,drSMgf3}.
Most recently even the full class of four-loop \order{\al\als^3}
contributions to $\De\rho$ became available~\cite{derhoalalscube}
and has been included, although the corrections turned out to be 
rather small numerically.


\subsubsection{Universal MSSM two-loop contributions}
\label{subsubsec:twoloop}
Within the MSSM, 
leading irreducible two-loop contributions to the $Z$~pole observables
are only available as universal corrections to $\De\rho$ in the
approximation where complex phases are neglected.
Reducible higher-order terms can be obtained in the same manner as in the
SM. 

Leading irreducible SUSY QCD corrections of \order{\al \als} entering
via the quantity $\De \rho$ arise from the diagrams 
shown in \reffi{fig:samplediagramsQCD}. They involve both 
gluon and gluino exchange in (s)top-(s)bottom loops and were first
evaluated in \citere{dr2lA}.  
Besides the \order{\al \als} contributions,  also the
leading electroweak two-loop corrections of
\order{\al_t^2}, \order{\al_b^2} and \order{\al_t\al_b}
to $\De\rho$ have become available~\cite{drMSSMal2B}.
These two-loop Yukawa coupling contributions are due to
MSSM Higgs and Higgsino exchange in (s)top-(s)bottom-loops, see
\reffi{fig:samplediagramsYuk}. 
In \citere{drMSSMal2B} the dependence of the
\order{\al_{t,b}^2} corrections on the lightest MSSM Higgs boson mass,
$\Mh$, was analysed. Formally, at this order
the approximation $\Mh = 0$ would have to be employed.
However, it was shown in \citere{drMSSMal2B} how a non-vanishing MSSM
Higgs boson mass can be consistently taken into account, including
higher-order corrections. 
Correspondingly we use the result of 
\citere{drMSSMal2B} for arbitrary $\Mh$ and employ the code
\fhtt~\cite{feynhiggs,mhiggsAEC,mhcMSSMlong} for the evaluation of the
MSSM Higgs sector parameters.

The final step is the inclusion of the complex MSSM parameters into 
the two-loop results. So far all generic two-loop results have been
obtained for real input parameters. 
Following \citere{MWweber},
we approximate the two-loop result of an observable
$O_Z=\Ga_f, \Ga_Z, \sweff^f,\dots$ for a certain value of phase 
$\phi$  by a simple interpolation,
based on the full phase dependence at the one-loop level and
the known two-loop results for real
parameters, $O_Z^{\textup{full}}(0)$, $O_Z^{\textup{full}}(\pi)$, 
\BEA
O_Z^{\rm full}(\phi) = O_Z^{\rm 1L}(\phi) 
 &+& \left[ O_Z^{\rm full}(0) - O_Z^{\rm 1L}(0) \right] \times 
     \frac{1+\cos\phi}{2} \non \\
 &+& \left[ O_Z^{\rm full}(\pi) - O_Z^{\rm 1L}(\pi) \right] \times 
     \frac{1-\cos\phi}{2}~.
\label{phases2L}
\EEA
Here $O_Z^{\rm 1L}(\phi)$ denotes the one-loop result, for which the
full phase dependence is known. The factors involving $\cos\phi$ 
ensure a smooth interpolation such that the known results 
$O_Z^{\rm full}(0)$, $O_Z^{\textup{full}}(\pi)$ are recovered for vanishing 
complex phase.
As a check the formula has been applied to the one-loop case. The numerical
difference between the approximated and the full one-loop result was at the
level of a typical SUSY two-loop contribution,
which is expected for a pure one-loop result.


\subsection{Leptonic observables} 
\label{sec:calclepobs}
We now combine the various contributions discussed in the previous
sections in order to obtain results for the leptonic observables 
at the $Z$~boson resonance.
\subsubsection{Effective leptonic weak mixing angle $\sweff$}
\label{subsec:EffmixAng}
In the SM the evaluation of higher-order corrections to
$\De \kappa_l$ and $\sweff$ is far advanced (here and in the following
we use the notation $\sweff\equiv\sin^2\theta_{\rm eff}^l$). 
Recently the complete \order{\al^2} contributions have become
available. 
They include fermionic
contributions~\cite{BayernFermionic,UliMeierFussballgottFermionic},
i.e.\ graphs with at least one 
closed fermion loop, as well as bosonic 
corrections~\cite{UliMeierFussballgottBosonic,dkappaSMbos2L}
without closed fermion loops.  
The  \order{\al\als} and \order{\al\als^2} results can be found in
\citere{MWSMQCD3LII} in terms of a leading top mass expansion. 
Universal three- and four-loop order corrections entering via
$\De\rho$ are incorporated following the prescription in
\refeq{leadingirredintozobs}. 
In total, the state-of-the-art expression for
$\De\kappa^{\textup{SM}}$  can be decomposed into
the following contributions (assuming lepton universality, 
we drop the index $l$ in the following) 
\begin{eqnarray}
\De\kappa^{\textup{SM}} &=& (\De\kappa^{(\al)})^{\textup{SM}} +
(\De\kappa^{(\al^2)})^{\textup{SM}} + (\De\kappa^{(\al\als)})^{\textup{SM}}\nn\\ 
&&+(\De\kappa^{(\al\als^2)})^{\textup{SM}}  
+\frac{\cw^2}{\sw^2}\left(\De\rho^{(\mt^4\gf^2\als)}
+\De\rho^{(\mt^6\gf^3)}+\De\rho^{(\al\als^3)}\right)^{\textup{SM}} .
\label{stateoftheartSMeffangle1}
\end{eqnarray}
Accordingly, the SM prediction for $\sweff$ is given by
\begin{equation}
\sweff^{\textup{SM}}|_{\al+\al^2+\al\als+\al\als^2+\al\als^3+\mt^4\gf^2\als+\mt^6\gf^3}
={(\sw^{\textup{SM}}})^2(1+\De\kappa^{\textup{SM}}).
\label{stateoftheartSMeffangle2}
\end{equation}
As already mentioned in \refse{subsec:effma}, it is crucial
to use the theoretical prediction for $\MW$ to calculate $\sw^2$ and
$\De\kappa$, employing in this way the more precisely measured experimental 
input parameter $\gf$ rather than $\MW^{\textup{exp}}$.
A simple parametrisation formula for
$(\De\kappa^{(\al^2)})^{\textup{SM}}$, which
approximates the full \order{\al^2} result to a precision
even below the anticipated ILC/GigaZ errors,
is given in \citere{dkappaSMbos2L}. In our analysis we use this
result for the SM prediction of $\De\kappa$ as well as 
the currently most precise SM prediction for 
$\MW$~\cite{MWSM}.

Using \refeqs{eq:obsSMSUSY} and (\ref{stateoftheartSMeffangle1}), our result
for $\De\kappa$ in the MSSM reads 
\begin{equation}
\De\kappa \equiv \De \kappa^{\textup{MSSM}}=
\De\kappa^{\textup{SM}}\big|_{\MHSM={M_{h_1}}} +
(\De\kappa^{(\alpha)})^{\textup{SUSY}} 
+\frac{\cw^2}{\sw^2}\left(({\De\rho^{(\al\al_s)}})^{\textup{SUSY}}
+(\De\rho^{(\al_f^2)})^{\textup{SUSY}}\right).
\label{DeltaKappaMSSM}
\end{equation}

In this way the complete \order{\al^2} contribution in the SM is
incorporated, as well
as the higher-order corrections 
which are either only known in the SM
or solely for real MSSM parameters. Accordingly, our prediction for 
$\sweff$ in the MSSM is given by 
\begin{equation}
\sweff|_{\al+\al^2+\al\als+\al\als^2+\al\als^3+\mt^4\gf^2\als+\mt^6\gf^3}
=\sw^2(1+\De\kappa),
\label{leptmixangleMSSM}
\end{equation}
where the most accurate  MSSM prediction for $\MW$~\cite{MWweber} is
used in the calculation of $\sw^2$ and~$\De\kappa$.


\subsubsection{Leptonic decay widths $\Ga_l$}
\label{subsec:LeptWidth}

In our calculations $\Ga_l$, the leptonic decay widths, are obtained
in terms of the form factors $\rho_l,\kappa_l$ as described in
\refse{subsec:dcw}.
The form factor
$\kappa_l$ can be directly related to the leptonic mixing angles from
\refeq{leptmixangleMSSM}.  
The overall normalisation of the decay width, $\rho_l$, is calculated
from the effective one-loop couplings $g^{l,(\al)}_{\{V,A\}}$ as described
in \refse{subsec:dcw}.  
It is supplemented with the leading universal SM and MSSM corrections
from \refse{subsec:higherorders} by applying
\refeq{eq:rhokappaexp}. Subleading corrections of \order{\al\als} are
also available in the literature~\cite{EWObsAlAls}. 
The charge renormalisation counterterm, as mentioned above, contains
the numerically sizable shift in the fine-structure constant,
$\De\al=\De\al^{(5)}_{\textup{had}}+\De\al_{\textup{lept}}$.
As these contributions are accounted for as part of
$g^l_{\{V,A\}}$, they are consequently also incorporated into $\rho_l$. 
In a
final step the above results are supplemented with the radiation
factors $R^l_{\{V,A\}}$. 
For leptonic final states these do not contain 
QCD corrections and are of the simple form (see, for example, \citere{Hollik:2000ap})
\begin{eqnarray}
R_V^l = \sqrt{1-4 \frac{m_l^2}{M_Z^2}} (1+2 \frac{m_l^2}{M_Z^2})  +
Q_l^2 \frac{3}{4} \frac{\al (M_Z)}{\pi},\nn\\ 
R_A^l = \sqrt{1-4 \frac{m_l^2}{M_Z^2}} (1-4 \frac{m_l^2}{M_Z^2})  +
Q_l^2 \frac{3}{4} \frac{\al (M_Z)}{\pi}, 
\label{leptradfac}
\end{eqnarray}
with the running electromagnetic coupling constant $\al$ at the scale 
$\MZ$. The inclusion of lepton masses $m_l$ yields  numerically
negligible effects.


\subsection{Hadronic observables} 
\label{sec:hadobs}
The calculation of the hadronic observables proceeds in a manner very
similar to the leptonic case.  
The discussion is therefore kept very short,
addressing only some additional features in the
hadronic sector.  
\subsubsection{Effective hadronic weak mixing angles $\sweff^q$}
\label{sec:effmixhad}

For the four lightest quarks (u, d, c, s) the SM calculation
is as advanced as in the leptonic sector (see
\refeq{stateoftheartSMeffangle2}) \cite{dkappaSMbos2L}. For the bottom
sector the calculation is more involved, as an additional mass scale
enters the calculation due to top-quark dependent two-loop vertex
graphs.
The resulting additional leading top mass dependent two-loop terms can be
accounted for via~\cite{toptermZbb,drSMgf2mt4B} 
\begin{eqnarray}
\kappa_b=\frac{\kappa_d}{1+\tau_b},
\end{eqnarray}
where
$\kappa_d$ is the form factor of the down-quark. The contribution
$\tau_b$ parametrises
the difference between $\kappa_d$ and $\kappa_b$. It is given
by~\cite{toptermZbb,drSMgf2mt4B}
\begin{equation}
\tau_b=-2x_t\left( 1+ x_t \tau_b^{(2)}\right), \ \ \ \ \ x_t =
\frac{\sqrt{2} \gf \mt^2}{16 \pi^2}. 
\label{eq:Zbbnonuni}
\end{equation}
The explicit expression for $\tau_b^{(2)}$ can be found in
\citere{toptermZbb,drSMgf2mt4B}. 

In our MSSM calculation we include the available higher-order SM and MSSM 
contributions in complete analogy to \refse{subsec:EffmixAng},
see in particular 
\refeq{DeltaKappaMSSM}%
\footnote{For the u,c,d- and s-quarks only parametrisation 
  formulas for $\sweff^q$ are
  available in the literature. We therefore extract
  $\De\kappa_q^{(\al^2)}$ from these formulas by applying the same
  strategy as in \citere{MWweber}.}.
In the hadronic process $Z\to q \bar q$, SUSY QCD contributions enter at the
one-loop level already (see \reffi{fig:vertexSUSYQCDoneloop}). As
explained above, we include these corrections into the form factors.
Numerically the SUSY QCD corrections only play a subleading role. 
The $Zb\bar b$ vertex graphs with virtual Higgs exchange contain
couplings that are enhanced by $\tb$. We resum the leading contributions
as described in \refse{sec:higgs}. 
We have furthermore included the two-loop SM
contributions to $\kappa_b$ of \refeq{eq:Zbbnonuni} (the one-loop terms
contained in \refeq{eq:Zbbnonuni} have been subtracted from the full
MSSM one-loop result to avoid double-counting).


\subsubsection{Hadronic decay widths}
\label{se:hadwiths}
As a first step
to calculate the hadronic partial widths we derive $\rho_q$ from the
effective one-loop couplings. The incorporation of SUSY QCD corrections, 
the resummation of $\tb$-enhanced contributions and the inclusion of
further higher-order corrections proceeds in the same way as described
above.
Leading non-universal corrections to the $Zb\bar b$ vertex are obtained
with~\cite{toptermZbb,drSMgf2mt4B} 
\begin{equation}
 \rho_b = \rho_d (1+\tau_b)^2,
\end{equation}
where  $\tau_b$ again parametrises the difference between down- and
bottom-quark couplings as in~\refeq{eq:Zbbnonuni}.  

The radiation factors $R^q_{V,A}$ are more involved than in the leptonic
case, since they incorporate both final state QED and
QCD interactions, as well as the
bottom-quark mass in the process $Z\to b \bar b$.
For light quarks with $m_q\approx0$ the form factors are of
comparatively simple form~\cite{KuhnQCD}
\begin{equation}
R^q_{V,A} = 1 + Q_q^2\frac{3}{4}\frac{\al(\MZ^2)}{\pi} +
\frac{\als(\MZ^2)}{\pi} + 1.41\left(\frac{\als(\MZ^2)}{\pi}\right)^2 
- 12.8 \left(\frac{\als(\MZ^2)}{\pi}\right)^3-Q_q^2 \frac{1}{4} 
 \frac{\al(\MZ^2)\als(\MZ^2)}{\pi^2}. 
\end{equation}
The explicit expressions for $R^q_{V,A}$, including also the case
$m_q\neq0$,
were implemented as given in \citeres{Bardin:1997xq,KuhnQCD}. 

A further correction to the $Z$~boson width available in the literature
arises from non-factorisable two-loop contributions.
The mixed non-factorisable QCD and
electroweak corrections for u,d,c,s-quarks are taken from
\citere{Czarnecki:1996ei}. For the bottom-quark we use results from
\citere{Fleischer:1999iq}. The non-factorisable corrections can
simply be added to the respective partial hadronic width.


\subsection{Decay width for $Z \to \neu{1} \neu{1}$}
\label{se:Zneuneuwidth}

As discussed above, as final ingredient for the computation of the total 
$Z$~boson width in the MSSM we evaluate the partial width for the process 
$Z\to\neu{1}\neu{1}$.
This is done in analogy to the leptonic case described above.
For $R_A^{\neu{1}}$ we use
\begin{equation}
R_A^{\neu{1}} = \sqrt{1-4 \frac{\mneu{1}^2}{M_Z^2}} 
(1-4\frac{\mneu{1}^2}{M_Z^2}) ,
\end{equation}
which is in accordance with the expression in \refeq{leptradfac} for
$Q_l = 0$.


\section{Numerical analysis}
\label{sec:numanal}

We now present our numerical results for 
the $W$~boson mass, $\MW$, and the most relevant $Z$~boson
observables: the effective leptonic weak mixing angle, $\sweff$,
the total $Z$~boson width, $\Ga_Z$, the ratios for the leptonic and
$b$-quark width of the $Z$~boson, $R_l$ and $R_b$, and the hadronic peak cross
section, $\si^0_{\rm had}$.
We do not explicitly discuss the effective hadronic weak mixing
angles defined in \refse{sec:effmixhad}, which nevertheless enter the hadronic decay
widths via the quantities $\kappa_f$, see
\refeq{eq:WidthalaHollik}~%
\footnote{
All $Z$~observables evaluated in this paper, including also 
forward-backward and left-right asymmetries, have already been 
included in a recent 
$\chi^2$~analysis in the constrained MSSM (CMSSM)~\cite{mastercode}.
}.
We start our numerical discussion 
with a detailed investigation
of the predictions for 
$\MW$, $\sweff$, $\Ga_Z$, $R_l$, $R_b$, and $\si^0_{\rm had}$ 
with respect to the different SUSY masses and
complex phases.
Then we discuss the results for
these observables in the MSSM 
for a choice of sample scenarios and discuss their decoupling
behaviour with respect to the SM limit. Specific scenarios such as the
CPX~scenario~\cite{cpx} and ``Split SUSY''~\cite{split} are
investigated. Finally a scan over all relevant SUSY parameters is
performed. Also the SUSY contributions to the invisible $Z$~boson width
are analysed. 

The numerical analysis of our analytical results for the $Z$~boson
observables, which were calculated as described above, and $\MW$, see
\citere{MWweber}, is performed with the help of a newly developed
Fortran program called  {\tt SUSY-POPE} (SUSY Precision Observables
Precisely Evaluated), which will be made publicly
available~\cite{pope}. 
Though built up from scratch,
for the calculation of the MSSM particle spectrum our code partially
relies on routines which are part of the {\tt FormCalc}~\cite{formcalc}
package. The Higgs sector parameters are obtained from
the program \fhtt~\cite{feynhiggs,mhiggsAEC,mhcMSSMlong}. 

If not stated otherwise, in the numerical analysis below for
simplicity we choose all soft 
SUSY-breaking parameters in the diagonal entries of the sfermion mass
matrices, \refeq{squarkmassmatrix}, to be the same,
\begin{equation}
\msusy  \equiv M_{\tilde F} = M_{\tilde F'} = \ldots~.
\label{eq:Msusy}
\end{equation}
In the chargino/ neutralino sector the GUT relation 
\begin{equation}
M_1 = \frac{5}{3} \frac{\sw^2}{\cw^2} M_2 
\label{eq:GUT}
\end{equation}
(for real values of $M_1$ and $M_2$) is often used to reduce the number 
of free MSSM
parameters. We have kept $M_1$ as a free
parameter in our analytical calculations, but will use the GUT relation 
to specify $M_1$ for our numerical analysis if not stated otherwise.

We have fixed the SM input parameters as
\BE 
\begin{aligned}
G_{\mu} &= 1.16637\times 10^{-5}, & \MZ &= 91.1875 \gev, & 
                                    \als(\MZ) &= 0.118 , \\
\alpha &= 1/137.03599911, & 
\De \al^{(5)}_{\textup{had}} &= 0.02758
\textup{\cite{DeltaAlfaMartin,Burkhardt:2005se}}, &  
\De \al_{\textup{lep}} &= 0.031498 \textup{\cite{DeltaAlfaStein}}, \\ 
\mt   &= 170.9 \gev \textup{\cite{mt1709}
} , &
\mb   &= 4.7 \gev, &
m_\tau&= 1.777\gev        , &\\
& & & & m_c &= m_s = \ldots = 0 .
\end{aligned} 
\label{eq:inputpars}
\EE
For the bottom-quark mass, $\mb$, 
we list the pole mass given above as a scheme-independent reference 
point.
Following \citere{Bardin:1997xq}, in the numerical calculations we use the corresponding running bottom quark
mass $\overline{m}_b(\MZ)\approx 2.8\gev$.

The results for physical observables are affected only
by certain combinations of the complex phases of the 
parameters $\mu$, the trilinear couplings $\At$, $\Ab$, \ldots, and the
gaugino mass parameters $M_1$, $M_2$,
$M_3$~\cite{MSSMcomplphasen,SUSYphases}.
It is possible, for instance, to rotate the phase $\phi_{M_2}$ away.
Experimental constraints on the (combinations of) complex phases 
arise in particular from their contributions to electric dipole moments of
heavy quarks~\cite{EDMDoink}, of the electron and 
the neutron (see \citeres{EDMrev2,EDMPilaftsis} and references therein), 
and of deuteron~\cite{EDMRitz}. While SM contributions enter 
only at the three-loop level, due to its
complex phases the MSSM can contribute already at one-loop order.
Large phases in the first two generations of (s)fermions
can only be accommodated if these generations are assumed to be very
heavy~\cite{EDMheavy} or large cancellations occur~\cite{EDMmiracle},
see however the discussion in \citere{EDMrev1}. 
Accordingly (using the convention that $\phi_{M_2} =0$, as done in this
paper), in particular 
the phase $\phi_\mu$ is tightly constrained~\cite{plehnix}, 
while the bounds on the phases of the third generation
trilinear couplings are much weaker.


\subsection{MSSM parameter dependence}
\label{subsec:MSSMpara}

We start by comparing our full MSSM result for the EWPO ($\MW$, $\sweff$,
$\Ga_Z$, $R_l$, $R_b$ and $\si^0_{\rm had}$) with the current experimental
results, which are listed in \refta{tab:expacc}. 
In the following we present results where 
the most important SM
and SUSY parameters are varied in order to identify the observables
with the highest sensitivity to SUSY loop effects.

\begin{table}[htb!]
\renewcommand{\arraystretch}{1.5}
\BC
\begin{tabular}{|c||c|c|c|c|}
\hline\hline
observable & central exp.\ value & $\si \equiv \si^{\rm today}$ &
             $\si^{\rm LHC}$ & $\si^{\rm ILC}$ \\ \hline \hline
$\MW$ [GeV] & $80.398$ & $0.025$ & $0.015$  & $0.007$ \\ \hline
$\sweff$    & $0.23153$ & $0.00016$ & $0.00020$--$0.00014$ & $0.000013$ 
                                                      \\ \hline
$\Ga_Z$ [GeV] & $2.4952$ & $0.0023$ & --- & 0.001 \\ \hline
$R_l$         & $20.767$ & $0.025$  & --- & 0.01 \\ \hline
$R_b$         & $0.21629$ & $0.00066$ & --- & 0.00014 \\ \hline
$\si^0_{\rm had}$ & $41.540$ & $0.037$ & --- & $0.025$ \\
\hline\hline
\end{tabular}
\EC
\renewcommand{\arraystretch}{1}
\caption{Summary of the electroweak precision observables that will 
be analysed in the following,
including their current experimental central values and 
  experimental errors, 
$\si \equiv \si^{\rm today}$~\cite{LEPEWWG,LEPEWWG2,TEVEWWG}. 
Also shown are the anticipated
experimental accuracies at the LHC, $\si^{\rm LHC}$ and at the ILC
(including the GigaZ option), $\si^{\rm ILC}$. Each number represents
the combined results of all detectors and channels at a given collider,
taking into account correlated systematic uncertainties, see
\citeres{blueband,PomssmRep,gigaz,moenig} for details. 
A recent review can be found  in \citere{Erler:2007sc}.
Non-existing analyses are referred to as ``---''.
}
\label{tab:expacc}
\end{table}


\subsubsection{Dependence on the sfermion mass scale}
\label{subsec:Msusy}
\begin{figure}[htb!]
\begin{center}
\includegraphics[width=7.7cm,height=6.4cm]{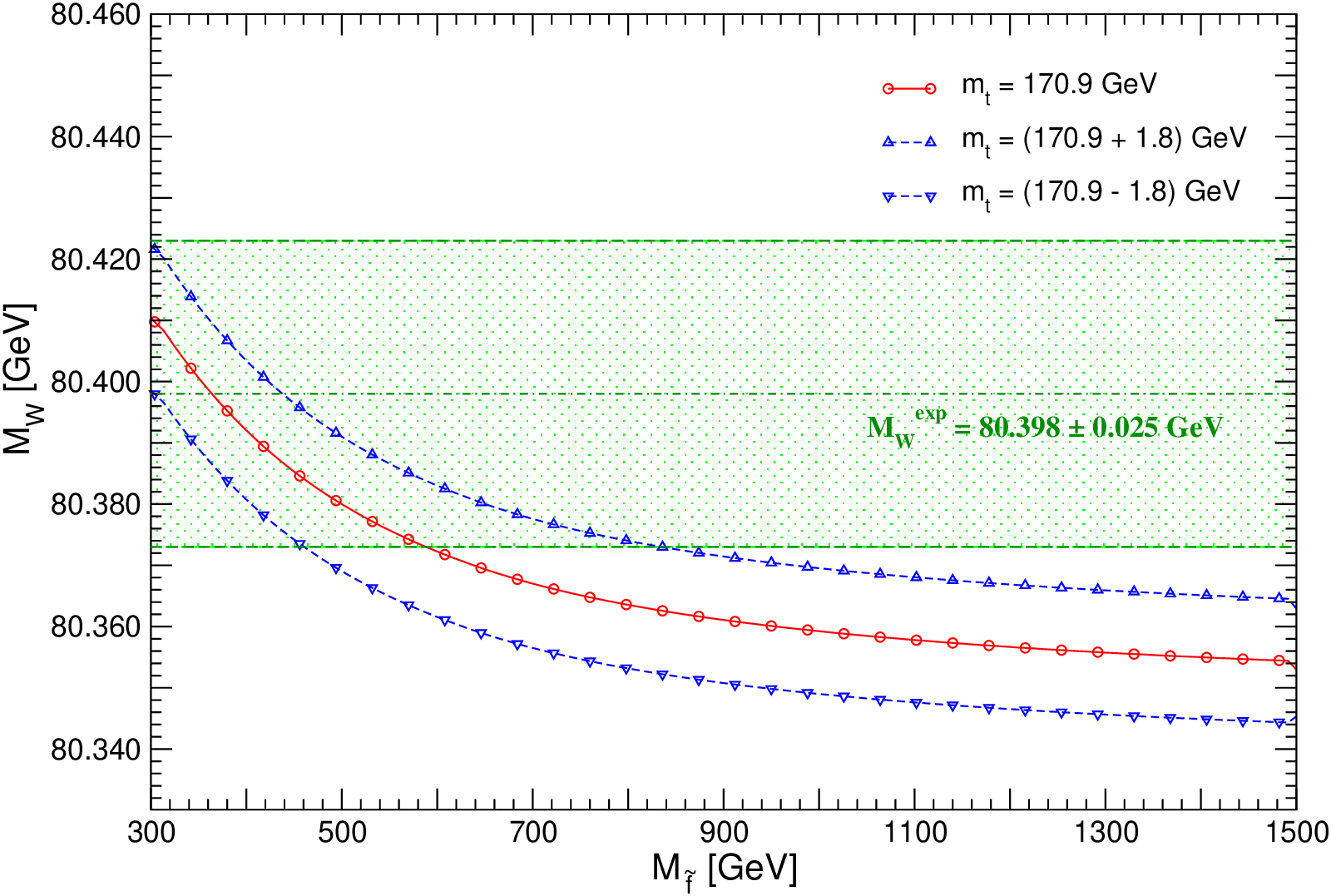}
\hspace{.3em}
\includegraphics[width=7.7cm,height=6.4cm]{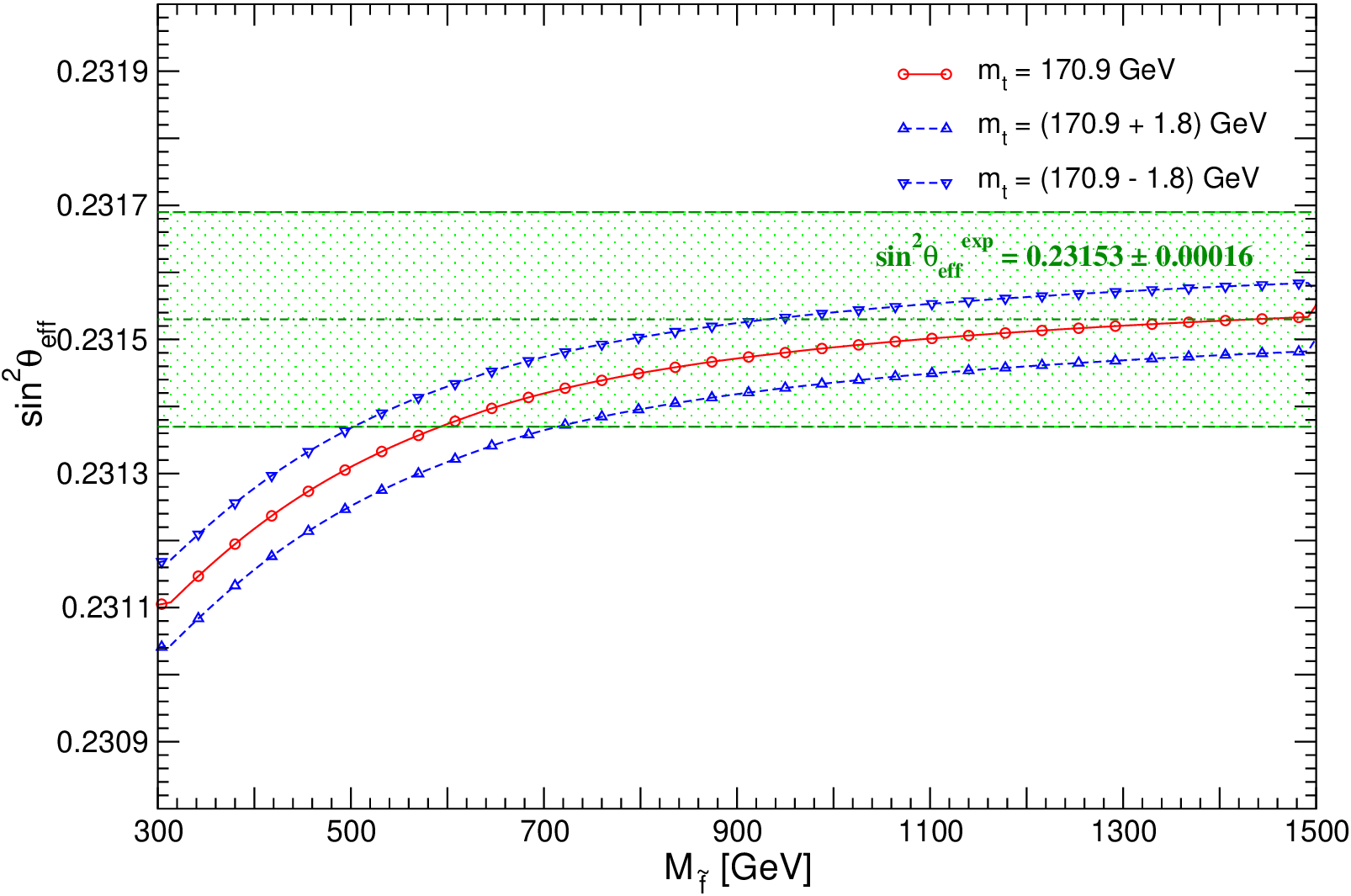}
\vspace{.3em}
\includegraphics[width=7.7cm,height=6.4cm]{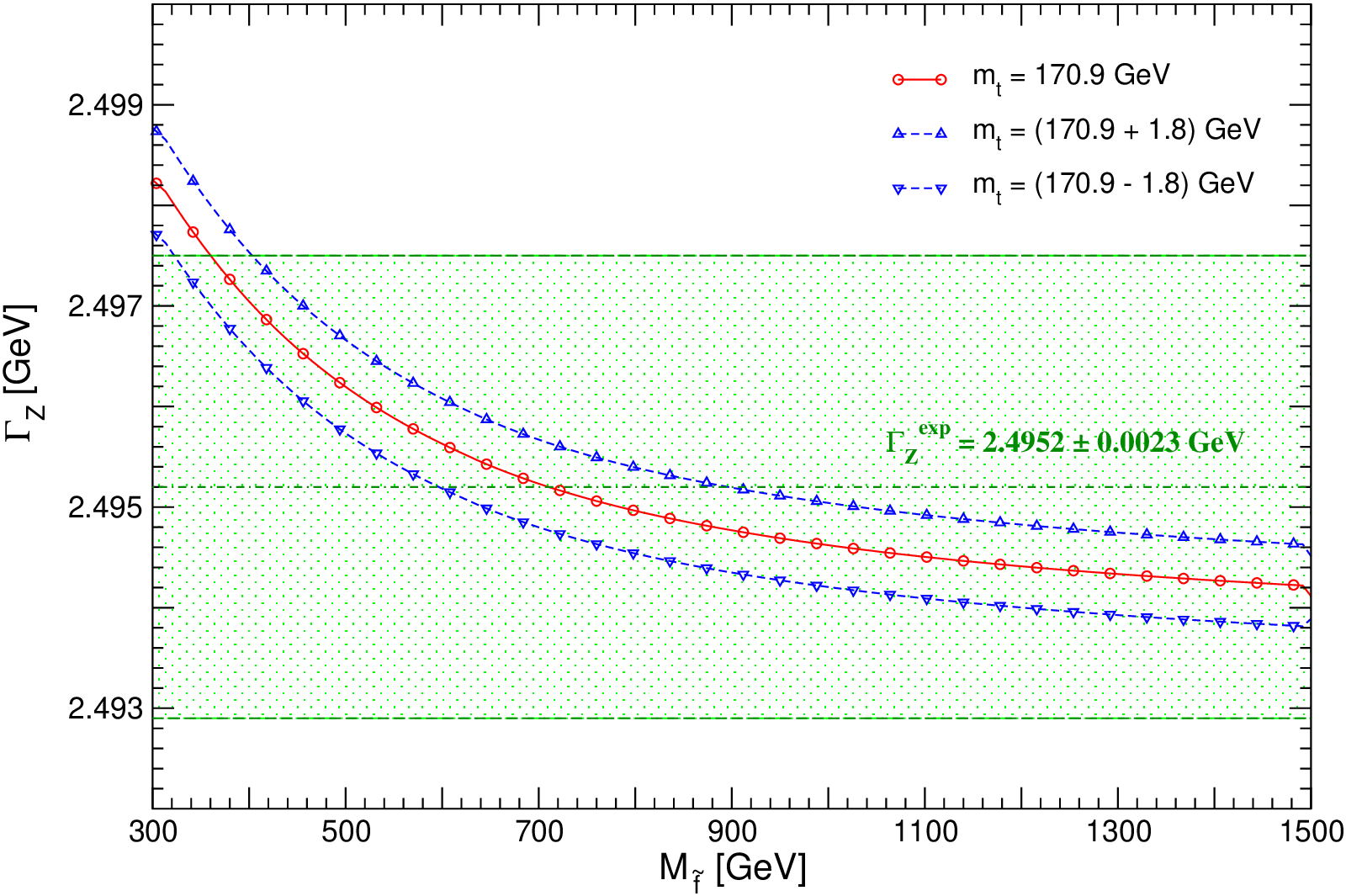}
\hspace{.3em}
\includegraphics[width=7.7cm,height=6.4cm]{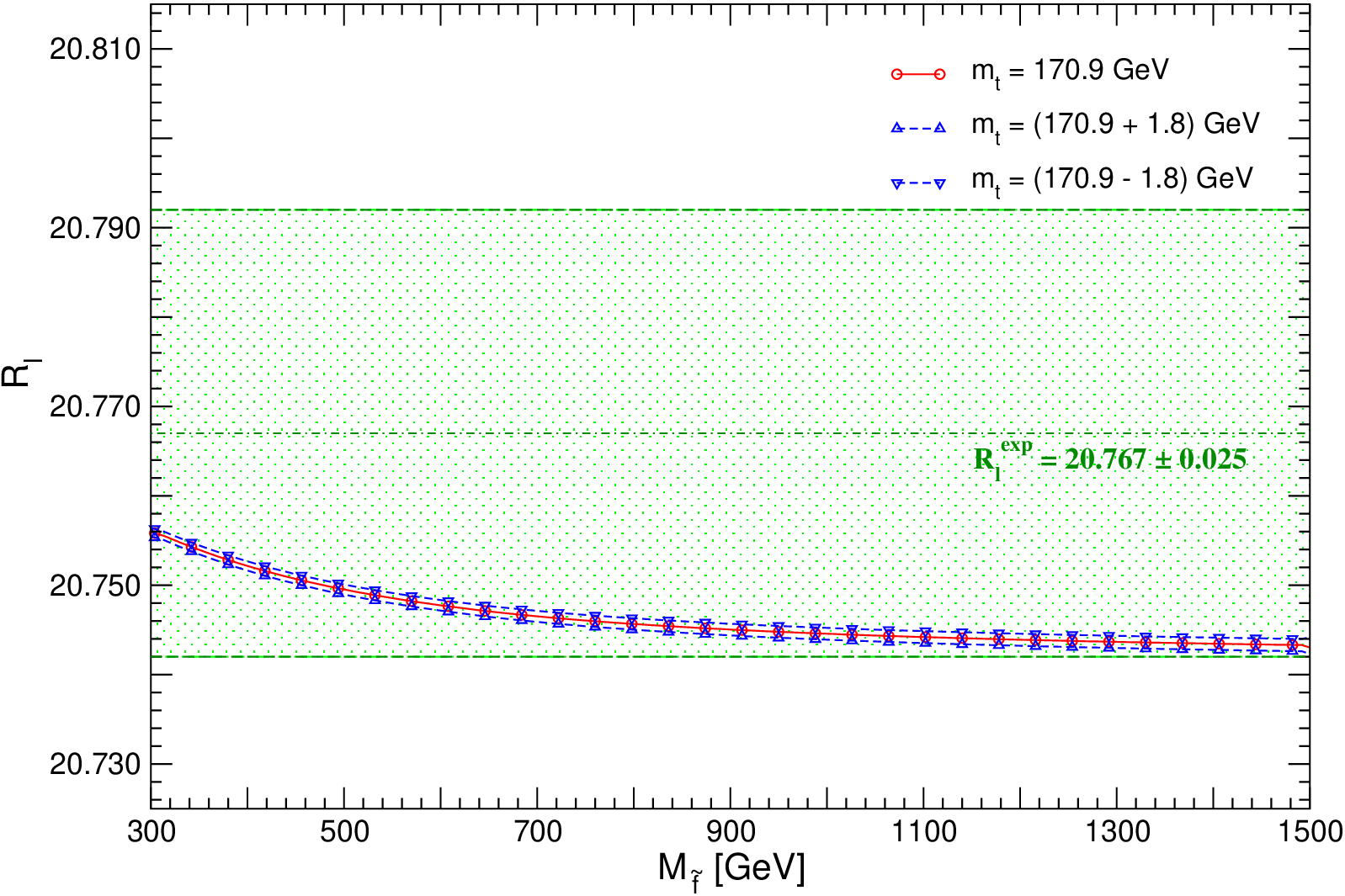}
\vspace{.3em}
\includegraphics[width=7.7cm,height=6.4cm]{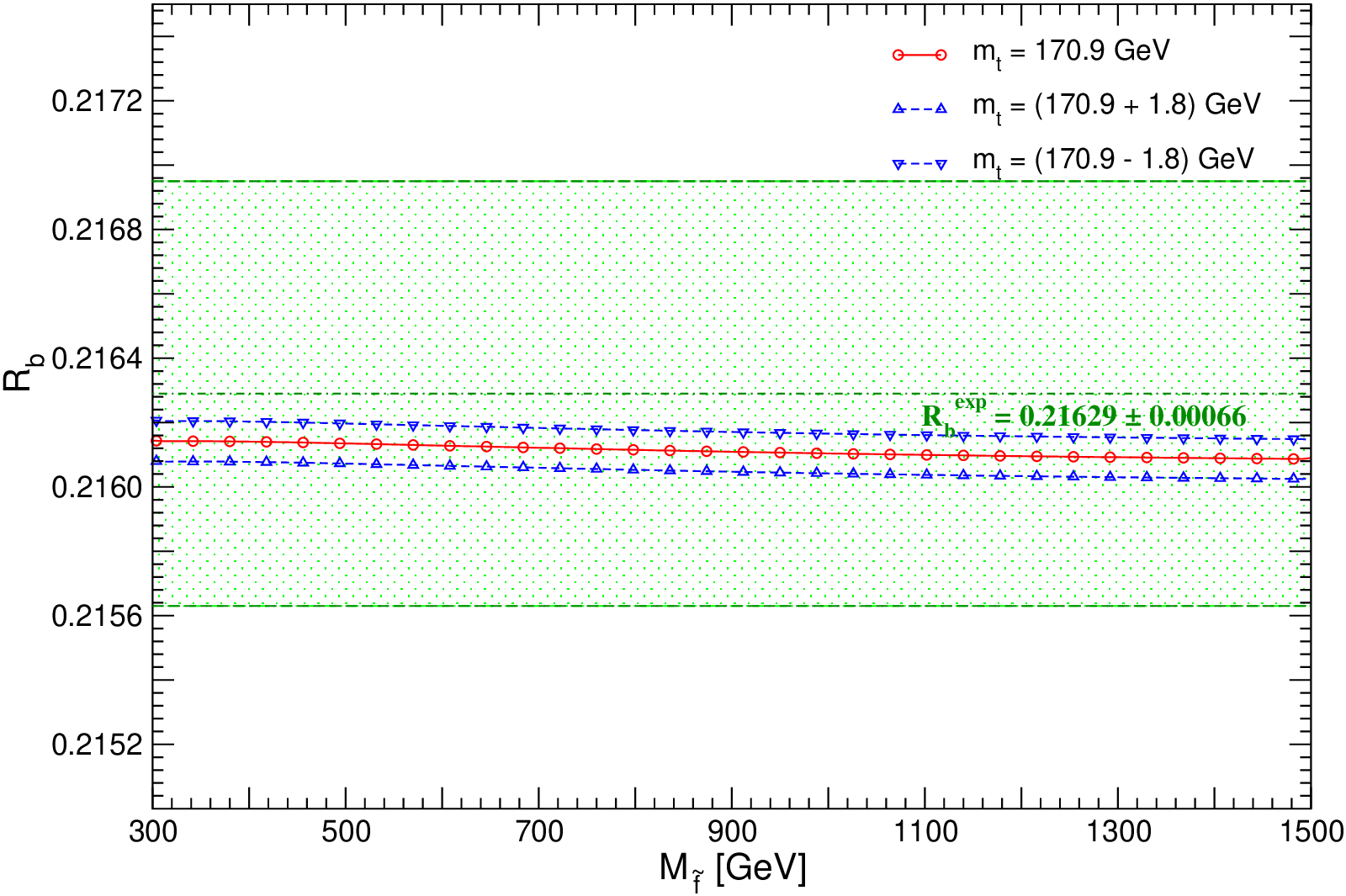}
\hspace{.3em}
\includegraphics[width=7.7cm,height=6.4cm]{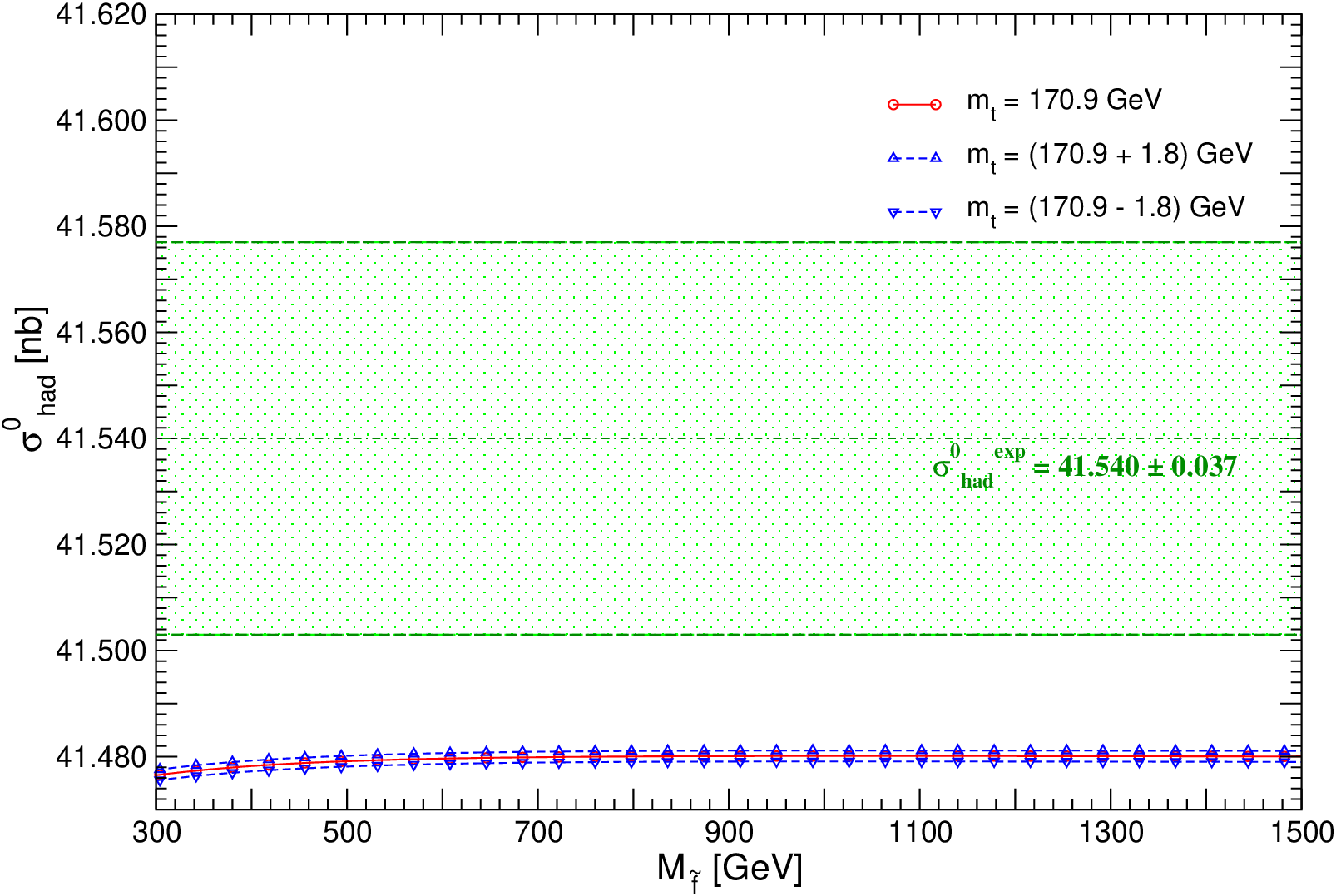}
\caption{Prediction for $\MW$, $\sweff$, $\Ga_Z$, $R_l$, $R_b$ and
  $\sigma_{\textup{had}}^0$ as function of the common sfermion mass scale
$\msusy$ for $\mt=(170.9 \pm 1.8)\gev$. The SUSY parameters are $\tb=10$,
  $A_{\tau}=A_{t}=A_b=2 \msusy$, $\MA=\mu=\mgl=M_2=300\gev$.}  
\label{fig:MSUSYdep} 
\end{center}
\end{figure}

In \reffi{fig:MSUSYdep} we show the prediction for the EWPO for real
parameters as a
function of $\msusy$ and indicate how this prediction changes if the
top-quark mass is varied within its experimental $1\si$~interval,
$\mt=(170.9 \pm 1.8)\gev$~\cite{mt1709}. 
The other parameters are $A_{t,b,\tau} = 2 \, \msusy$, 
$\mu = \MA = \mgl = M_2 = 300 \gev$ and $\tb = 10$.
The result is compared
with the current experimental values, see \refta{tab:expacc}. 
It can be seen in \reffi{fig:MSUSYdep} that 
only $\MW$, $\sweff$ and $\Ga_Z$ exhibit a pronounced sensitivity to
$\msusy$. While $\MW$ shows a mild preference for light $\msusy$,
$\sweff$ is in better agreement with the experimental value for large
$\msusy$.  
In interpreting the latter result it should be noted that the
world-average on $\sweff$ involves independent measurements that differ
from each other by more than three standard deviations~\cite{LEPEWWG}.
An experimental resolution of this issue will most likely require an
ILC with a GigaZ option.
The prediction for $\Ga_Z$ lies within its observed $1\si$~error for
most of the parameter space. 
The dependence of $R_l$, $R_b$, and $\si^0_{\rm had}$ on $\msusy$ is
nearly flat, where 
the first two observables are within the $1\si$~band, and the latter
is slightly below. The impact of varying $\mt$ within its experimental
error is non-negligible only for $\MW$, $\sweff$, and $\Ga_Z$. It
results in the following shifts%
\footnote{
The parametric uncertainties were
evaluated for the SPS1a$'$ benchmark point, see \refse{subsec:sps}, to
allow for a comparison with the 
theoretical errors estimated in \refse{sec:theounc}. However, the
uncertainties are only weakly dependent on the mass scale of the SUSY
particles.
}%
\begin{align}
\label{eq:dMWdmt}
&\de\MW^{{\rm para},\mt} = 11\mev, \\
\de\mt^{\rm exp} = 1.8 \gev~\textup{\cite{mt1709}} \quad \Rightarrow \quad
&\de\sweff^{{\rm para},\mt} = 5.4 \times 10^{-5},\\
&\de\Ga_Z^{{\rm para},\mt} = 0.43\mev.
\end{align}
Similarly we have analysed the impact of a shift in $\De \al_{\rm had}^{(5)}$, 
\begin{align}
&\de\MW^{{\rm para},\De\al_{\rm had}^{(5)}} = 6.3 \mev, \\
\de(\De\al_{\rm had}^{(5)}) = 3.5 \times 10^{-4} 
  \textup{\cite{LEPEWWG,DeltaAlfaMartin,Burkhardt:2005se}}\quad 
  \Rightarrow \quad
&\de\sweff^{{\rm para},\De\al_{\rm had}^{(5)}} = 12 \times 10^{-5},\\
&\de\Ga_Z^{{\rm para},\De\al_{\rm had}^{(5)}} = 0.32 \mev.
\label{eq:dGZdal}
\end{align}
It should be noted that the parametric uncertainty in $\sweff$ induced
by $\de(\De\al_{\rm had}^{(5)})$ is of similar size as the current
experimental uncertainty. A significant improvement of the uncertainty 
in $\De\al_{\rm had}^{(5)}$ is clearly very desirable in order to be
able to fully exploit future progress in the experimental measurements 
of the EWPO as well as in their theoretical predictions, see also
\refse{sec:theounc}.

\subsubsection{Dependence on $\mu$ and $M_2$}
\label{subsec:cndep}

\begin{figure}[htb!]
\begin{center}
\includegraphics[width=7.7cm,height=6.4cm]{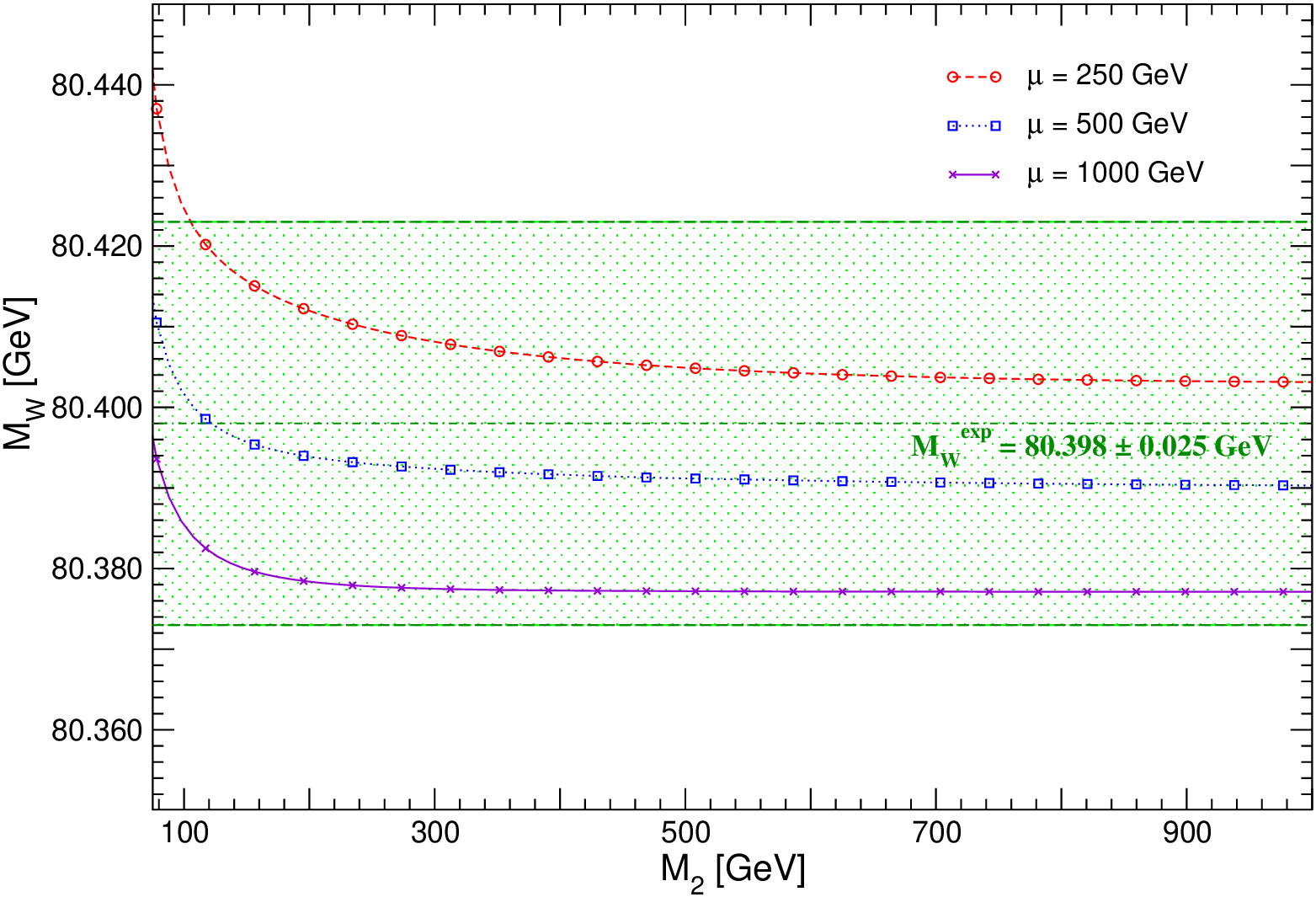}
\hspace{.3em}
\includegraphics[width=7.7cm,height=6.4cm]{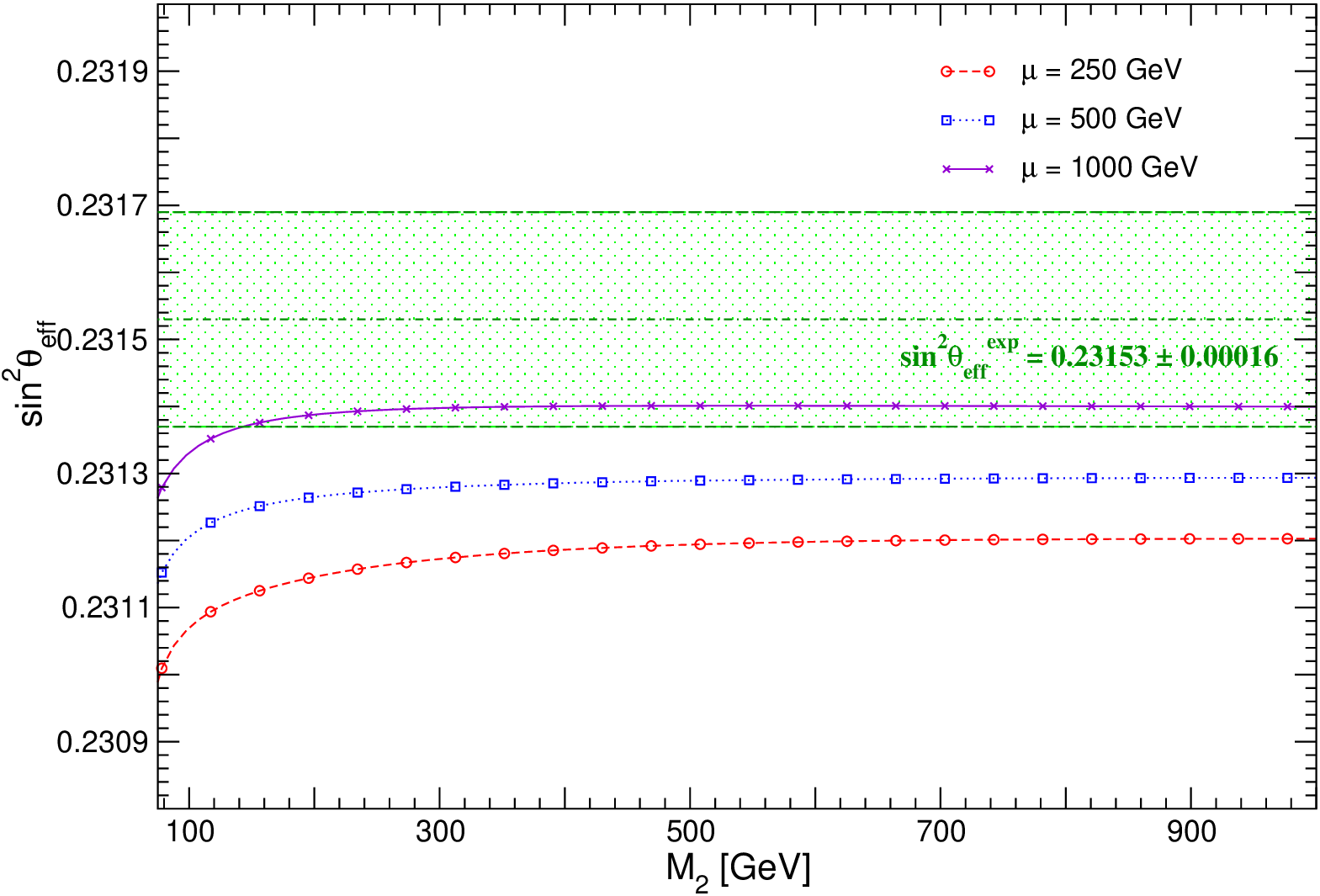}
\vspace{.3em}
\includegraphics[width=7.7cm,height=6.4cm]{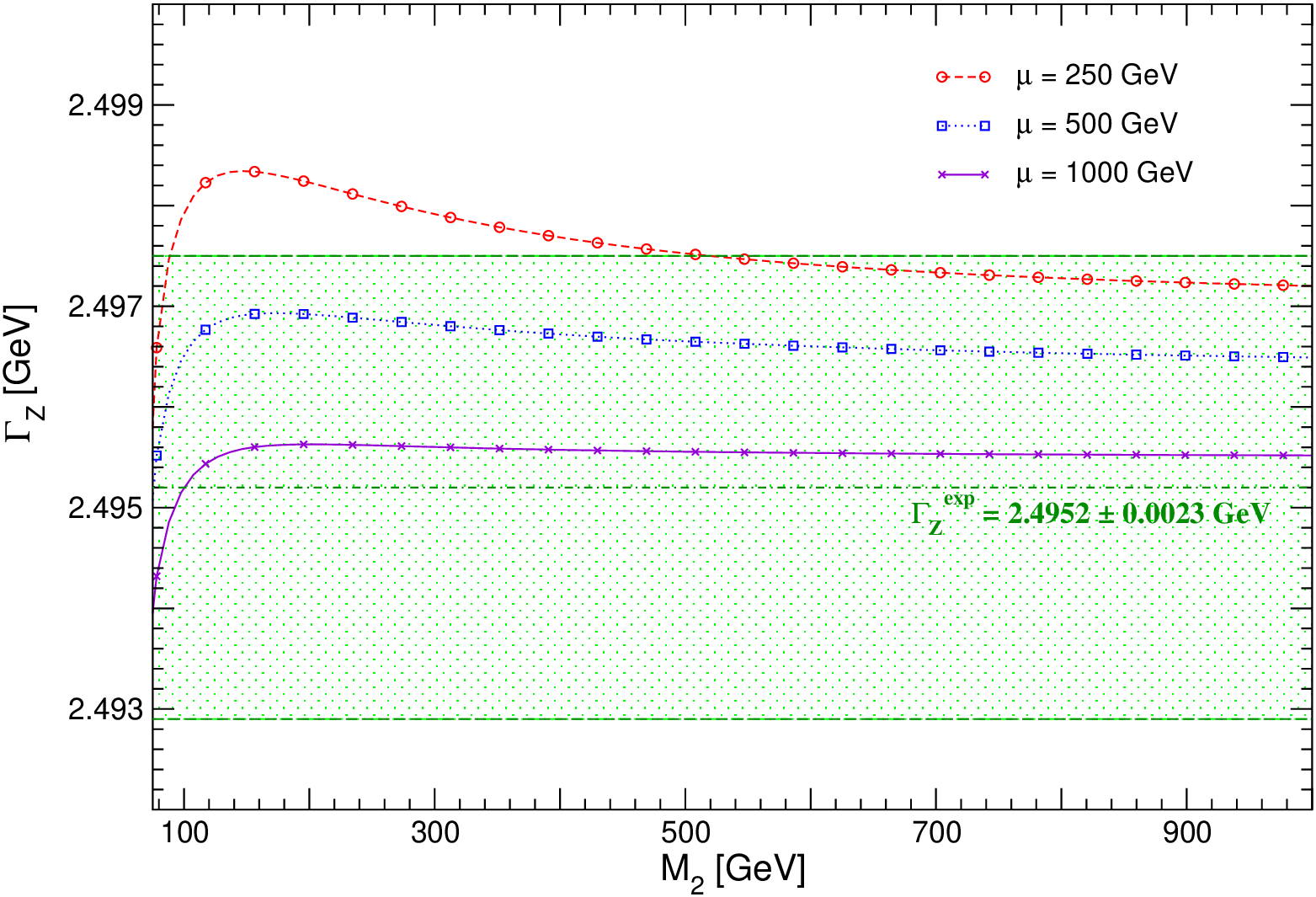}
\hspace{.3em}
\includegraphics[width=7.7cm,height=6.4cm]{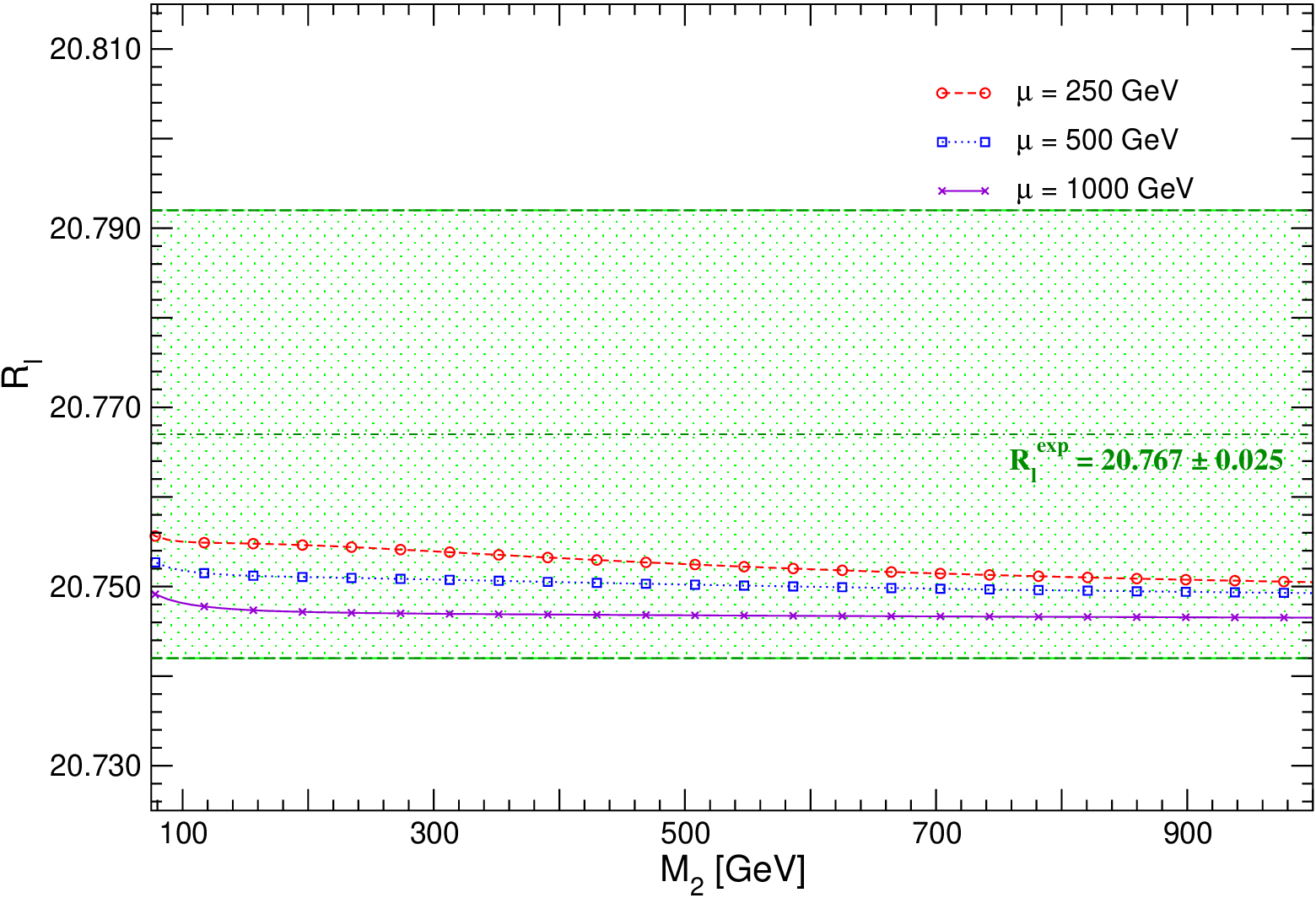}
\vspace{.3em}
\includegraphics[width=7.7cm,height=6.4cm]{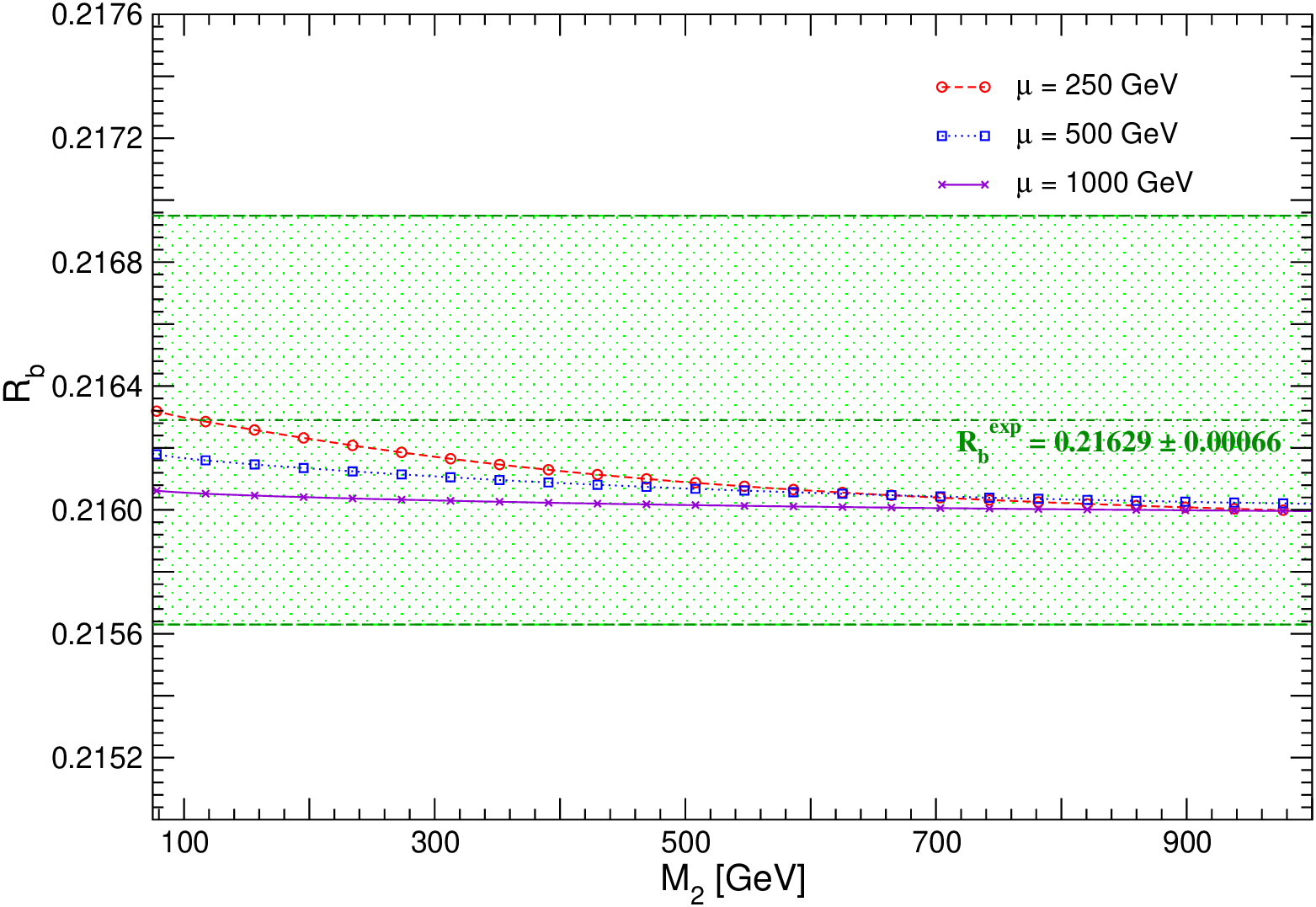}
\hspace{.3em}
\includegraphics[width=7.7cm,height=6.4cm]{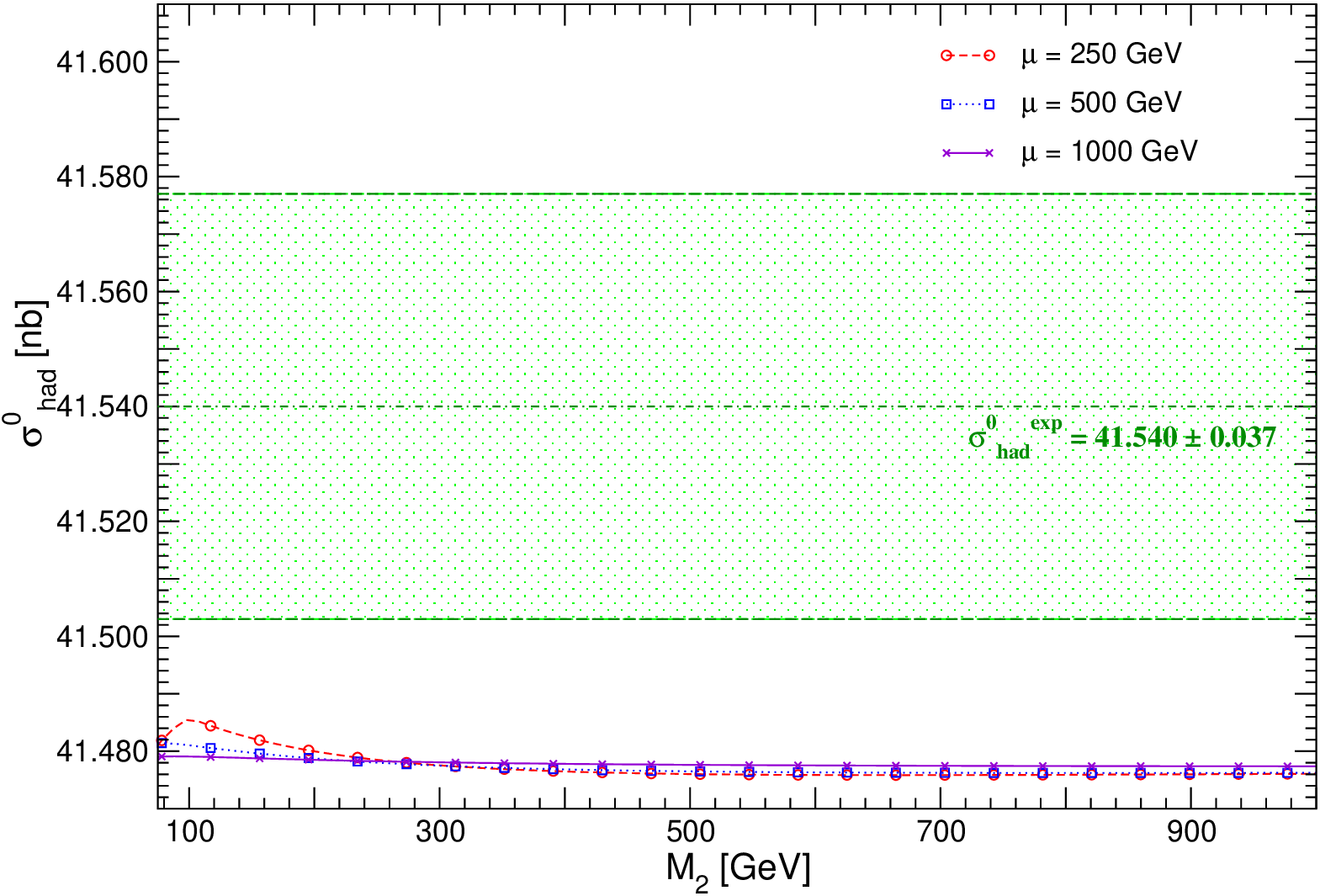}
\caption{Prediction for $\MW$, $\sweff$, $\Ga_Z$, $R_l$, $R_b$ and
  $\sigma_{\textup{had}}^0$ as function of $M_2$ for
  $\mu=250,500,1000\gev$. The remaining SUSY parameters are $\tb=10$,
  $\msusy=300\gev$, $A_{\tau}=A_{t}=A_b=2 \msusy$, $\mgl=600\gev$,
  $\MA=1000\gev$.}  
\label{fig:M2dep} 
\end{center}
\end{figure}

The variation of the EWPO with the parameters from the
chargino/neutralino sector is investigated in \reffi{fig:M2dep}. We show
the prediction for the six EWPO as a function of $M_2$ for $\mu = 250,
500,$ $1000 \gev$ (and $M_1$ is chosen according to the GUT
relation). The other parameters are set to $\msusy = 300 \gev$,
$A_{t,b,\tau} = 2 \msusy$, $\MA = 1000 \gev$, $\tb = 10$ and $\mgl = 600
\gev$. As expected, also in this case the observables with the largest
sensitivity to the variation of the SUSY parameters are $\MW$, $\sweff$
and $\Ga_Z$. The impact of varying $M_2$ and $\mu$ on the  other three
observables is negligible. The variation with $\mu$ results in shifts in
$\MW$, $\sweff$ and $\Ga_Z$ at the $1\,\si$ level. A sizable variation
with $M_2$ can only be observed for $M_2 \lsim 200 \gev$ for this set of
parameters. While $\MW$ ($\sweff$) shows a monotonous decrease
(increase) by $\sim 1 \si$ with $M_2$, $\Ga_Z$ exhibits a strong
increase up to $M_2 \lsim 150 \gev$ and then slowly decreases for
further increasing $M_2$. Good agreement between the $\Ga_Z$ prediction
and the experimental value is found for small  $M_2$, see also
\refse{subsec:splitsusy}.

\subsubsection{Dependence on complex phases}
\label{subsec:phases}

For the analysis of the dependence of the complex phases we focus on the
three EWPO
that show the strongest variation with the SUSY parameters,
$\MW$, $\sweff$ and $\Ga_Z$, see the two previous subsections. Since the
dependence on the sfermion mass parameters is much stronger than on the
chargino/higgsino parameters we only investigate the dependence on the
phases of $\At$ and $\Ab$.
As for $\De r/\MW$~\cite{MWweber} we find that
the effective one-loop couplings~$g^f_{\{V,A\},(1)}$  
depend  only on the absolute values $|\Xt|$, $|\Xb|$
of the off-diagonal entries in the $\Stop$~and $\Sbot$~mass matrices,
where $\Xt = \At - \mu/\tb$, $\Xb = \Ab - \mu\,\tb$. Thus, the phases of 
$\mu$, $\At$ and $\Ab$ enter only in the combinations 
$(\phi_{A_{t,b}} + \phi_{\mu})$, giving rise to modifications of the squark
masses and mixing angles. It furthermore follows that the impact of $\phiat$
($\phiab$) on the sfermion 
masses (see \refeq{sfermmasses}) is stronger for low (high) $\tb$.

\begin{figure}[htb!]
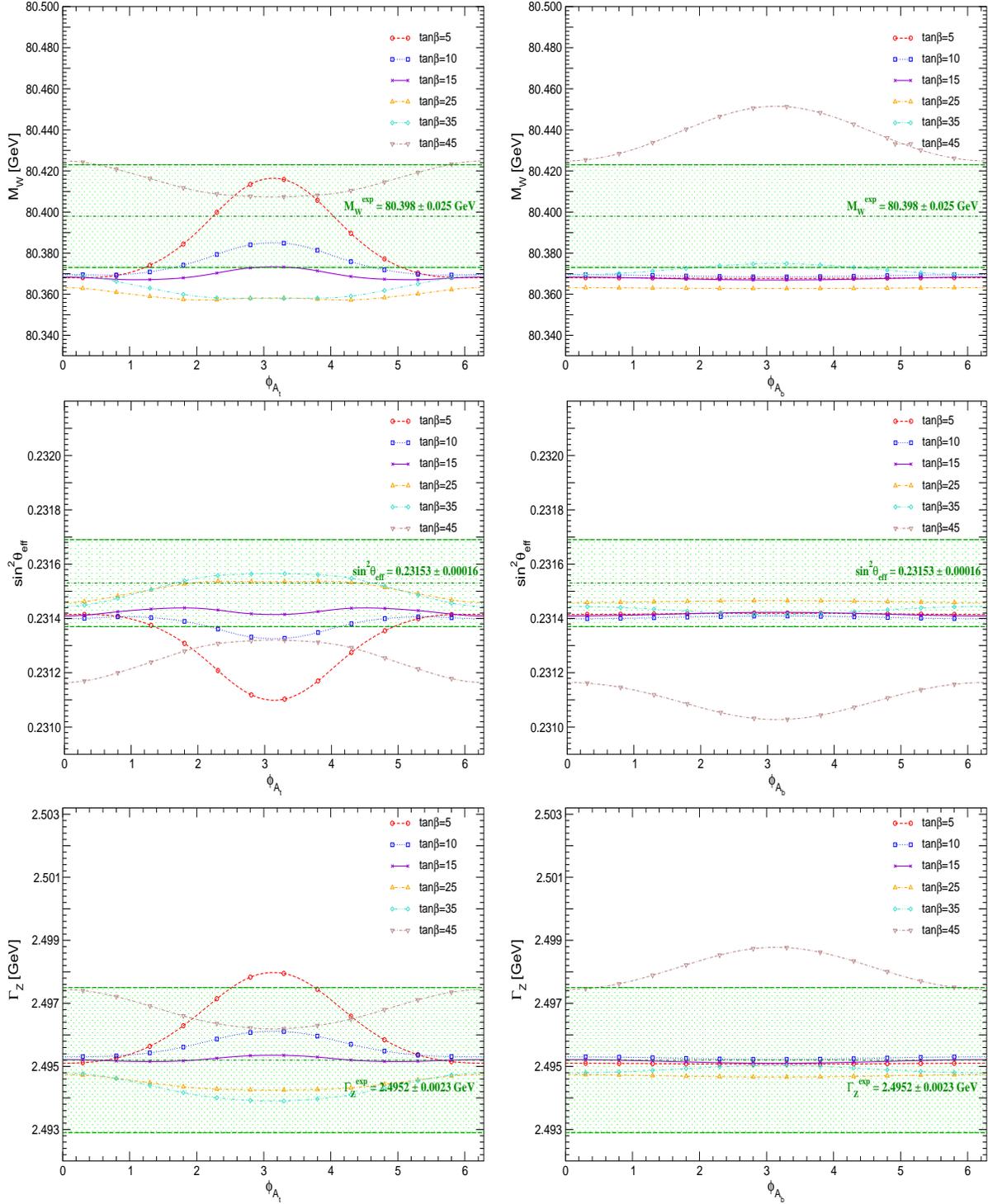

\begin{center}
\includegraphics[width=7.7cm,height=6.4cm]{MW-Atphase.eps}
\hspace{.3em}
\includegraphics[width=7.7cm,height=6.4cm]{MW-Abphase.eps}
\vspace{.3em}
\includegraphics[width=7.7cm,height=6.4cm]{SinTheta-Atphase.eps}
\hspace{.3em}
\includegraphics[width=7.7cm,height=6.4cm]{SinTheta-Abphase.eps}
\vspace{.3em}
\includegraphics[width=7.7cm,height=6.4cm]{GammaZ-Atphase.eps}
\hspace{.3em}
\includegraphics[width=7.7cm,height=6.4cm]{GammaZ-Abphase.eps}
\caption{Prediction for $\MW$, $\sweff$, and $\Ga_Z$ as function of the
  phase of the trilinear coupling $A_t$ (plots on the left hand side) and
  $A_b$ (plots on the right hand side). The other SUSY parameters are:
  $\msusy= \MHp= M_2 = \mgl=500\gev, A_\tau=A_t=A_b=\mu=1000\gev,
 \phi_{A_\tau}= \phi_{\mu}=\phi_{M_1} = \phigl=0,\phi_{A_b}=0$ 
  (plots on the left), 
  $\phi_{A_t}=0$ (plots on the right).} 
\label{fig:phases} 
\end{center}
\vspace{-2em}
\end{figure}

In \reffi{fig:phases} we show the three EWPO
as a function of 
$\phiat$ (with $\phimu = \phiab = 0$, left plots) and 
$\phiab$ (with $\phimu = \phiat = 0$, right plots)
for different values of $\tb$ (varied from $\tb = 5$ to $\tb = 45$).
The other parameters are set to  
$\msusy = \MHp = M_2 = \mgl = 500 \gev$, 
$|A_{t,b,\tau}| = |\mu| = 1000 \gev$, $\phi_{A_\tau}=\phiMe  = \phigl=0$. 
As expected, the dependence of $\MW$, $\sweff$ and $\Ga_Z$ on $\phiat$ 
is most pronounced for small $\tb$ (see left panel of
\reffi{fig:phases}). The variation of $\phiat$ in this 
case gives rise to a shift in the three precision observables by
$1$--$2\si$. The effect becomes smaller for increasing $\tb$, up to $\tb=15$. 
On the other hand, for high $\tb$ the lighter $\Sbot$~mass becomes
rather small for the
parameters chosen in \reffi{fig:phases}, reaching values as low as about 
$100 \gev$ for $\tb=45$. This leads to a sizable shift of $\sim
1$--$2\si$ in the EWPO already for vanishing phases.
The slight rise in the dependence on $\phiat$ for  $\tb\geq25$ is due to
the overall enlarged SUSY contributions which occur for  large $\tb$ and
the resulting low sbottom masses.

The dependence of the precision observables on $\phiab$ (plots on the 
right-hand side of \reffi{fig:phases}) is rather small, except for the
highest $\tb$ value shown in  \reffi{fig:phases}, $\tb = 45$. This is
again related to the sizable correction induced by the  
lighter $\Sbot$~mass. In this scenario the variation of $\phiab$ 
yields a shift in $\MW$, $\sweff$ and $\Ga_Z$ at the $1\si$ level.


\subsection{Impact of $Z \to \neu{1} \neu{1}$}
\label{subsec:LightNeu1}
As next step in our numerical analysis we investigate the impact of the
decay $Z \to \neu{1}\neu{1}$ on the invisible $Z$~boson width,
$\Ga_{\textup{inv}}$. A sizable
$Z\neu{1}\neu{1}$ coupling in combination with a light neutralino,
$\mneu{1} \lsim \MZ/2$, results in an additional contribution to the 
invisible width of the $Z$~boson, in addition to the SM decays into
neutrinos. Since the experimental result for the invisible width of the 
$Z$~boson is somewhat below the SM prediction~\cite{LEPEWWG}, additional
contributions from new physics are tightly constrained. It is of
interest in how far the possibility of a light neutralino is affected by
the precision measurement of the invisible width of the $Z$~boson.

The mass of the lightest neutralino is determined by $M_1$, $M_2$ and
$\mu$, see \refeq{Nmatrix} (here we assume all parameters to be real). 
If $M_1$ or $M_2$ or $\mu$ is much smaller than the other two, 
the lightest neutralino is mostly a bino, zino, or higgsino,
respectively.
Its mass is to a large extent determined by this smallest
mass value.
If the condition
\begin{equation}
M_1= M_1^{(0)} \equiv 2 \MZ^2\sw^2 \frac{\tb}{(1+\TQb)}\frac{M_2}{
      ( \mu M_2 - 2\MZ^2 \tb/(1+\TQb)\cw^2 )}
\label{masslessNino}
\end{equation}
was exactly fulfilled, the lightest neutralino, $\neu{1}$, would be 
massless and almost entirely bino-like.
\begin{figure}[htb!]
\begin{center}
\includegraphics[width=7.7cm,height=7.4cm]{GammaInv-LightNeutralino-M1.eps}
\hspace{.3em}
\includegraphics[width=7.7cm,height=7.4cm]{GammaInv-LightNeutralino-MNeu1.eps}
\end{center}
\caption{The invisible $Z$~boson width $\Gamma_{\textup{inv}}$ 
in- and excluding the process
  $Z\to\neu{1}\neu{1}$ is shown as a function of $M_1$ (left) and
  $\mneu{1}$ (right). The SUSY parameters are chosen
  to be: $\msusy = 250\gev, A_{\tau}=A_t=A_b=\mu=\mgl=500\gev, \MA=500\gev,
  M_2 = 200\gev$. A light, almost entirely bino-like, $\neu{1}$ is obtained
  by varying $M_1$ around the value $M_1^{(0)}$ defined in \refeq{masslessNino}
  within the range of $-100$ to $+100\gev$.} \label{fig:LightNeu1}   
\end{figure}
%
\begin{figure}[htb!]
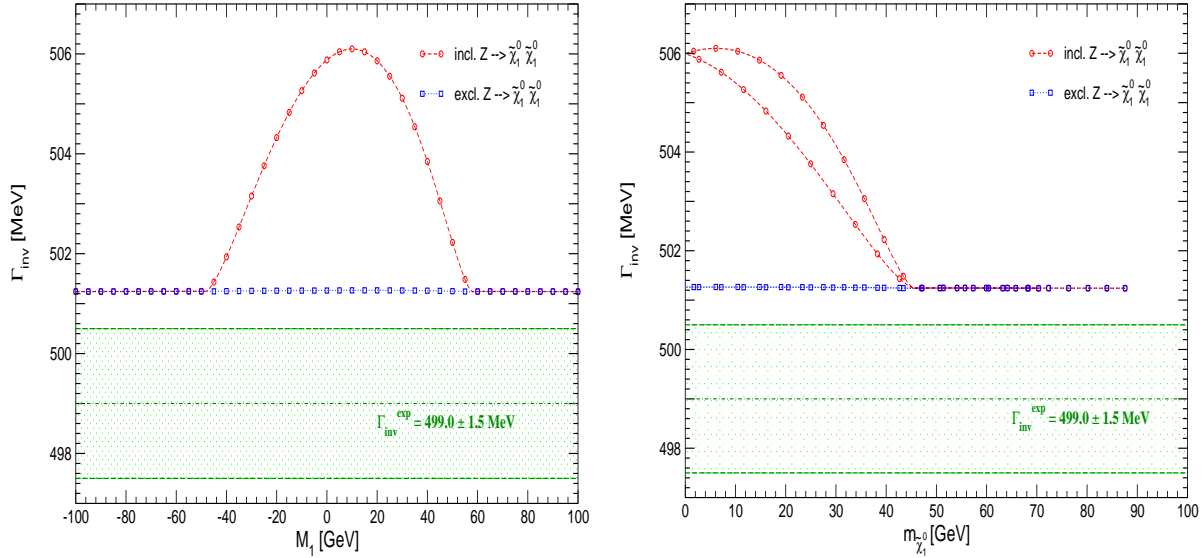

\vspace{2em}
\begin{center}
\includegraphics[width=7.7cm,height=7.4cm]{GammaInv-LightNeutralino2-M1.eps}
\hspace{.3em}
\includegraphics[width=7.7cm,height=7.4cm]{GammaInv-LightNeutralino2-MNeu1.eps}
\end{center}
\caption{The invisible $Z$~boson width $\Gamma_{\textup{inv}}$ 
in- and excluding the process
  $Z\to\tilde\chi^0_1\tilde\chi^0_1$ is shown as a function of $M_1$
(left) and $\mneu{1}$ (right). The SUSY parameters are chosen
  as follows: $\msusy = A_{\tau}=A_t=A_b=\MA=\mgl=600\gev,  \mu=125\gev,M_2 =
  200\gev$,  $M_1=-100\dots100\gev$.} \label{fig:LightNeu12}   
\end{figure}

In \reffi{fig:LightNeu1} we analyse a scenario with a light neutralino
(see also the discussion in \citere{lightninopaper}). 
We do this by varying $M_1$ around the value $M_1^{(0)}$ defined in 
\refeq{masslessNino}, i.e.\
\begin{equation}
M_1= M_1^{(0)} + \De M ,
\label{eq:lightNino}
\end{equation}
where we let $\De M$ run from $-100$ to $+100 \gev$.
The other parameters in \reffi{fig:LightNeu1} are $\msusy = 250 \gev$, 
$A_{t,b,\tau} = \mu = \mgl = \MA = 500 \gev$, $M_2 = 200 \gev$, $\tb=10$.
In the left (right) plot of \reffi{fig:LightNeu1} the results for
$\Ga_{\textup{inv}}$
are shown as a function of $M_1$ ($\mneu{1}$). 
The curves in the plot on the right hand side of \reffi{fig:LightNeu1}
consist of two branches for a given value of $\mneu{1}$. 
This is a consequence of the fact that different values of $M_1$ in
general lead to different values of $\Ga_{\textup{inv}}$ although the
value of $\mneu{1}$ may be the same. 
Comparing with the plot on the left one can see that the upper branch
corresponds to $M_1\gsim0$, the lower branch to $M_1\lsim0$.  
The coupling of a nearly pure bino to $Z$~bosons is very weak, resulting
in a visible, but negligible effect on $\Ga_{\textup{inv}}$.
The entire range of $\Ga_{\textup{inv}}$
shown in \reffi{fig:LightNeu1} is outside its $1\si$ error. As mentioned
above, this is a consequence of the well-known fact 
that the measured value of the invisible $Z$~width is below the SM
prediction~\cite{LEPEWWG}.
As the predictions including and excluding the process
$Z\to\neu{1}\neu{1}$ differ at most by $\sim0.02\mev$, 
compared to an experimental range of
$\Ga_{\textup{inv}}^{\textup{exp}}=499.0\pm1.5\mev$, the inclusion of
the decay into neutralinos $\neu{1}$ clearly does not give rise to
additional constraints on $M_1$ or the mass of the lightest
neutralino~$\mneu{1}$ in this scenario. 

A different scenario with $\mu \sim M_1$, but $M_1 \lsim \edz M_2$
(in coarse agreement with the GUT relation) is analysed in
\reffi{fig:LightNeu12}. The parameters are set to 
$\msusy = A_{t,b,\tau} = \MA = \mgl = 600 \gev$, 
$\mu = 125 \gev$, $M_2 = 200 \gev$, $\tb=10$, and $M_1$ is varied from 
$-100 \gev$ to $+100 \gev$. In this scenario the lightest neutralino can have
a sizable higgsino component. 
In the left (right) plot of \reffi{fig:LightNeu12} the results for 
$\Ga_{\textup{inv}}$
are shown as a function of $M_1$ ($\mneu{1}$). 
The lower branches in \reffi{fig:LightNeu12}~(plot on the right) again
correspond to $M_1\lsim0$, the upper branches to $M_1\gsim0$. 
For values $\mneu{1} \lsim \MZ/2$ the $\Ga_{\neu{1}}$ contribution to  
$\Ga_{\textup{inv}}$ becomes
sizable. In this case the deviation between the MSSM prediction for
the invisible $Z$~width and the measured value, which is slightly above 
$1 \si$ if the decay into neutralinos is not open, raises above the 
$2\si$~level for
$|M_1| \lsim 35\gev$ or  $\mneu{1} \lsim 35 \gev$.

As a consequence, if the SUSY parameters are such that $\neu{1}$
has a sizable coupling to the
$Z$~boson (e.g.\ $M_2 > M_1 \gg \mu$), the theoretical prediction for
$\Gamma_{\textup{inv}}$ can easily exceed the experimental central value
by several standard 
deviations (provided that $\mneu{1}<\MZ/2$). 
In such a scenario the contribution to the invisible width
arising from $\Ga_{\neu{1}}$ yields interesting constraints on the
parameters $\mu$, $M_1$, and $M_2$.


\subsection{The SPS benchmark scenarios}
\label{subsec:sps}

\begin{figure}[htb!]
\begin{center}
\includegraphics[width=7.7cm,height=6.4cm]{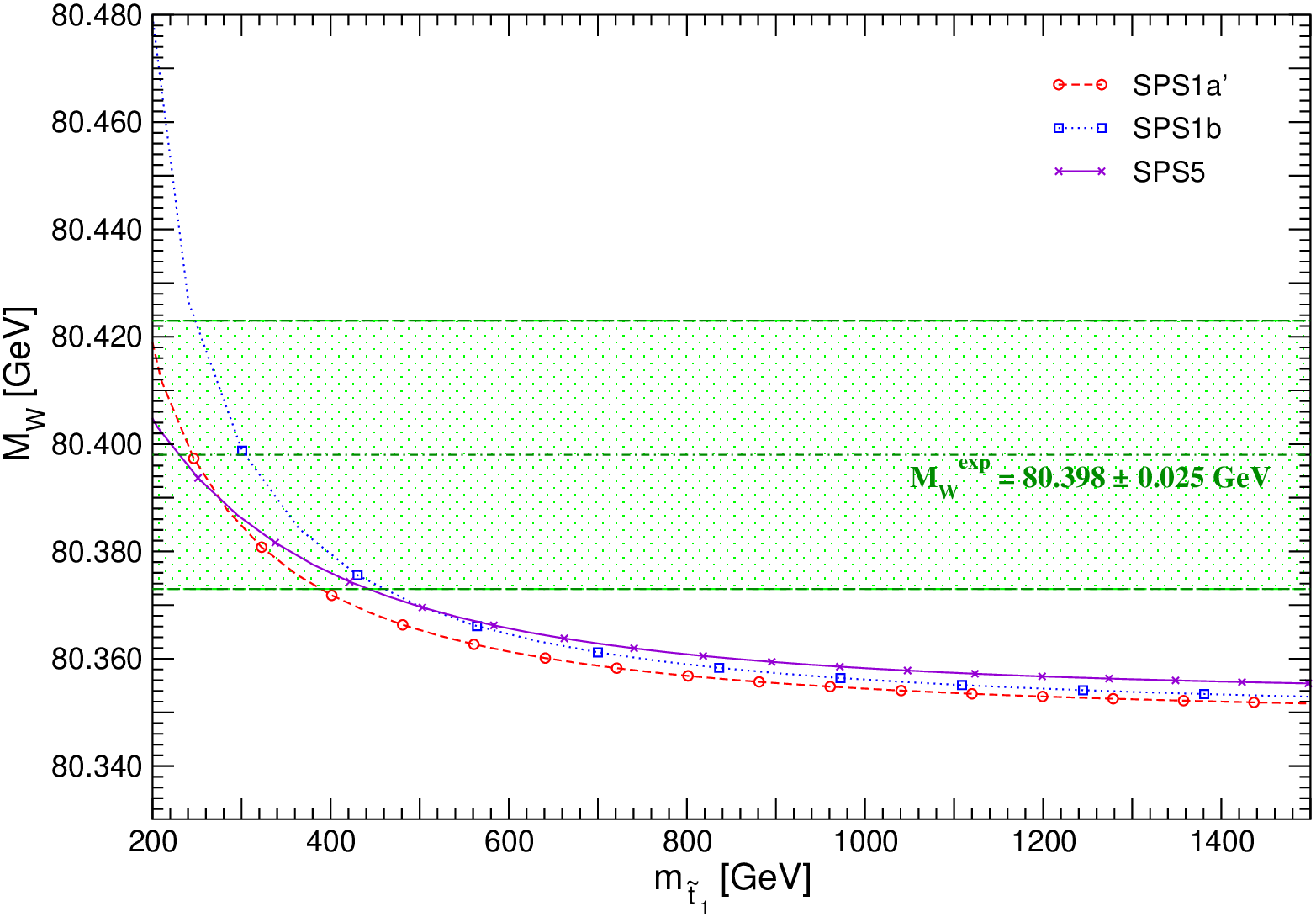}
\hspace{.3em}
\includegraphics[width=7.7cm,height=6.4cm]{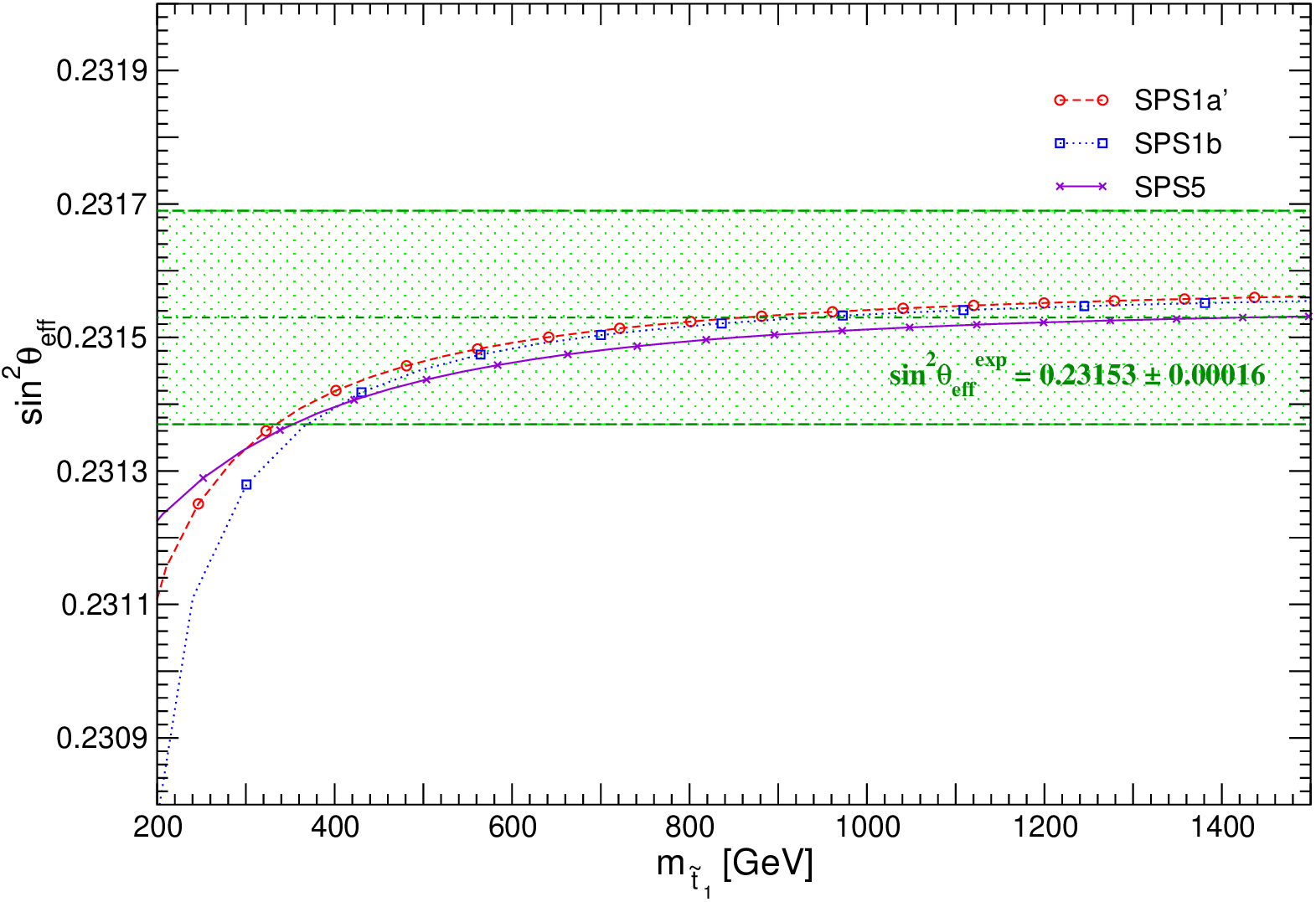}
\vspace{.3em}
\includegraphics[width=7.7cm,height=6.4cm]{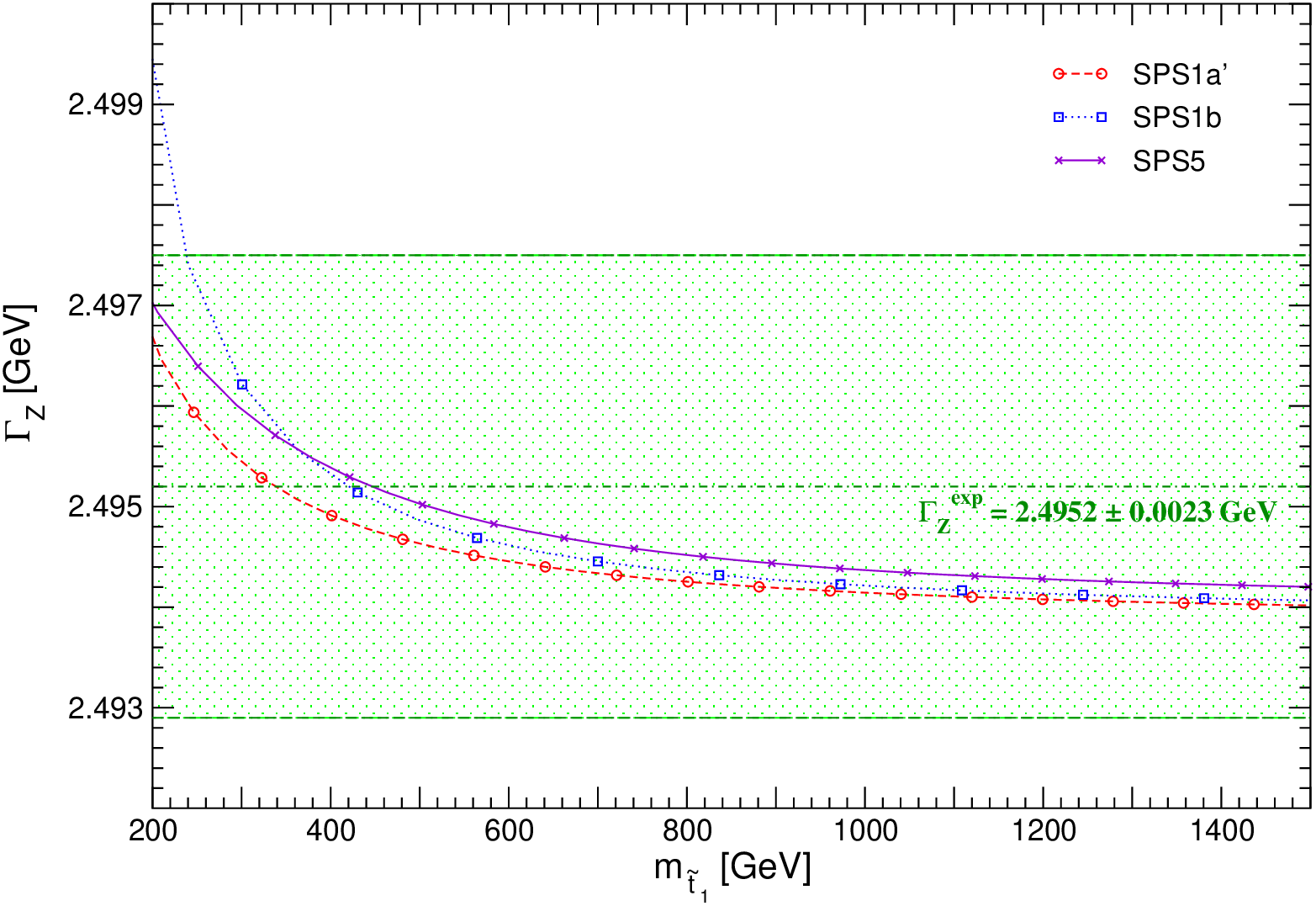}
\hspace{.3em}
\includegraphics[width=7.7cm,height=6.4cm]{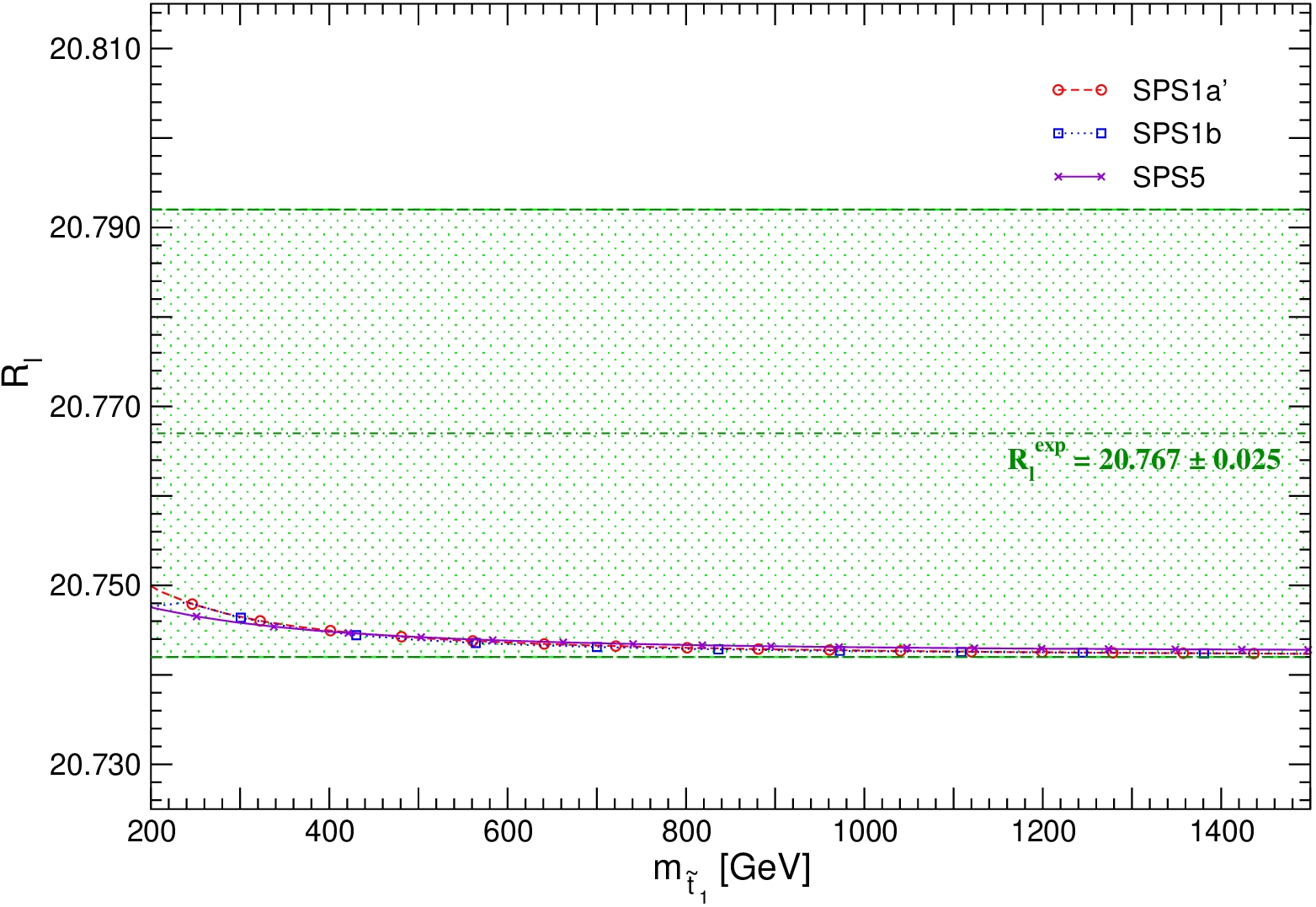}
\vspace{.3em}
\includegraphics[width=7.7cm,height=6.4cm]{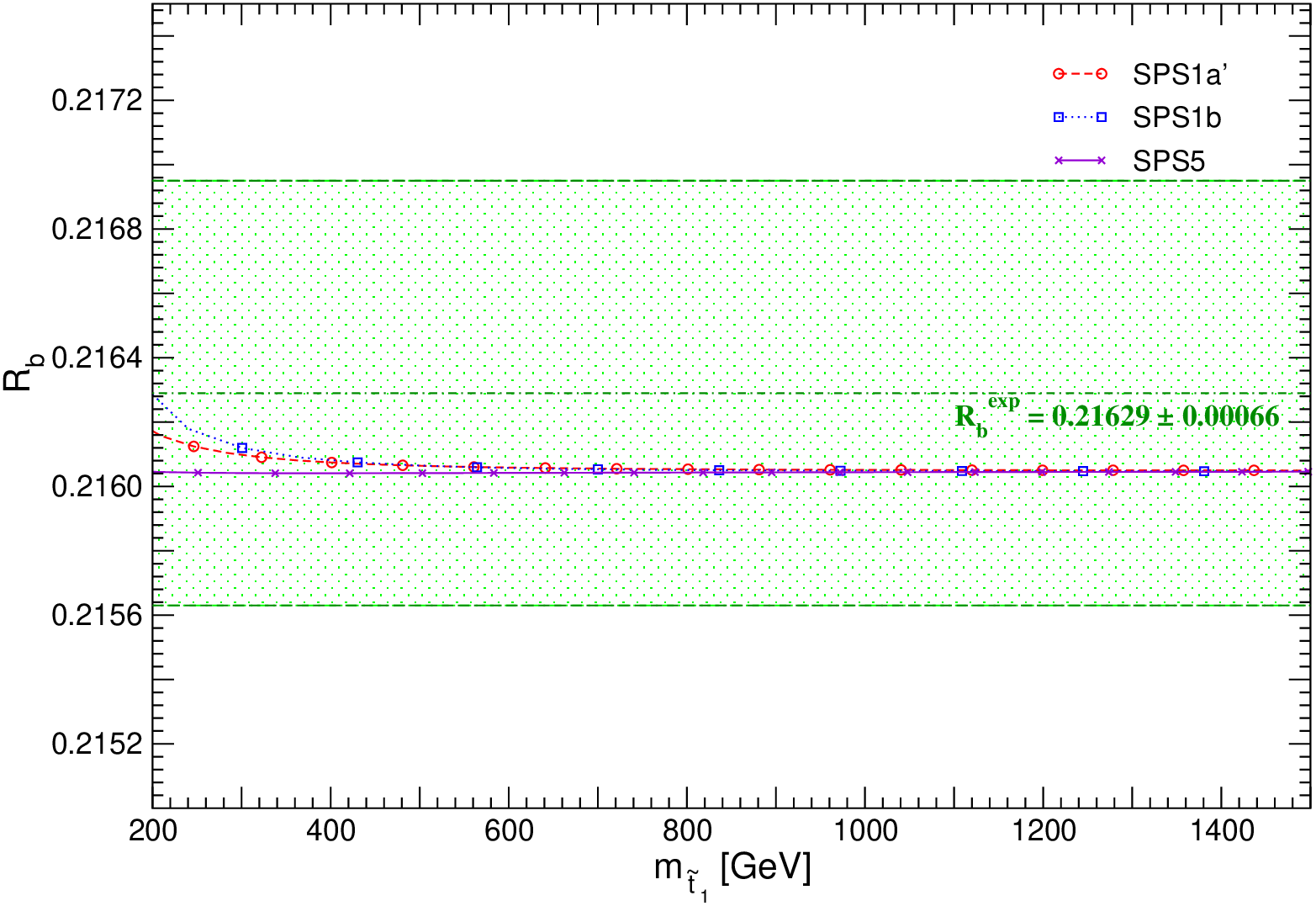}
\hspace{.3em}
\includegraphics[width=7.7cm,height=6.4cm]{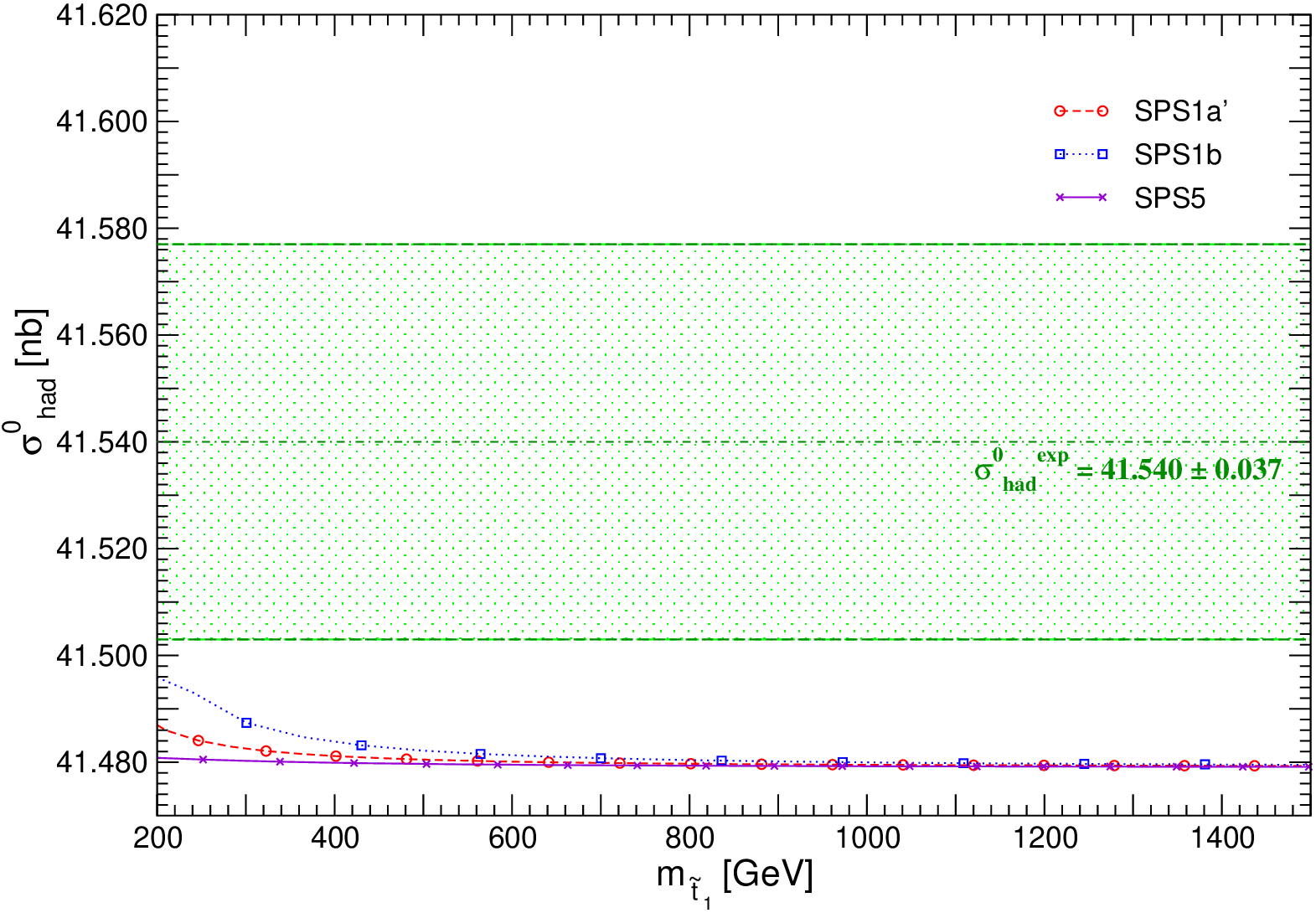}
\caption{Predictions for $\MW$, $\sweff$, $\Ga_Z$, $R_l$, $R_b$ and
  $\sigma_{\textup{had}}^0$ within the SPS1a$'$, SPS1b and SPS5  
  scenarios. The observables are shown as a function of $\mste$, the
  mass of the lighter of the two scalar top-quarks. The SPS 
  parameters of mass dimension are varied with the
  scale of supersymmetry as described in the text.} 
\label{fig:SPSMSf133} 
\end{center}
\vspace{-2em}
\end{figure}

In this section we analyse the six EWPO 
$\MW$, $\sweff$, $\Ga_Z$, $R_l$, $R_b$, and $\si^0_{\rm had}$
in the SPS~1a$'$, SPS~1b and SPS~5
benchmark scenarios~\cite{sps,spa}. Our analysis extends the results of 
\citere{MWweber} to the $Z$~pole observables. 
In order to
analyse the dependence of the EWPO on the scale of supersymmetry we
scale for each SPS point all SUSY parameters carrying mass dimension 
by a common factor, i.e.\
$\MA=(\textup{scalefactor})\times \MA^{\textup{SPS}}$, 
$M_{{\tilde F},{\tilde F'}}=(\textup{scalefactor})\times 
                            M_{{\tilde F},{\tilde F'}}^{\textup{SPS}}$,
$A_{t,b,\tau}=(\textup{scalefactor})\times A_{t,b,\tau}^{\textup{SPS}}$,
$\mu=(\textup{scalefactor})\times\mu^{\textup{SPS}}$,
$M_{1,2,3}=(\textup{scalefactor})\times M_{1,2,3}^{\textup{SPS}}$.  
In \reffi{fig:SPSMSf133} we show $\MW$, $\sweff$, $\Ga_Z$, $R_l$, $R_b$,
and $\si^0_{\rm had}$ as a function of the lighter $\Stop$~mass,
$\mste$,
 in the three SPS scenarios. As before, only $\MW$, $\sweff$ and
$\Ga_Z$ show a sizable variation with $\mste$ (i.e., the scalefactor).
For all the EWPO the (loop-induced) 
variation between the three SPS scenarios is relatively small.
$\MW$ shows best agreement with the experimental results for low
$\mste$. $\sweff$ lies in the $\pm 1\,\si$ range for 
$\mste \gsim 400 \gev$. $\Ga_Z$ shows a variation with $\mste$ of 
$\sim 1\,\si$ around the experimental value.

\begin{figure}[htb!]
\begin{center}
\includegraphics[width=7.7cm,height=6.4cm]{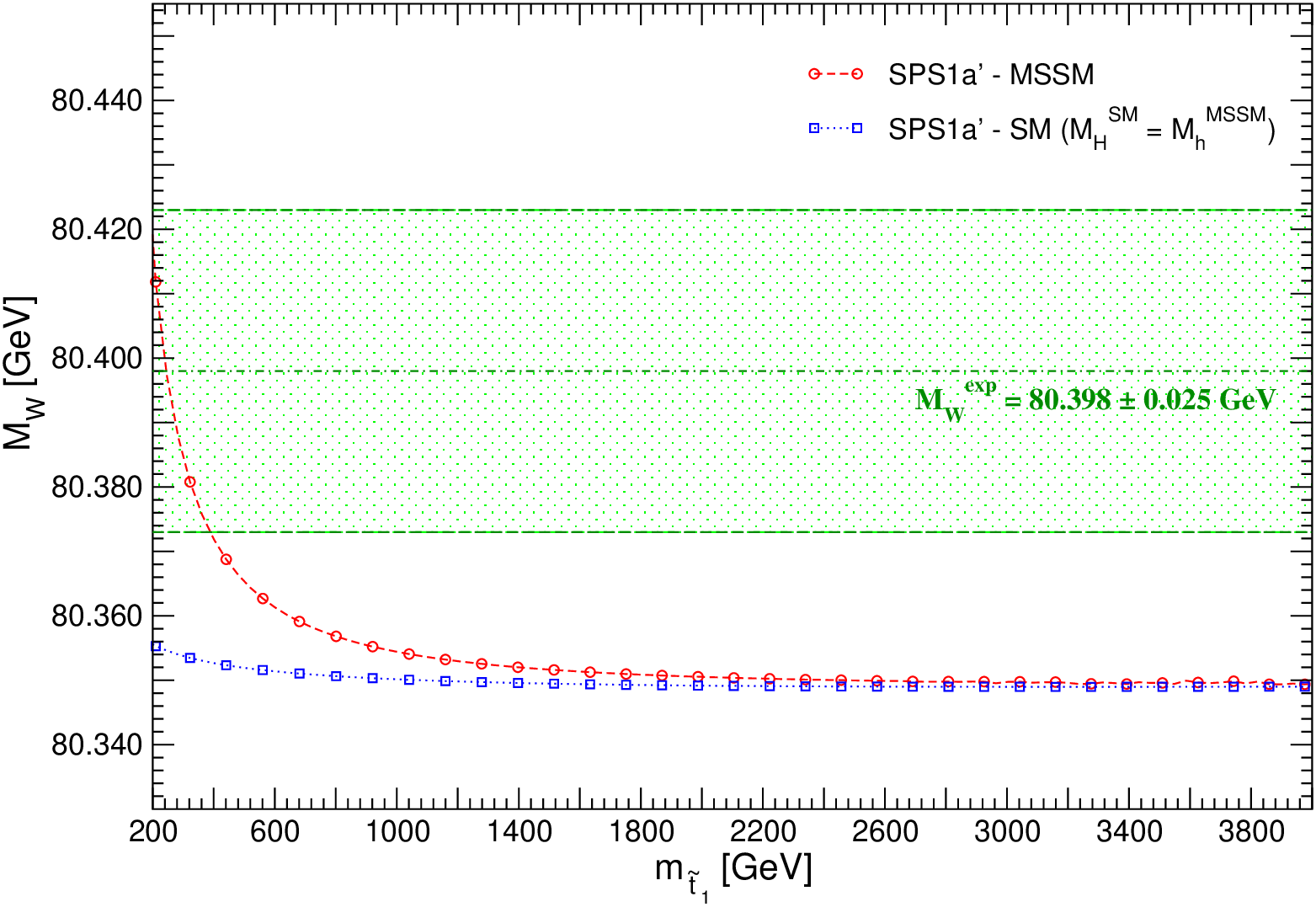}
\hspace{.3em}
\includegraphics[width=7.7cm,height=6.4cm]
                                 {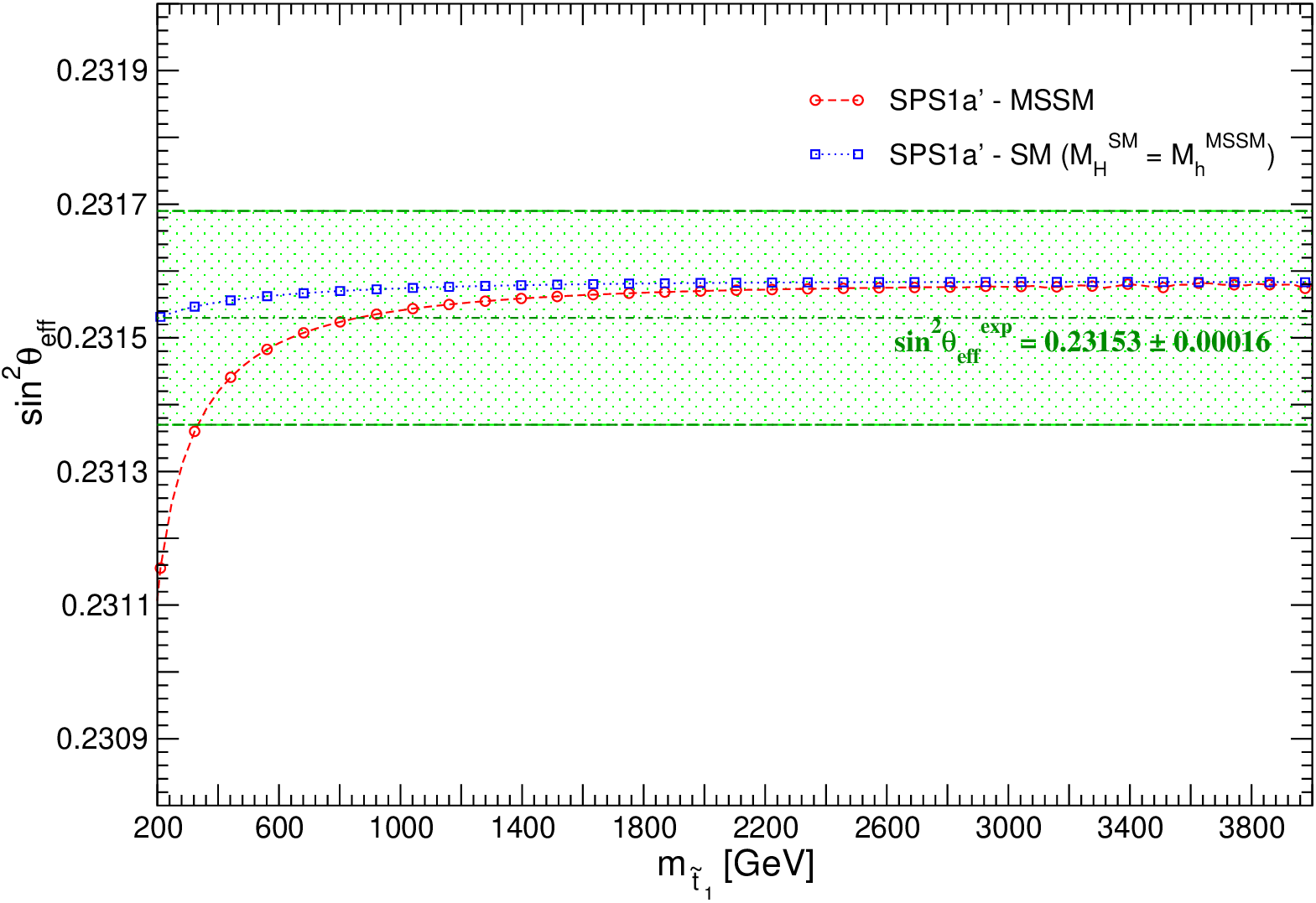}
\vspace{.3em}
\includegraphics[width=7.7cm,height=6.4cm]{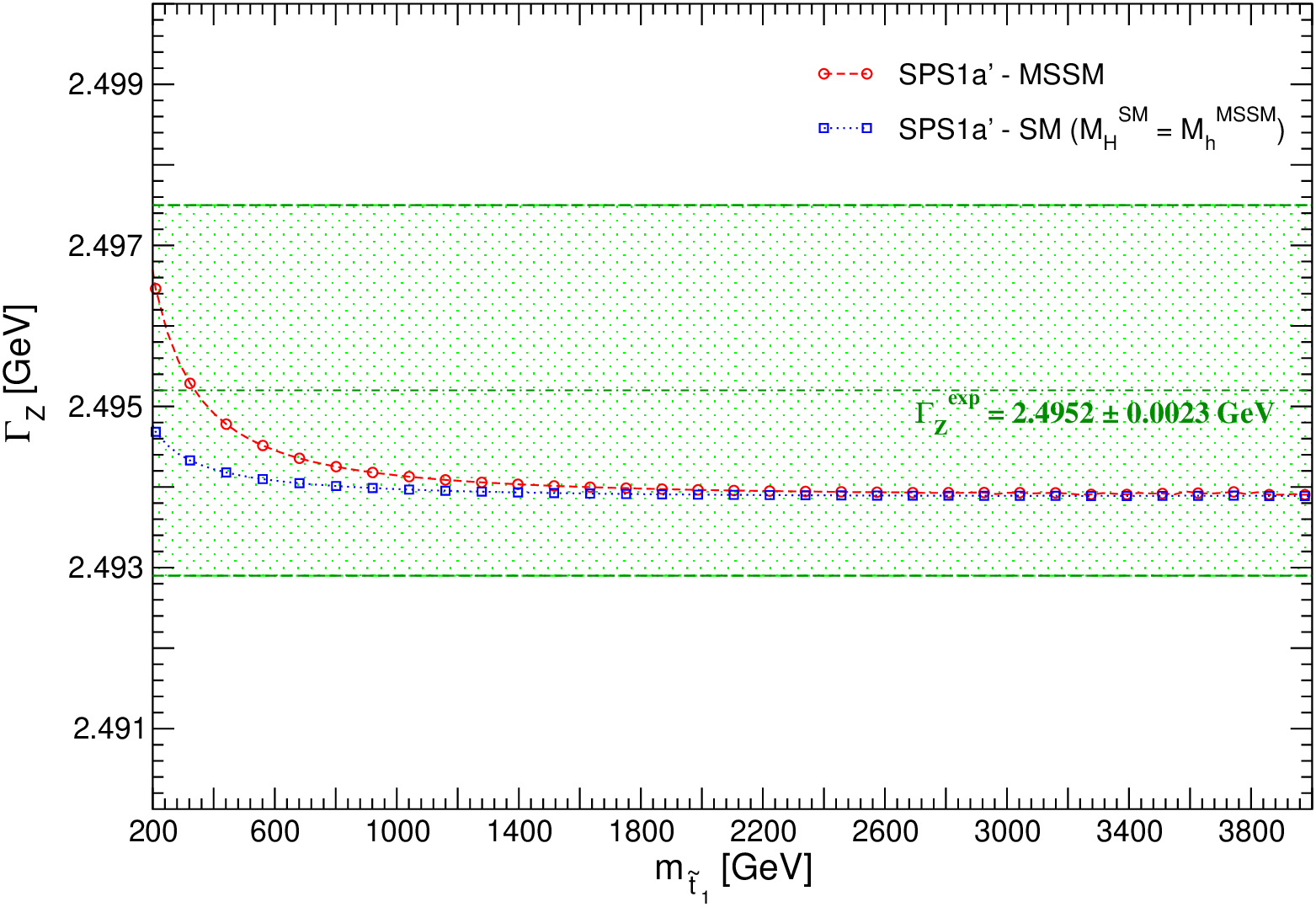}
\hspace{.3em}
\includegraphics[width=7.7cm,height=6.4cm]{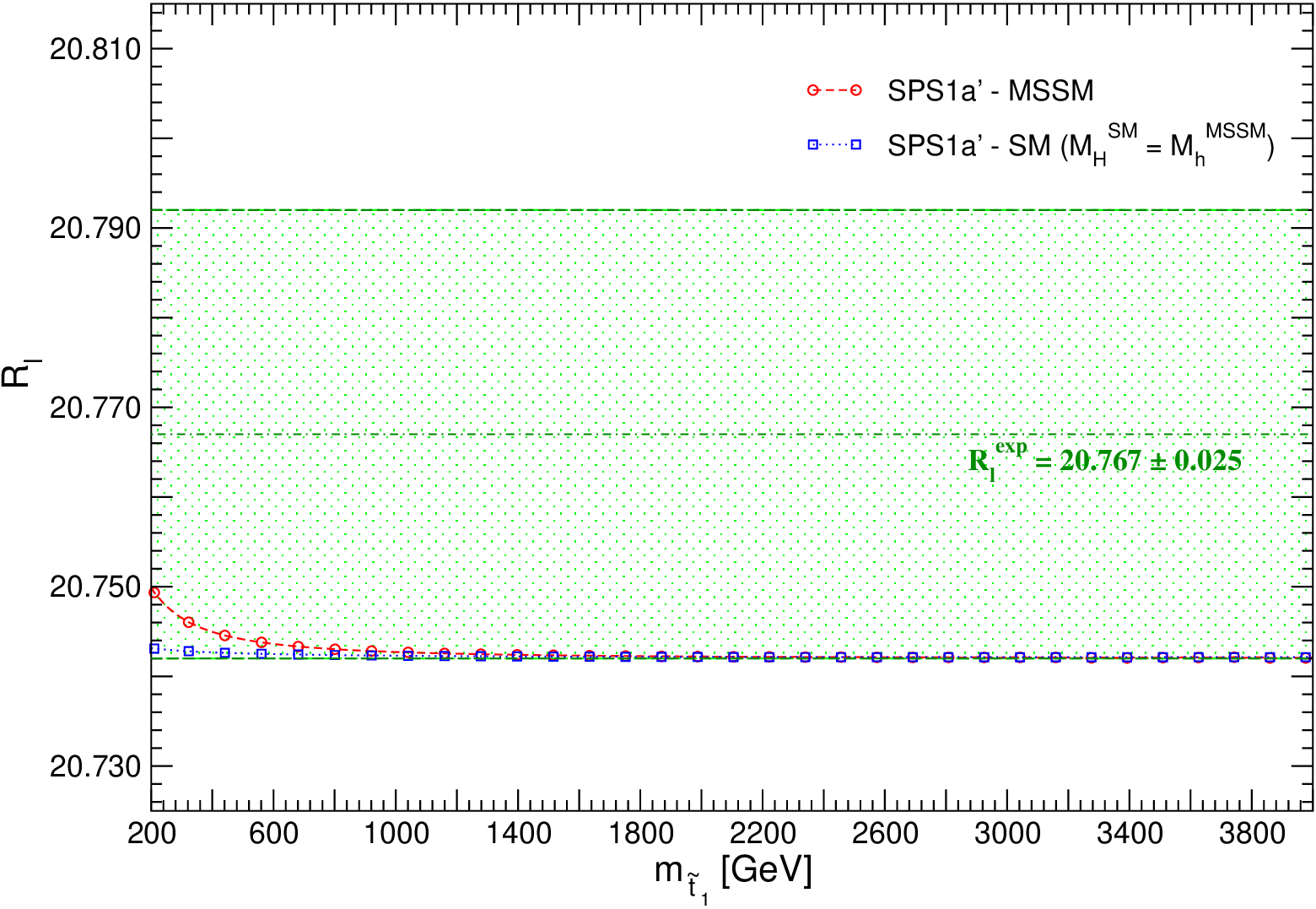}
\vspace{.3em}
\includegraphics[width=7.7cm,height=6.4cm]{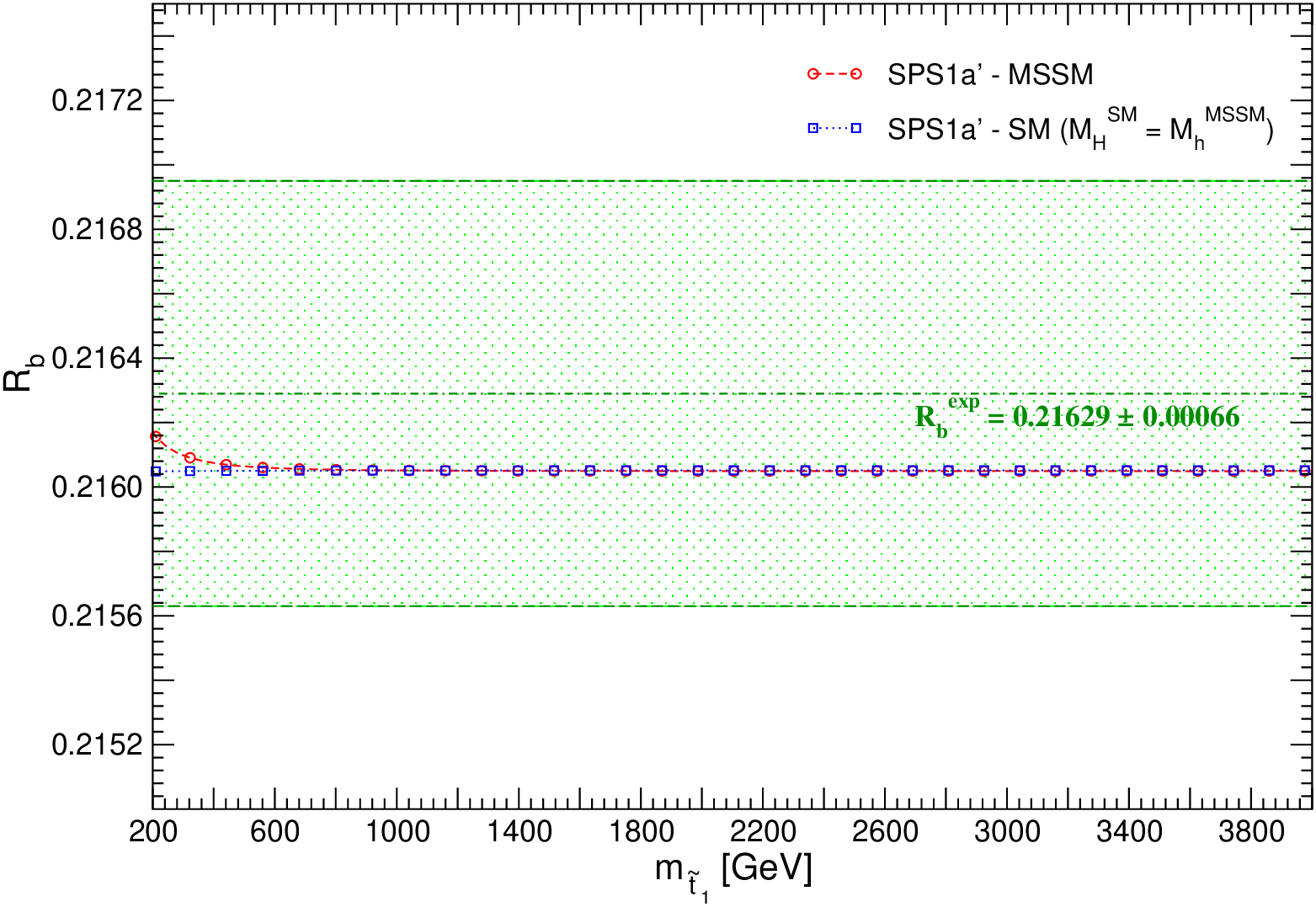}
\hspace{.3em}
\includegraphics[width=7.7cm,height=6.4cm]{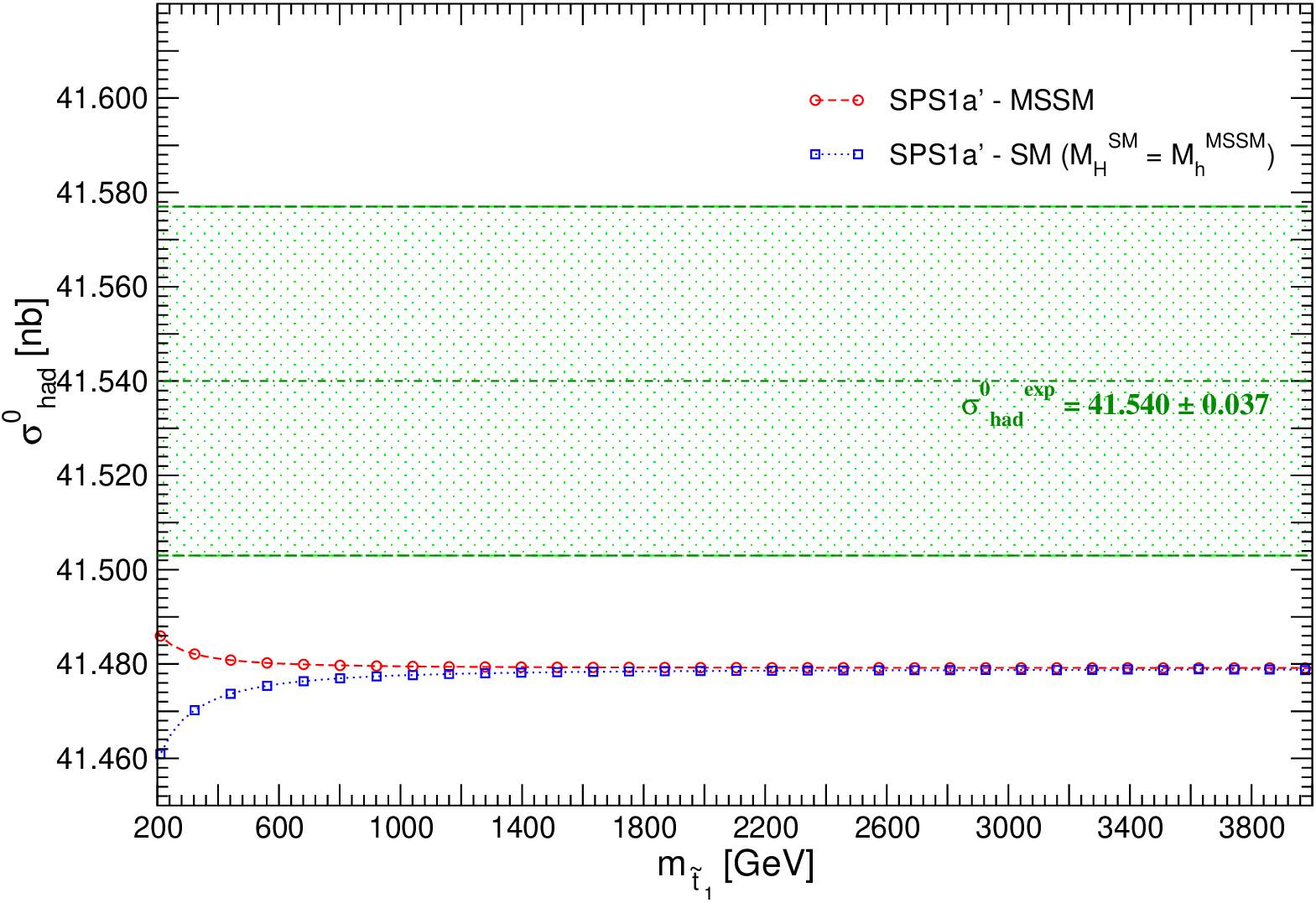}
\caption{Predictions for $\MW$, $\sweff$, $\Ga_Z$, $R_l$, $R_b$ and
  $\sigma_{\textup{had}}^0$ within the SPS1a$'$ scenario in comparison with
  the SM result calculated for $\MHSM=M_h$. The
  observables are shown as a function of $\mste$, the mass of the
  lighter of the two scalar top-quarks. The SPS 
  parameters of mass dimension are varied with the
  scale of supersymmetry as described in the text.} 
\label{fig:SPSMSf133Decoupling} 
\end{center}
\vspace{-2em}
\end{figure}

\reffi{fig:SPSMSf133Decoupling} shows for the example of the 
SPS~1a$'$ scenario that, as expected, the MSSM predictions for the EWPO
approach the corresponding predictions in the SM 
(for $\MHSM = \Mh^{\rm MSSM}$) for large values of the SUSY mass scale.
The predictions within the MSSM and the SM (for $\MHSM = \Mh^{\rm MSSM}$) 
are shown as a function of $\mste$ (as before, all parameters carrying
mass dimension are scaled by a common factor). The variation of the SM
prediction with the SUSY scale is induced by the corresponding change in the
light $\cp$-even Higgs-boson mass.
The difference between the MSSM and
the SM predictions becomes negligible for $\mste \gsim 2000 \gev$ for
$\MW$ and $\sweff$. For the other EWPO the decoupling of the
supersymmetric contributions 
occurs 
for even lower values of $\mste$. For the hadronic peak cross section,
$\si^0_{\rm had}$,  
the change with the SUSY scale in the MSSM and the SM prediction has opposite
signs (on the other hand,
for the individual factors entering $\si^0_{\rm had}$, 
see \refeq{peakcross}, the dependence of the MSSM and SM predictions on 
the SUSY scale goes into the same direction).
The fact that we
recover the most up-to-date SM prediction for the EWPO in
the decoupling limit is a consequence of the procedure
described in \refse{subsec:higherordersSM} (see \refeq{eq:obsSMSUSY}).

\newpage
\subsection{The CPX benchmark scenario}
\label{subsec:CPX}

In this section we analyse the prediction of the EWPO in the
CPX~scenario. This scenario is defined as~\cite{cpx}
\BEA
&& \msusy = 500 \gev, |\At| = 1000 \gev, \At = \Ab = A_\tau \non \\
\label{cpxpar}
&& M_2 = 200 \gev, |\mgl| = 1000 \gev, \mu = 2000 \gev\\
&& \phi_{A_{t,b,\tau}} = \phi_{\gl} = \pi/2~, \non
\label{eq:cpx}
\EEA
with the aim to indicate the possible size of $\cp$-violating effects.

The LEP Higgs searches have been interpreted within the CPX
scenario~\cite{LEPHiggsMSSM}. A parameter region for intermediate $\tb$
and and light $\MHe$ could not be excluded at the 95\% C.L., so that no
lower limit on the mass of the lightest Higgs boson could be set. The reason
for the appearance of these ``CPX holes'' is a strong suppression of the
coupling of the lightest Higgs boson to gauge bosons, while the second
lightest Higgs boson (also having a somewhat reduced coupling to gauge
bosons), which may also be within the kinematic reach of
LEP, decays predominantly into the lightest Higgs boson. The latter
leads to a rather difficult topology of the final state. 

\begin{figure}[htb!]
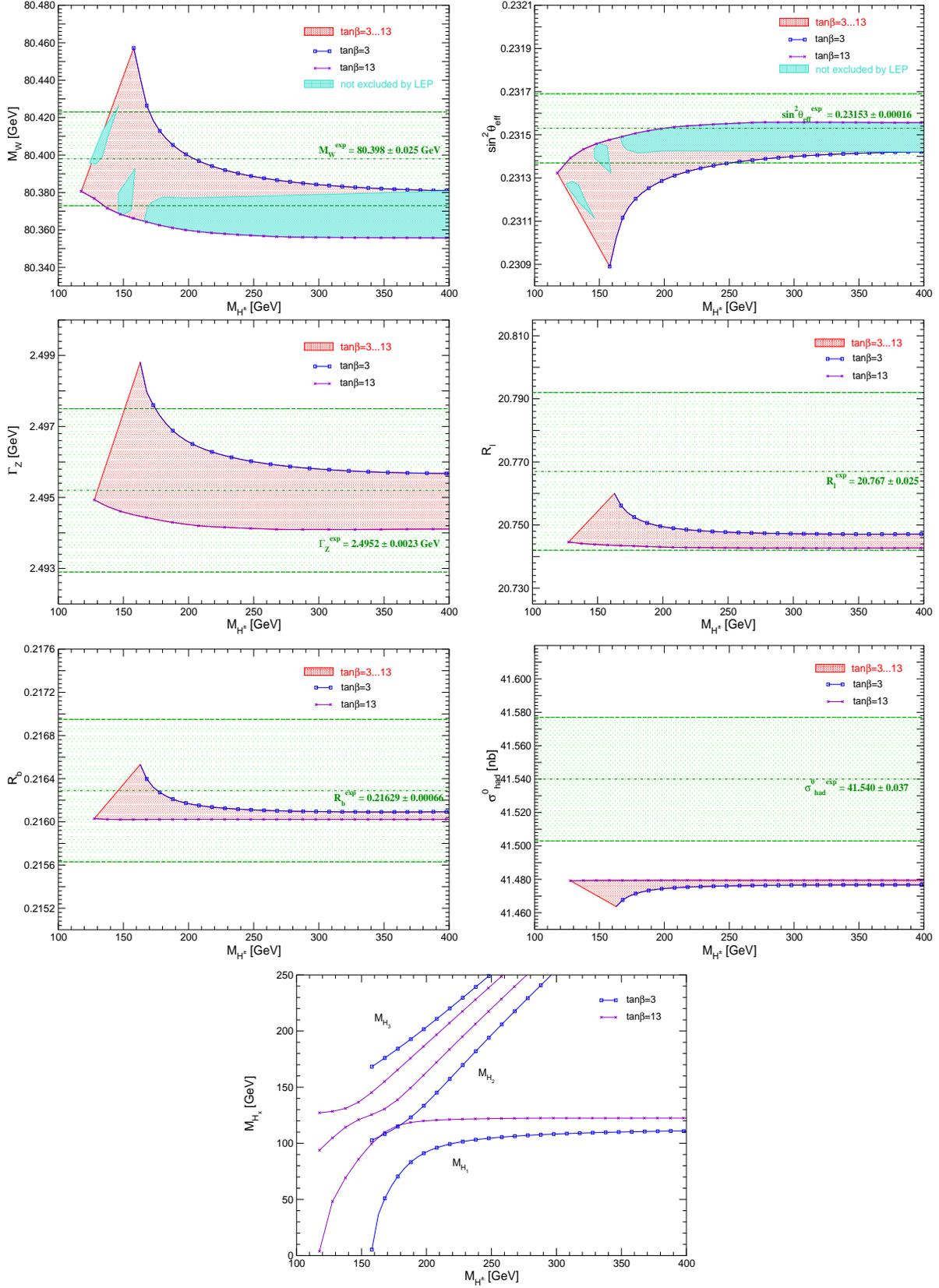

\begin{center}
\includegraphics[width=7.7cm,height=5.4cm]{MWCPXMHp.eps}
\hspace{.3em}
\includegraphics[width=7.7cm,height=5.4cm]{EffLeptMixAngleCPXMHp.eps}
\vspace{.3em}
\includegraphics[width=7.7cm,height=5.4cm]{GammaZCPXMHp.eps}
\hspace{.3em}
\includegraphics[width=7.7cm,height=5.4cm]{RlCPXMHp.eps}
\vspace{.3em}
\includegraphics[width=7.7cm,height=5.4cm]{RbCPXMHp.eps}
\hspace{.3em}
\includegraphics[width=7.7cm,height=5.4cm]{PoleCrosshadMHpCPX.eps}
\hspace{.3em}
\includegraphics[width=7.7cm,height=5.4cm]{MH123CPXMHp.eps}
\caption{Predictions for $\MW$, $\sweff$, $\Ga_Z$, $R_l$, $R_b$ and
  $\sigma_{\textup{had}}^0$ within the CPX benchmark scenario for 
  $\tb = 3\dots13$ as a function of $\MHp$, the mass of the charged
  Higgs boson. 
   In the last row the masses $\MHe$, $\MHz$, $\MHd$ of the three
   neutral Higgs bosons in the $\cp$-violating case are shown. }  
\label{fig:CPX} 
\end{center}
\vspace{-7em}
\end{figure}

As a consequence of its strongly suppressed coupling to gauge bosons,
the impact of the contributions of a light Higgs boson associated with
the ``CPX holes'' on the EWPO is expected to be rather small. 
In \reffi{fig:CPX} we investigate whether the current measurements of
the EWPO, see \refta{tab:expacc}, allow to put constraints on the
parameter space of the ``CPX holes''.
The consistent inclusion of the loop-corrected Higgs boson masses and
couplings is crucial for this analysis.

In \reffi{fig:CPX} we show the predictions of the six EWPO as a band
for the range $\tb = 3$--$13$, where the blue curves (boxes) correspond
to $\tb=3$, the purple curves (crosses) to $\tb=13$.  
The band is to the left cut off for the lowest physically allowed values
of~$\MHp$. (For illustrative purposes we also show
the values of the three neutral Higgs boson masses in the lowest plot of
\reffi{fig:CPX} as obtained with
\fh~\cite{feynhiggs,mhiggsAEC,mhcMSSMlong}.)
The regions unexcluded by the LEP Higgs searches (starting at 
$\tb \approx 3.6$) are indicated by a light blue shading for the 
two
observables $\MW$ and $\sweff$ 
(for simplicity, we have omitted
them for the other observables). As explained above,
our one-loop result contains the full complex phase dependence, 
whereas at the two-loop level we use the approximation formula given in
\refeq{phases2L}. While this method works for $\tb = 13$,
for $\tb = 3$ the value for $\phi = \pi$ (see \refeq{phases2L}) cannot
be evaluated since (at least) one of the squark mass squares turns
negative. Therefore for $\tb = 3$ we use
$O(\phi) = O^{\rm 1L}(\phi) + (O^{\rm full}(0) - O^{\rm 1L}(0))$, 
$O = \MW$, $\sweff$, $\Ga_Z$, $R_l$, $R_b$, $\si^0_{\rm had}$. 
This results in a higher theoretical uncertainty of the prediction for
the $\tb = 3$ border shown in \reffi{fig:CPX}, which should be kept in mind
for the interpretation of the figure. 

\reffi{fig:CPX} shows that the predictions for the EWPO in this
parameter region of the CPX scenario are in general in good agreement
with the experimental results. Only for low $\tb$ (i.e., values close to
the boundary of the region indicated by the contour with $\tb = 3$)
and small $\MHp$ sizable deviations can be observed, most notably for 
$\sweff$. On the other hand, the parameter regions corresponding to the 
``CPX holes'' show only small deviations from the EWPO. Most of the
corresponding regions are even within the 1~$\si$ intervals of the
experimental values of $\MW$ and  $\sweff$.
Thus, more
precise EWPO measurements combined with improved theoretical predictions 
(and correspondingly smaller intrinsic uncertainties in the EWPO 
calculations) would be needed to reach the sensitivity for probing the 
``CPX hole'' regions via their effects on EWPO%
\footnote{
In case that a sizable deviation between the predictions in the CPX
scenario and the measurements of the EWPO would occur, it would be 
important to check whether changes in the CPX
parameters (e.g.\ 
shifts in the mass parameters of the first two families) could
bring the EWPO prediction into agreement with the experimental results,
while not affecting the LEP Higgs analyses (which are mostly affected by
the parameters of the third family).
}%
.


\subsection{Scenario where no SUSY particles are observed at the LHC}
\label{sec:ILCscen}

It is interesting to investigate whether the high accuracy achievable at
the GigaZ option of the ILC would provide sensitivity to indirect effects of
SUSY particles even in a scenario where the (strongly interacting) 
superpartners are so heavy that they escape detection at the LHC.

\begin{figure}[htb!]
\begin{center}
\includegraphics[width=12.7cm,height=10.7cm]{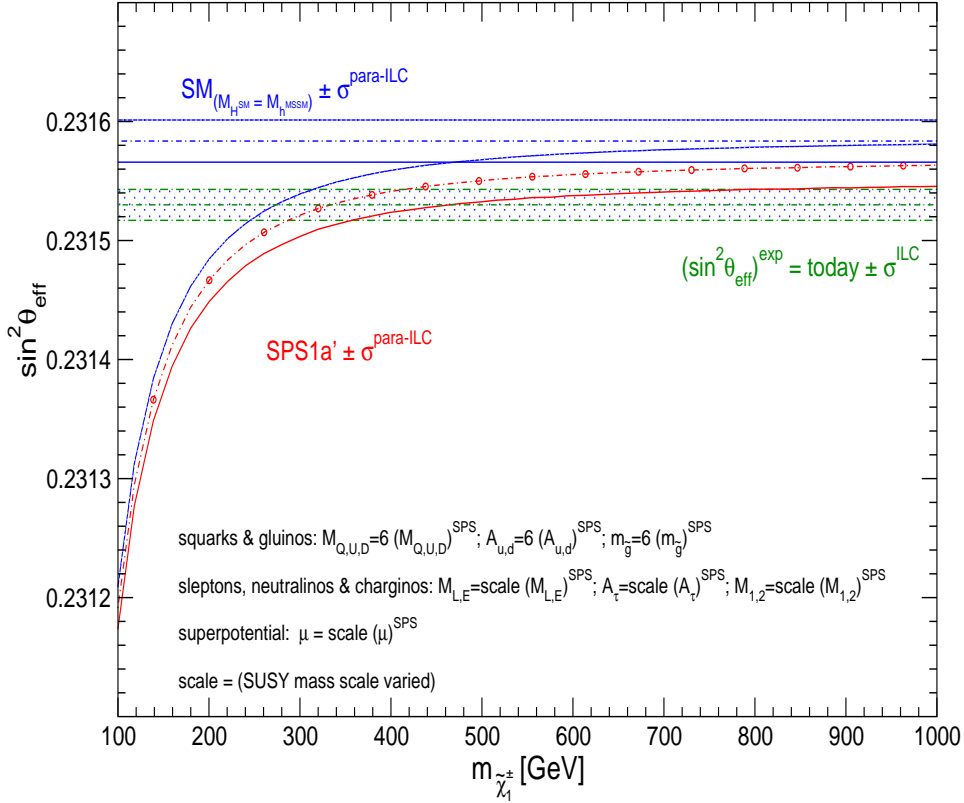}
\hspace{.3em}
\caption{
Theoretical prediction for $\sweff$ in the SM and the MSSM (including
prospective parametric theoretical uncertainties) compared to
the experimental precision at the ILC with GigaZ option.  
An SPS~1a$'$ inspired scenario is used, where the squark and gluino
mass
parameters
are fixed to 6~times their SPS~1a$'$ values. The other mass 
parameters
are 
varied with a common scalefactor, see
\refse{subsec:sps}.}  
\label{fig:ILC} 
\end{center}
\end{figure}

We consider in this context a scenario with very heavy squarks and a 
very heavy gluino. It is based on SPS~1a$'$, but the squark and gluino
mass parameters
are fixed to 6~times their SPS~1a$'$ values. The other masses are 
scaled with a common scale factor as described in \refse{subsec:sps},
except $\MA$ which we keep fixed at its SPS~1a$'$ value. In this scenario 
the strongly interacting particles are too heavy to be detected at the
LHC, while, depending on the scale-factor, some colour-neutral particles
may be in the ILC reach (the reach for the direct production of
colour-neutral particles at the
LHC will be rather limited~\cite{lhc}).
In \reffi{fig:ILC} we show the prediction for
$\sweff$ in
this SPS~1a$'$ inspired scenario as a function of the lighter chargino
mass, $\mcha{1}$. The prediction includes the parametric
uncertainty, $\si^{\rm para-ILC}$, induced by the ILC measurement of $\mt$, 
$\de\mt = 100 \mev$~\cite{mtdet1,mtdet2}, and the numerically more
relevant prospective future uncertainty on $\De\al^{(5)}_{\textup{had}}$,
$\de(\De\al^{(5)}_{\textup{had}})=5\times10^{-5}$~\cite{fredl}. 
The MSSM prediction for $\sweff$
is compared with the experimental resolution with GigaZ precision,
$\si^{\rm ILC} = 0.000013$, using for simplicity the current
experimental central value. The SM prediction (with 
$\MHSM = \Mh^{\rm MSSM}$) is also shown, applying again the parametric 
uncertainty $\si^{\rm para-ILC}$.

Despite the fact that no coloured SUSY 
particles would be observed at the LHC in this scenario, the ILC with
its high-precision 
measurement of $\sweff$ in the GigaZ mode could resolve indirect effects
of SUSY up to $m_{\tilde\chi^\pm_1} \lsim 500 \gev$. This means that the
high-precision measurements at the ILC with GigaZ option could be
sensitive to indirect effects of SUSY even in a scenario where SUSY
particles have {\em neither \/} been directly detected at the LHC nor the
first phase of the ILC with a centre of mass energy of up to $500 \gev$.

\subsection{Heavy scalar masses}
\label{subsec:splitsusy}

The scenario discussed in the previous section is characterised by
rather heavy scalar quarks (and a heavy gluino). Other scenarios 
with heavy scalar masses that found attention in recent years are the 
so-called ``split SUSY'' scenario~\cite{split} and the 
``focus point region''~\cite{focus} of the CMSSM. For completeness, we
also briefly discuss the sensitivity of the EWPO to these scenarios.

In the split SUSY scenario the fermionic masses 
(i.e.\ the chargino, neutralino, and gluino masses) are relatively light
and retain their GUT-induced hierarchy, resulting in a heavier gluino
and lighter charginos and neutralinos. The scalar mass parameters, on
the other hand, are set to very large values, larger than 
$10^6$--$10^9 \gev$, i.e., they decouple from the predictions of the EWPO.
Consequently, 
only a small deviation in the EWPO prediction from the SM limit is to be
expected in this scenario. We focus here on the two most sensitive
EWPO, $\MW$ and $\sweff$ (the results given for $\MW$ are an update of
\citere{MWweber}).

\begin{figure}[htb!]
\begin{center}
\includegraphics[width=10.7cm,height=10.7cm]{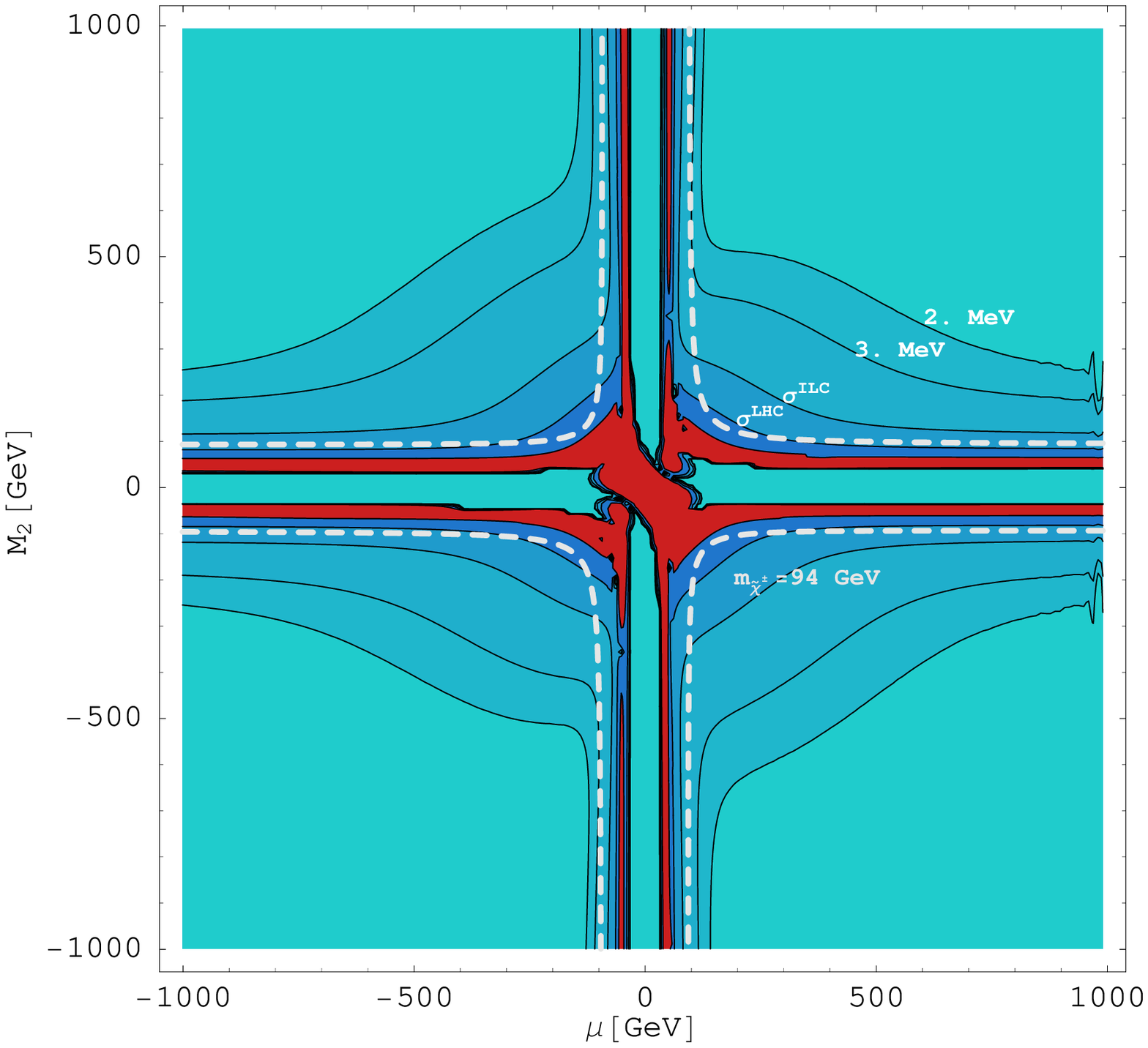}
\includegraphics[width=10.7cm,height=10.7cm]{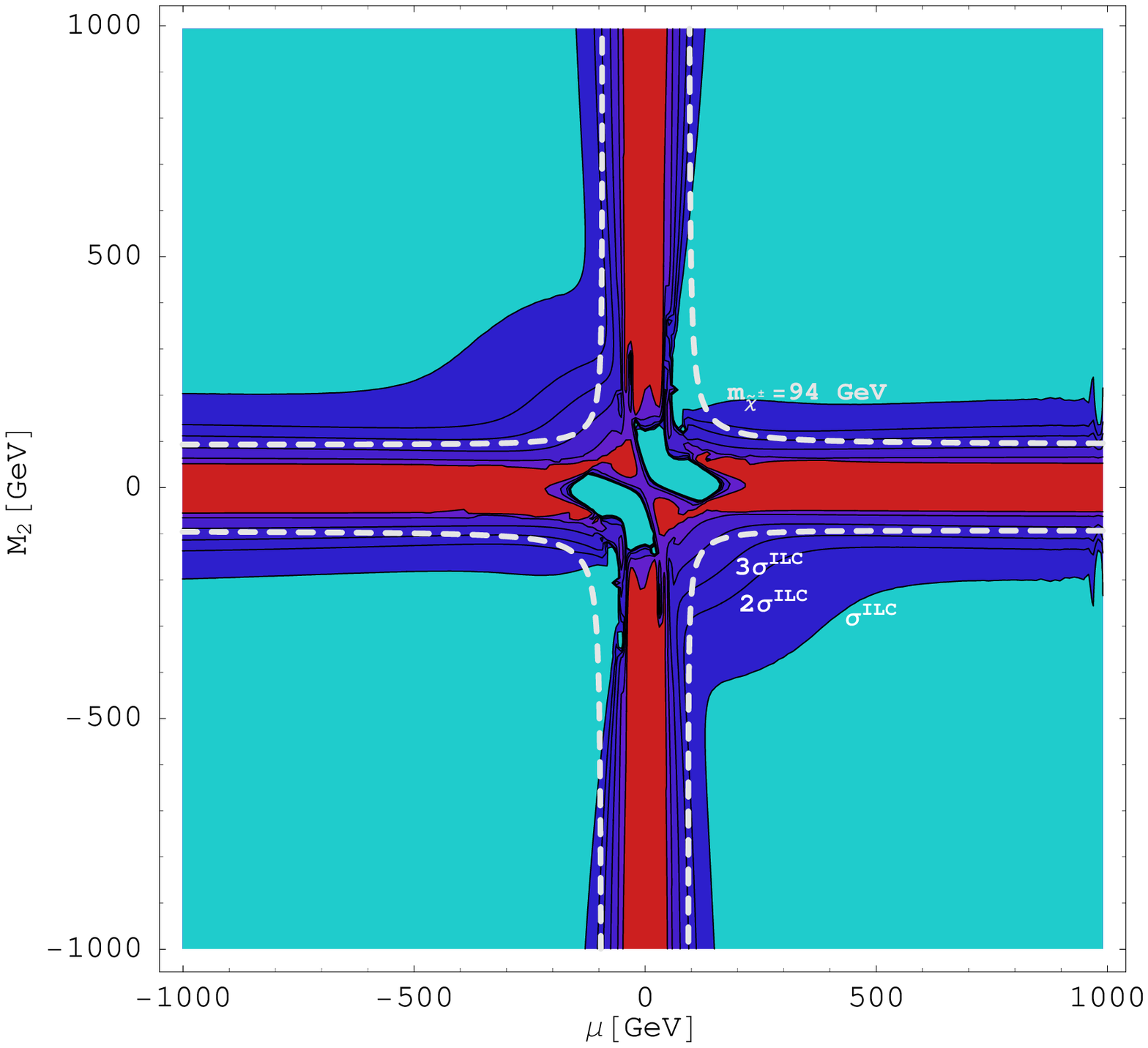}
\vspace{-1.5em}
\caption{$\de\MW=(\MW^{\textup{MSSM}}-\MW^{\textup{SM}})$ (upper plot) and
  $\de\sweff=(\sweff^{\textup{MSSM}}-\sweff^{\textup{SM}})$ (lower plot)
  in a split SUSY scenario. The SUSY parameters are 
  $\msusy = 5000\gev, \MA=2500\gev,
  \tb=10, A_\tau=A_t=A_b=2 \msusy,\mgl=500\gev$. Contour lines are drawn for
  $\de\MW=\big\{30\mev$, $\sigma^{\textup{LHC}}=15\mev$,
    $\sigma^{\textup{ILC}}=7\mev$, $3\mev$, $2\mev \big\}$ and
  $\de\sweff=\big\{30\times 10^{-5}$, $15\times 10^{-5}$,
    $3\sigma^{\textup{ILC}}=3.9\times  10^{-5}$,
    $2\sigma^{\textup{ILC}}=2.6\times  10^{-5}$, 
  $\sigma^{\textup{ILC}}=1.3\times  10^{-5} \big\}$.
} 
\label{fig:splitsusy} 
\end{center}
\vspace{-6.5em}
\end{figure}

In \reffi{fig:splitsusy} we show in a split SUSY scenario
the SUSY contribution, i.e., the
difference between the SUSY result and the corresponding SM result, 
to $\MW$ (upper plot) and to $\sweff$ (lower
plot). This is done 
by choosing a large value for $\msusy$ and subtracting the
SM result with $\MHSM = \Mh^{\rm MSSM}$ from the result obtained in the
MSSM. We have chosen
$\msusy = 5 \tev$ and $\MA = 2.5 \tev$. Choosing even higher 
values for the scalar mass parameters would only lead to 
negligible shifts as compared to the results shown
\reffi{fig:splitsusy}.  The results are displayed in \reffi{fig:splitsusy}
in the $\mu$--$M_2$ plane for $\tb = 10$. 
The gluino mass has been fixed to $\mgl = 500 \gev$ (the results are
insensitive to this choice). 
The results are given in terms of contour lines representing the
different collider precisions, see \refta{tab:expacc}.  
The region excluded by chargino searches at LEP~\cite{pdg} is inside
the dashed white lines. 
As can be seen in the figure, deviations from the SM prediction 
of more than $30 \mev$ in $\MW$
and $3 \times 10^{-4}$ in $\sweff$ 
occur only in those experimentally excluded areas. The parameter regions
leading to a shift as large as the GigaZ precisions in 
$\MW$ of $1\si^{\rm ILC} = 7 \mev$ and in $\sweff$ of 
$1\si^{\rm ILC} = 0.000013$ are significantly larger. In particular, a 
1~$\si$ effect in $\sweff$ can occur for $|\mu|, |M_2| \lsim 400 \gev$
(for negative $(\mu M_2)$ only).
(Similar results were obtained in \citere{splitMW}, see also
\citere{splitMW2}.) 

\medskip
Another scenario with heavy scalar masses is the 
focus point region~\cite{focus} in the CMSSM (the CMSSM is
characterised by a common scalar mass parameter, $m_0$, a common
fermionic mass parameter, $m_{1/2}$, and a common trilinear coupling,
$A_0$, at the GUT scale, supplemented by the low-scale parameter $\tb$
and the sign of the parameter $\mu$). In the focus point region
$m_{1/2}$ is relatively small, while $m_0$ is rather large, 
$m_0 \sim \mbox{ few} \tev$, and also $\tb$ is relatively large,
$\tb \gsim 40$.
We have investigated (using the program {\tt ISAJET~7.71}~\cite{isajet}) the
predictions for $\MW$, $\sweff$ and $\Ga_Z$ in the focus point region.
The results for $\sweff$ and $\Ga_Z$ follow the pattern observed in 
\citere{MWweber}, where $\MW$ had been analysed.
The main contributions to the shifts in the EWPO arise from
the chargino and neutralino sector (see \refse{subsec:cndep}) and only
very small effects arise from the scalar fermion contributions.
For $\tb = 50$, $\mu > 0$, $m_{1/2} = 250 \gev$, $m_0 = 1500 \gev$, 
$A_0 = -250 \gev$, corresponding to a point with the currently lowest
value of $m_{1/2}$ in the focus point region
for which the dark matter density is allowed
by WMAP and other cosmological data (see, for example, \citere{ehoww}
for a more detailed discussion), we find an effect of the SUSY
contribution of 
$6 \times 10^{-5}$ in $\sweff$ and $1 \mev$ in $\Ga_Z$. Thus, even for
the low-$m_{1/2}$ region of the focus point scenario the GigaZ precision
for $\sweff$ will be needed to gain sensitivity to indirect effects in
the EWPO.


\subsection{Higgs sector at higher orders}
\label{subsec:HSatHO}
We next investigate the impact of higher-order contributions associated
with the Higgs sector of the MSSM. 
As mentioned above, since the Higgs sector enters the EWPO only via loop
corrections, in order to evaluate predictions for the EWPO at one-loop
order it would formally be sufficient to treat the Higgs sector in
leading order, i.e., at tree~level. However, at the tree level the
predicted value for the mass of the light $\cp$-even Higgs boson is so
low that it is below the exclusion limit from the Higgs searches at 
LEP~\cite{LEPHiggsMSSM,LEPHiggsSM}. As a consequence, treating the 
MSSM Higgs sector at tree level in the predictions for the EWPO would
lead to artificially large contributions to the EWPO from the
light MSSM Higgs boson. Therefore a consistent incorporation of
higher-order contributions in the Higgs sector, as described in 
\refse{sec:higgs}, is crucial in order to be able to use realistic
mass values for the light MSSM Higgs boson and to take into account
potentially large higher-order effects.

\begin{figure}[htb!]
\begin{center}
\includegraphics[width=7.7cm,height=6.4cm]{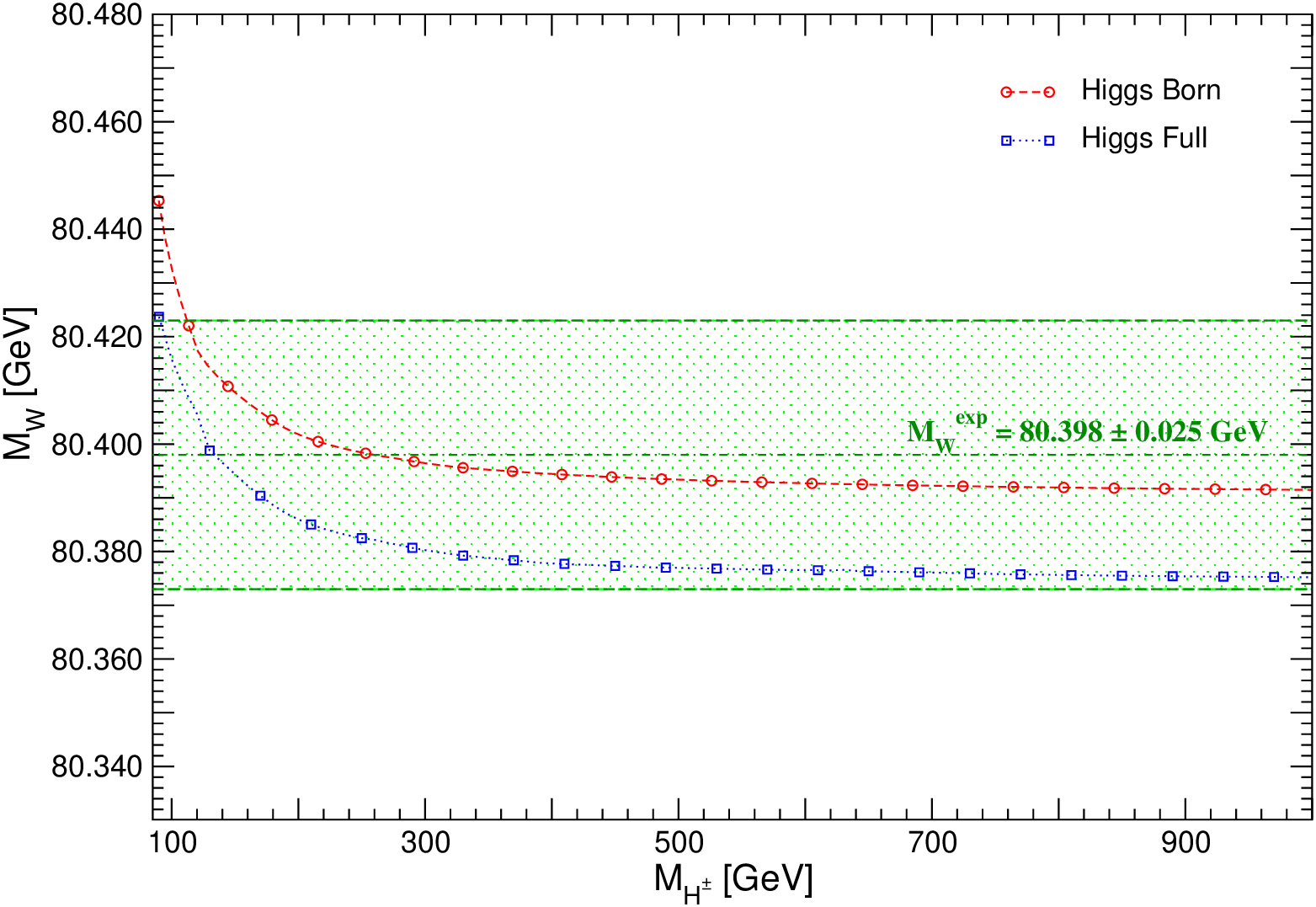}
\hspace{.3em}
\includegraphics[width=7.7cm,height=6.4cm]{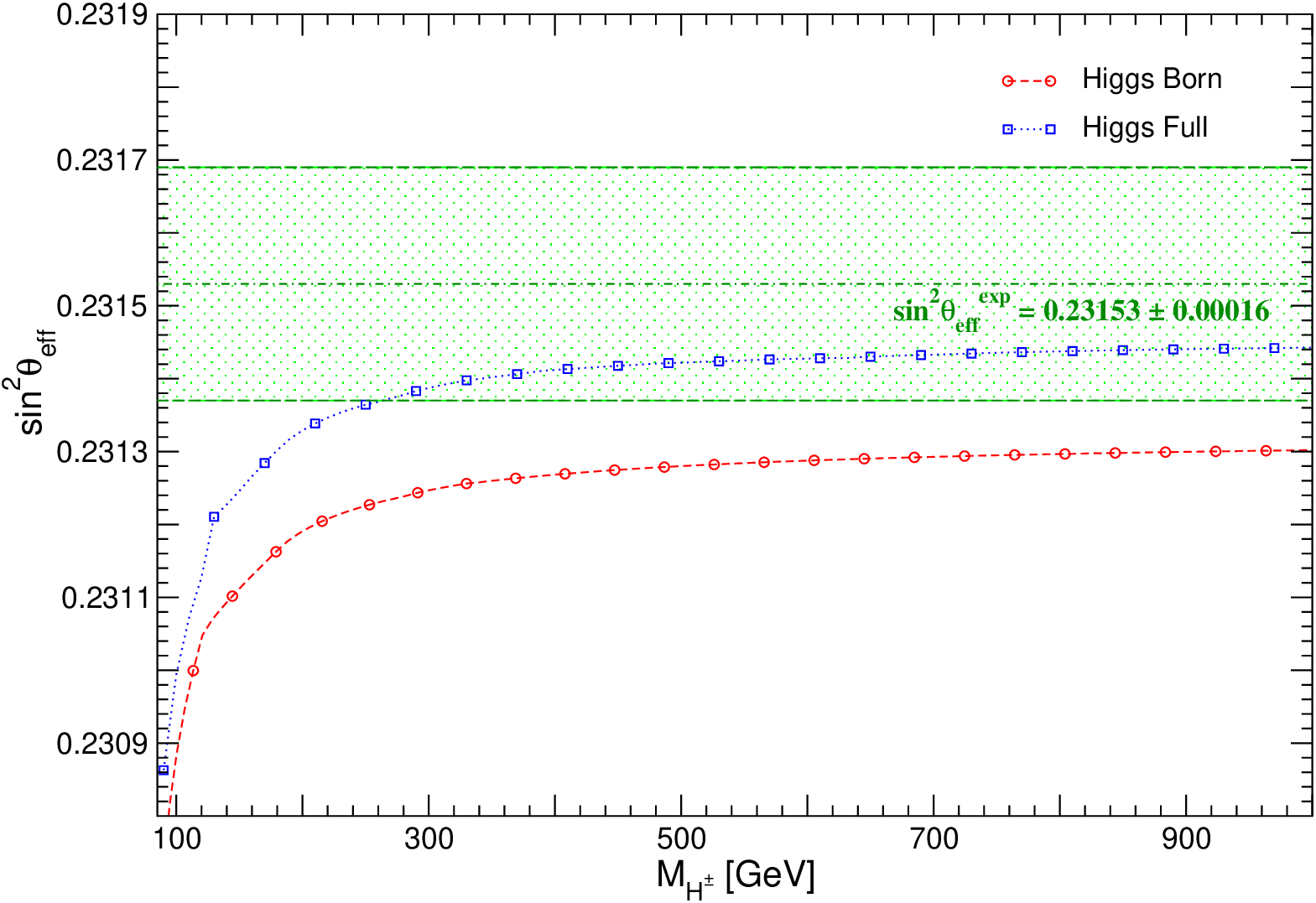}
\vspace{.3em}
\includegraphics[width=7.7cm,height=6.4cm]{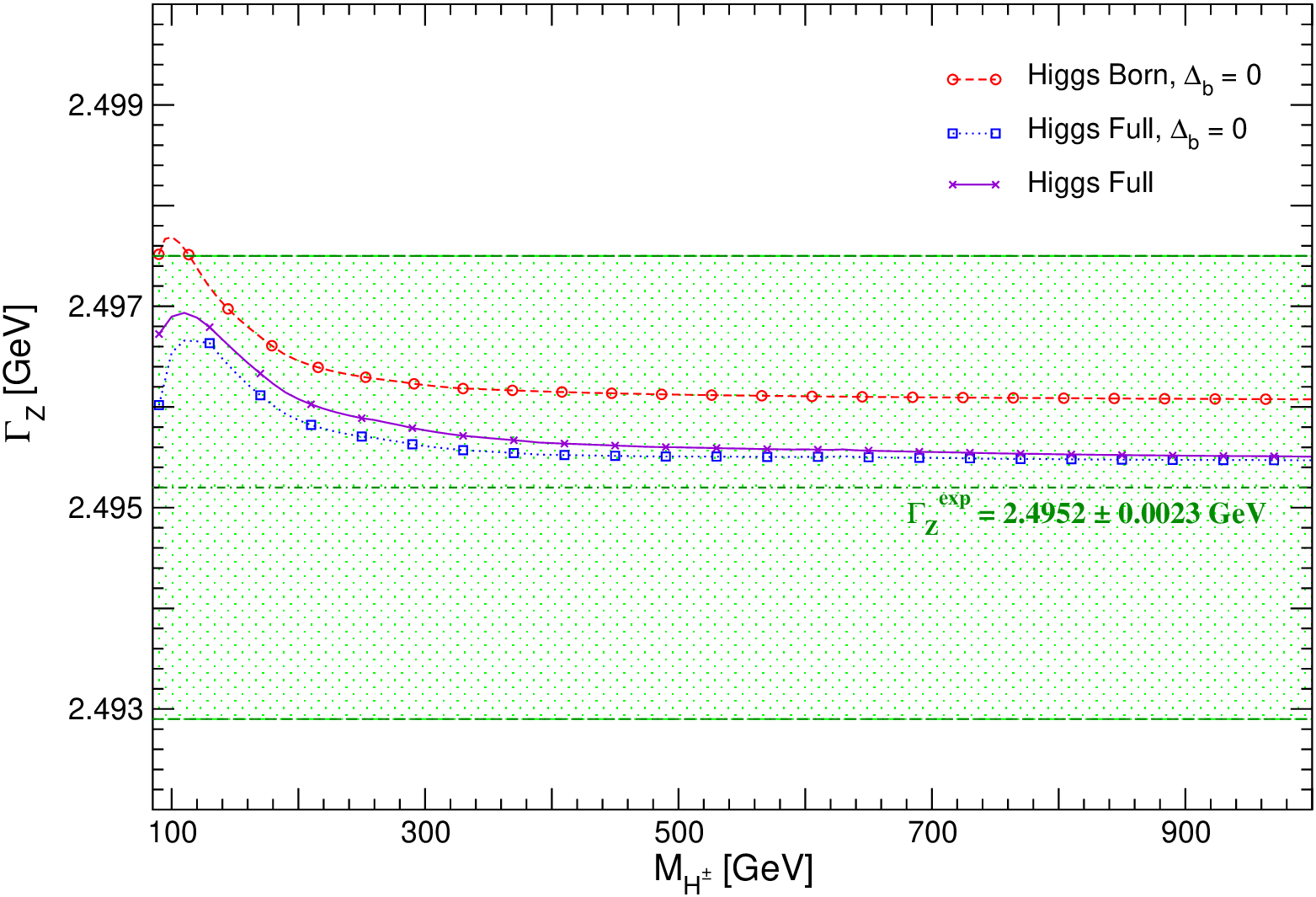}
\hspace{.3em}
\includegraphics[width=7.7cm,height=6.4cm]{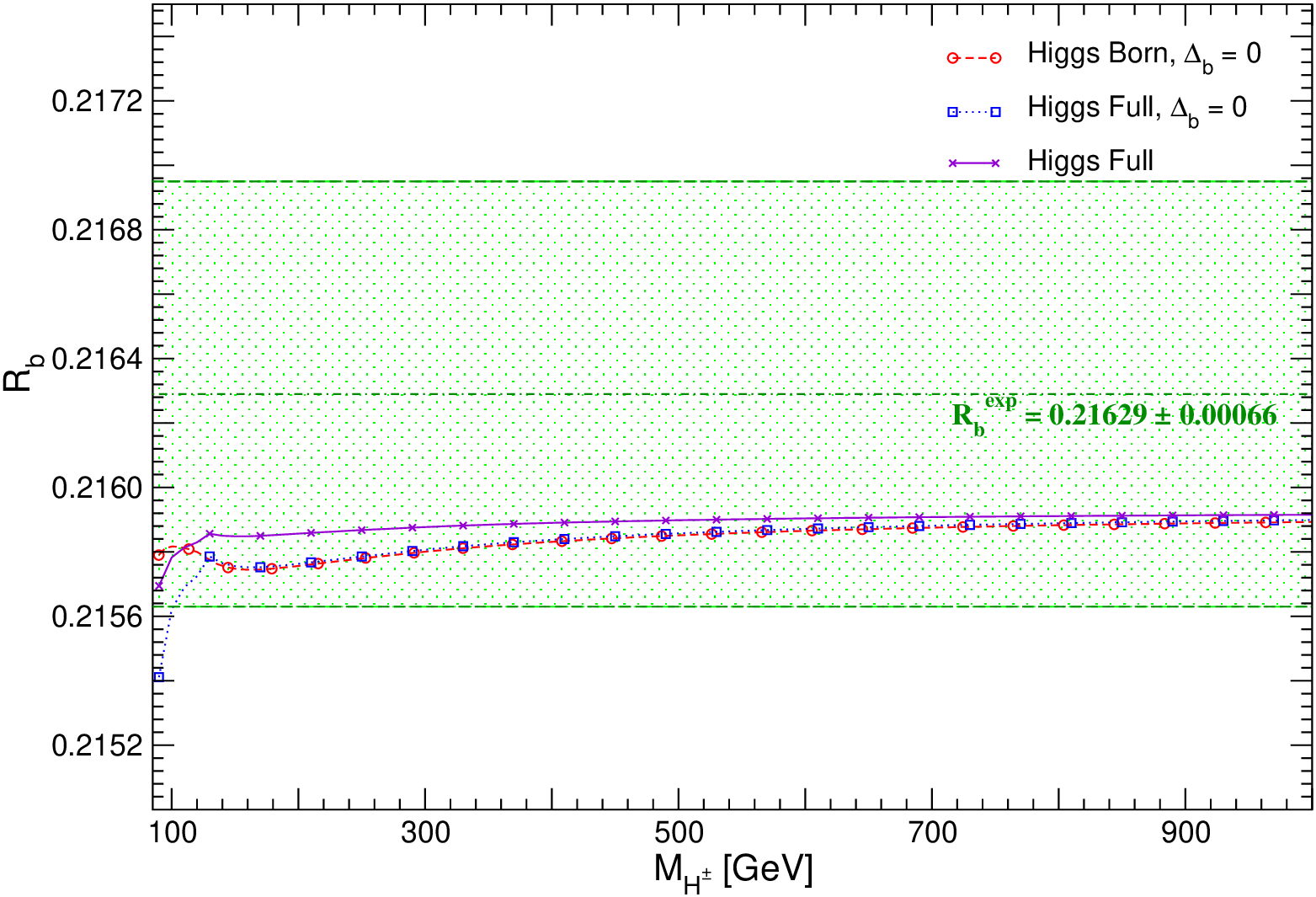}
\caption{Predictions for $\MW$, $\sweff$, $\Ga_Z$, and $R_b$ as a
function of $\MHp$. The MSSM
  parameters are $\tb =50$, $M_2=300\gev$, $\mgl=600\gev$, 
  $\msusy=300\gev$, $\mu = 300\gev$,
  $|\At|=|\Ab|=|\Atau|=2 \msusy$, $\phi_{\At}=\pi/2$,
  $\phi_{\Atau}=\phi_{\Ab}=\phi_{\mu}=\phigl=0$, 
  $\MHp = 85\gev \dots 1000 \gev$.
  ``Higgs Born'' labels the results calculated for Born-level Higgs
  sector parameters, while ``Higgs Full'' refers to the results
  calculated for the loop-corrected
  Higgs sector parameters as described in \refse{sec:higgs}. 
  For $\Gamma_Z$ and $R_b$ (second row) also the result without
  resummation of $\tb$ enhanced contributions to the bottom Yukawa
  coupling is shown, labeled as ``Higgs Full, $\De_b = 0$''.
    }  
\label{fig:HiggsBornvsFH} 
\end{center}
\end{figure}

To analyse the numerical effects of higher-order contributions in the
Higgs sector we
study $\MW$, $\sweff$, $\Gamma_Z,$ and $R_b$ for the MSSM parameter set 
$\tb = 50$, $M_2 = 300 \gev$,  $\mgl=600\gev$, 
$\msusy = 300 \gev$, $\mu = 300 \gev$,
$|\At| = |\Ab| = |\Atau| = 2 \msusy$, $\phi_{\At} = \pi/2$, 
$\phi_{\Atau} = \phi_{\Ab} = \phi_{\mu} = \phi_{M_1}= \phigl = 0$. $\MHp$ is
varied from $85\gev$ to $1000\gev$. A rather large value for $\tb$ was chosen
to further analyse a possible numerical impact of the resummation of leading
$\tb$ enhanced Higgs (s)bottom couplings in the process $Z\to b\bar b$,
entering via $\db$, see \refeq{def:dmb}. Our
results are shown in \reffi{fig:HiggsBornvsFH}, where ``Higgs Born'' labels
the numerical results where only Born-level Higgs masses and couplings were
used. ``Higgs Full'' are the results which take into account the
implementation  described in  \refse{sec:higgs}, i.e., they
account for loop-corrected Higgs masses and mixing angles, mixing of
$\cp$ eigenstates in the presence of complex MSSM parameters, and $\tb$
enhanced Higgs (s)bottom couplings, included via $\db$.

In the upper row of \reffi{fig:HiggsBornvsFH} we show the results for $\MW$
and $\sweff$. The impact of the higher-order corrections in the Higgs
sector, corresponding to the difference between the two results shown in
\reffi{fig:HiggsBornvsFH}, amounts to a shift in $\MW$ and $\sweff$ of
about $1\si$. Thus, this contribution, which is formally of two-loop
order, has a sizable numerical impact and should be
taken into account in order to arrive at a precise
prediction for $\MW$ and $\sweff$. The lower row of \reffi{fig:HiggsBornvsFH}
displays the results for $\Ga_Z$ and $R_b$, where we also show the impact of
the $\db$ corrections alone by comparing with the result where 
$\db = 0$ (labeled as ``Higgs Full, $\De_b = 0$'').
The numerical impact of the higher-order corrected Higgs boson sector on
the observables $\Ga_Z$ and $R_b$ 
relative to their current experimental errors is less pronounced
as compared to $\MW$ and
$\sweff$. Setting $\db = 0$ yields only a relatively small
shift from the full result.
It should be kept in mind that $\db$, see \refeq{def:dmb}, approximates
the leading contribution from the sbottom sector only for large SUSY
mass scales. If the relevant particle masses are simultaneously small,
$\msusy \lsim \mt$, the theoretical uncertainties in the predictions of
the EWPO can be slightly larger.

\subsection{MSSM parameter scans}
\label{subsec:scans}

As a final step of our numerical analysis we investigate the behaviour 
of the two EWPO that are most sensitive to higher-order effects in the
MSSM, $\MW$ and $\sweff$, by
scanning over a broad range of the SUSY parameter space. The following SUSY
parameters are varied independently of each other in a random parameter scan
within the given range:
\begin{eqnarray}
 {\rm sleptons} &:& M_{{\tilde F},{\tilde F'}}= 100\dots2000\gev, \non \\
 {\rm light~squarks} &:& M_{{\tilde F},{\tilde F'}_{\textup{up/down}}}
                   = 100\dots2000\gev, \non \\
 \Stop/\Sbot {\rm ~doublet} &:& 
                         M_{{\tilde F},{\tilde F'}_{\textup{up/down}}}
                    = 100\dots2000\gev,\non\\
 && A_{\tau,t,b} = -2000\dots2000\gev, \non \\
 {\rm gauginos} &:& M_{1,2}=100\dots2000\gev, \non \\
 && \mgl=195\dots1500\gev, \non \\
 && \mu = -2000\dots2000\gev,\non \\ 
 {\rm Higgs} &:& \MA=90\dots1000\gev, \non \\
 && \tb = 1.1\dots60.
\label{scaninput}
\end{eqnarray}
Those parameters which may in general be complex are taken to be real,
as the dominant effects of the complex parameters only enter via the
shifts they induce in the sparticle masses and mixings (see the
discussion in \refsubse{subsec:phases}). Effects of this kind are thus
covered by scanning over the parameters and ranges given
\refeq{scaninput}. 
Performing the scans, only the constraints on the MSSM parameter space
from the LEP Higgs searches~\cite{LEPHiggsMSSM,LEPHiggsSM} and the lower
bounds on the SUSY particle masses from direct searches as given 
in \citere{pdg} were taken into account.
Apart from these constraints no other restrictions on the MSSM parameter
space were made.

\begin{figure}[t!]
\begin{center}
\includegraphics[width=12.7cm,height=9.5cm]{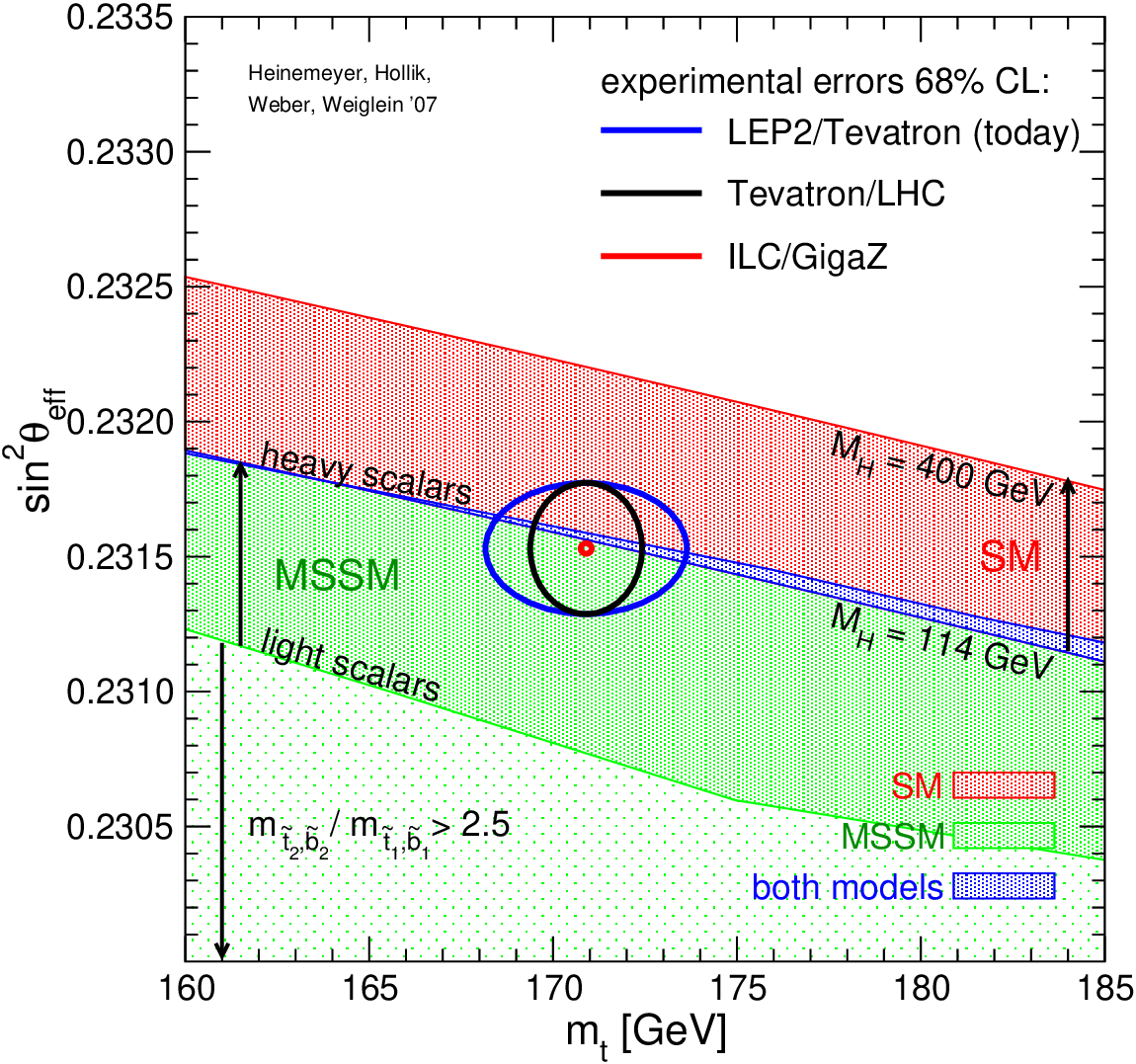}
\begin{picture}(0,0)
\CBox(-290,235)(-195,255){White}{White}
\end{picture}
\end{center}
\vspace{-2.0em}
\caption{MSSM parameter scan for $\sweff$ as a function of $\mt$ over the
  ranges given in \refeq{scaninput}.  Todays 68\%~C.L.\ ellipses
  as well as future precisions, drawn around todays central value,  are
  indicated in the plot.}  
\label{fig:Scans2} 
\vspace{-1.0em}
\end{figure}
%
\begin{figure}[b!]
\begin{center}
\includegraphics[width=12.7cm,height=9.5cm]{MWMT.eps}
\begin{picture}(0,0)
\CBox(-120,033)(-19,048){White}{White}
\end{picture}
\end{center}
\vspace{-2.0em}
\caption{MSSM random parameter scan for $\MW$ as a function of $\mt$ over the
  ranges given in \refeq{scaninput}. Todays 68\%~C.L.\ ellipses 
  as well as future precisions, drawn around todays central value,  are
  indicated in the plot.}  
\label{fig:Scans1} 
\vspace{-1.0em}
\end{figure}

\begin{figure}[htb!]
\begin{center}
\includegraphics[width=12.7cm,height=9.5cm]{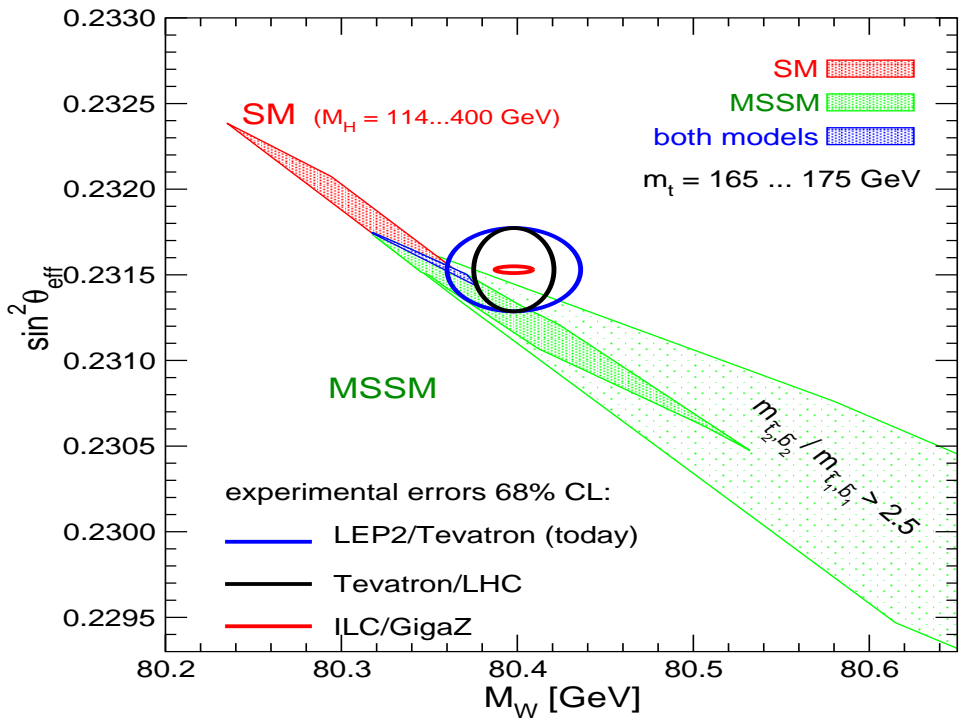}
\begin{picture}(0,0)
\CBox(-290,238)(-195,257){White}{White}
\end{picture}
\end{center}
\caption{MSSM random parameter scan over the ranges given in
\refeq{scaninput}. The top-quark mass is treated as an additional free
parameter and is varied in the range $\mt=165\dots175\gev$. Shown is the
combination of $\MW$ and $\sweff$. Todays 68\%~C.L.\ ellipses  as well
as future precisions, drawn around todays central value,  are  indicated
in the plot.}   
\label{fig:Scans3} 
\end{figure}

In \reffi{fig:Scans2} we
compare the SM and the MSSM predictions for $\sweff$
as a function of $\mt$ as obtained from the scatter data. 
The predictions within the two models 
give rise to two bands in the $\mt$--$\sweff$ plane with only a relatively 
small overlap region (indicated by a dark-shaded (blue) area).
The allowed parameter region in the SM (the medium-shaded (red)
and dark-shaded (blue) bands) arises from varying the only free parameter 
of the model, the mass of the SM Higgs boson, from $\MHSM = 114\gev$, the LEP 
exclusion bound~\cite{LEPHiggsSM}
(lower edge of the dark-shaded (blue) area), to $400 \gev$ (upper edge of the
medium-shaded (red) area).
The very light-shaded (green), the light shaded (green) and the
dark-shaded (blue) areas indicate allowed regions for the unconstrained
MSSM. In the very light-shaded region at least one of the ratios
$\mstz/\mste$ or $\msbz/\msbe$ exceeds~2.5,%
\footnote{
We work in the convention that $\msfe \le \msfz$.
}%
~while the decoupling limit with SUSY masses of \order{2 \tev}
yields the upper edge of the dark-shaded (blue) area. Thus, the overlap 
region between the predictions of the two models corresponds in the SM
to the region where the Higgs boson is light, i.e., in the MSSM allowed
region ($\Mh \lsim 130 \gev$~\cite{feynhiggs,mhiggsAEC}). In the MSSM it
corresponds to the case where all 
superpartners are heavy, i.e., the decoupling region of the MSSM.
The 68\%~C.L.\ experimental results
for $\mt$ and $\sweff$ are indicated in the plot. As can be seen from
\reffi{fig:Scans2}, the current experimental 68\%~C.L.\ region for 
$\mt$ and $\sweff$ is in good agreement with both models and does not 
indicate a preference for one of the two models.
The prospective accuracies for the Tevatron/LHC 
($\de\sweff^{\rm Tevatron/LHC} = 0.00016$, 
$\de\mt^{\rm Tevatron/LHC} = 1 \gev$) and the ILC with GigaZ option 
($\de\sweff^{\rm ILC/GigaZ} = 0.000013$,
$\de\mt^{\rm ILC/GigaZ} = 0.1 \gev$) 
are also shown in the plot (using the current central values),
indicating the strong potential for a significant improvement of the
sensitivity of the electroweak precision tests~\cite{gigaz}.

In \reffi{fig:Scans1} we compare the SM and the MSSM predictions for $\MW$
as a function of $\mt$ as obtained from the scatter data%
\footnote{
The plot shown here is an update of 
\citeres{MWMSSM1LA,PomssmRep,MWweber}.
}%
.~The ranges of the varied parameters and the band structure of the SM
and MSSM predictions are analogous to the ones 
in \reffi{fig:Scans2}. 
The experimental value for $\MW$
includes the latest CDF~measurement~\cite{MWcdf}, resulting in the 
world average of $\MW = (80.398 \pm 0.025) \gev$~\cite{MWcdf,LEPEWWG2}. 
The 68\% C.L.\ region for $\mt$ and $\MW$ exhibits a preference for
the MSSM over the SM.
The prospective accuracies for the Tevatron/LHC 
($\de\MW^{\rm Tevatron/LHC} = 15 \mev$) and the ILC with GigaZ option 
($\de\MW^{\rm ILC/GigaZ} = 7 \mev$) 
are also shown in the plot (using the current
central value).

Finally in \reffi{fig:Scans3} we show the combination
of $\MW$ and $\sweff$ with the top-quark mass varied in the range of 
$165 \gev$ to $175 \gev$.
The ranges of the other varied parameters and the colour coding 
are the same as in \reffis{fig:Scans2}, \ref{fig:Scans1}.
The current 68\%~C.L.\ experimental results
for $\MW$ and $\sweff$ are indicated in the plot. The region of the SM
prediction
inside todays 68\%~C.L.\ ellipse corresponds to relatively large $\mt$ values,
outside the current experimental range of 
$\mt=(170.9\pm1.8)\gev$~\cite{mt1709}.
Thus, the combination of $\MW$ and $\sweff$ exhibits a slight preference
for the MSSM over the SM. Again also shown are the anticipated future
improvements in the measurements of $\MW$ and $\sweff$.



\section{Remaining higher-order uncertainties for 
   \boldmath{$\sweff$}}
\label{sec:theounc}
Following the discussion in~\citere{MWweber} we now estimate the
missing higher order uncertainties in the prediction of $\sweff$. As
seen in \refse{sec:numanal}, besides the $W$~boson mass, $\sweff$ is the
observable with the most pronounced dependence on higher-order SUSY
contributions. Thus, it is important to reduce its intrinsic theoretical
uncertainty from unknown higher-order (SM and SUSY) correctins 
sufficiently below its experimental error and its parametric theoretical
uncertainties (induced by the experimental errors of the input
parameters).
The remaining SM theory uncertainty was estimated to
be~\cite{BayernFermionic,dkappaSMbos2L}
\begin{equation}
\de\sweff^{\rm SM} = 4.7 \times 10^{-5}.
\label{eq:SMunc}
\end{equation}
This corresponds to the theory uncertainty of our prediction for
$\sweff$ in the decoupling limit of the MSSM, since we have
incorporated all known SM higher order contributions into our result
(see \refse{sec:calclepobs}). 

As detailed in \citeres{MWweber,drMSSMal2B},
additional theoretical uncertainties arise from higher-order
corrections involving supersymmetric particles in the loops.  
Depending on the overall SUSY scale these uncertainties were estimated
to be~\cite{drMSSMal2B} 
\begin{eqnarray}
\de\sweff &=& 4.7 \times 10^{-5} \mbox{ for } \msusy < 500 \gev , \non \\
\de\sweff &=& 1.5 \times 10^{-5} \mbox{ for } \msusy = 500 \gev ,  
                                                   \label{eq:sw2effunc}\\
\de\sweff &=& 1.3 \times 10^{-5} \mbox{ for } \msusy = 1000 \gev . \non 
\end{eqnarray}
Adding SM and SUSY uncertainties in quadrature one finds
$\de\sweff=(4.9 \, - 6.6) \times 10^{-5}$, depending on the SUSY mass
scale~\cite{drMSSMal2B}. 

Additional theory uncertainties arise for complex parameters, as we
only include the full phase dependence at the one-loop level. However,
MSSM two-loop terms which are only known for real MSSM parameters are
incorporated into the full prediction for $\sweff$ via the
interpolation relation \refeq{phases2L}.  
Following the prescription in \citere{MWweber}, where the corresponding
error in the $\MW$ evaluation had been obtained, we estimate here the
maximal error over the $\phiat\in[0,\pi]$ interval to be%
\footnote{ As representative SUSY scenarios we have chosen SPS~1a$'$,
SPS~1b, and SPS~5, each for $\msusy = 1000\gev$, $500\gev$, and for
$\msusy<500\gev$.
The lowest values considered for $\msusy$ are roughly 300, 300, 400 $\gev$
for SPS~1a$'$, SPS~1b, SPS~5, respectively. 
For lower values the parameter points are 
excluded by Higgs mass constraints. The light stop mass for the SPS~5
point lies considerably below $400\gev$.
}%
\begin{eqnarray}
\de\sweff &=& 2.6 \times 10^{-5} \mbox{ for } \msusy < 500 \gev , \non \\
\de\sweff &=& 1.7 \times 10^{-5} \mbox{ for } \msusy = 500 \gev ,  
                                                 \label{eq:sw2effunc2}\\
\de\sweff &=& 0.6 \times 10^{-5} \mbox{ for } \msusy = 1000 \gev . \non 
\end{eqnarray}
Similar values are found in an independent approach: the approximation
formula \refeq{phases2L} can be applied to the one-loop case, see
\refse{subsubsec:twoloop}. The difference between the approximation and
the full phase dependence at one-loop order is scaled to the two-loop
level by applying a conservative factor of 0.2 (obtained from the
analyses of the SUSY contributions to the
$\rho$~parameter~\cite{dr2lA,drMSSMal2B}). The result for the
estimate of the uncertainty induced by the approximation formula for the
phase dependence at the two-loop level is similar (slightly below) to
the numbers in \refeq{eq:sw2effunc2}.

The full theoretical uncertainty from unknown higher-order corrections
in the MSSM with complex parameters can
now be obtained by adding in 
quadrature the SM uncertainties from \refeq{eq:SMunc}, the theory
uncertainties from \refeq{eq:sw2effunc} and the additional 
SUSY uncertainties from \refeq{eq:sw2effunc2}. This yields 
$\de\sweff=(4.9 - 7.1) \times 10^{-5}$ depending on the SUSY mass scale.

The other source of theoretical uncertainties, besides the one from
unknown higher-order corrections, 
is the parametric uncertainty induced by the experimental errors of
the input parameters. The corresponding uncertainties for $\de\mt =
1.8 \gev$ and $\de(\De \al_{\rm had}^{(5)})$ are
given in \refeqs{eq:dMWdmt} -- (\ref{eq:dGZdal}). 
The uncertainty in $\mt$ will decrease during the next years as a
consequence of  
a further improvement of the accuracies at the Tevatron and the
LHC. Ultimately it will be reduced by more than an order of magnitude at
the ILC~\cite{mtdet1,mtdet2}, see also the phenomenological analysis in
\citere{deltamt}. For $\De\al_{\rm had}^{(5)}$ one can hope for 
an improvement down to $5 \times 10^{-5}$~\cite{fredl}, reducing the
parametric uncertainty to the $1.8 \times 10^{-5}$ level (for a
discussion of the parametric uncertainties  
induced by the other SM input parameters see, for example, \citere{PomssmRep}).
In order to reduce the theoretical uncertainties from unknown
higher-order corrections to the level of $1.8 \times 10^{-5}$, further
results on SM-type corrections beyond two-loop order and higher-order
corrections involving supersymmetric particles will be necessary.


\section{Conclusions}
\label{sec:conclusions}

We have presented the currently most accurate evaluation of $Z$~pole
observables in the MSSM. These comprise the effective weak mixing
angle, $\sweff$, $Z$~decay widths to SM fermions, 
$\Ga(Z \to f \bar f)$, the invisible and total width, $\Ga_{\rm inv}$ and
$\Ga_Z$, forward-backward and left-right asymmetries, $A_{\rm FB}$ and
$A_{\rm LR}$, and the total hadronic cross section, $\si^0_{\rm had}$. 
The calculation includes the complete one-loop results, for the first time 
taking into account the full complex phase dependence. It furthermore
includes all existing higher-order corrections available in the MSSM. 
We have also incorporated all available SM contributions
that go beyond the existing MSSM corrections. These corrections can be
numerically important since 
the evaluations of higher-order corrections in the SM are more
advanced than in the MSSM. As a result, in the decoupling limit 
(i.e., if all SUSY mass parameters are large) our calculation 
reproduces the currently most up-to-date SM results for the
$Z$~pole observables.
Concerning the $Z$~boson decay widths we
obtained for the first time a full one-loop result for 
$\Ga(Z \to \neu{1}\neu{1})$ that contributes to the invisible width of
the $Z$~boson if this decay is kinematically possible.
A public computer code based on the results for electroweak precision
observables (the $Z$~pole observables and the $W$~boson
mass~\cite{MWweber}) is in preparation~\cite{pope}.

We presented a detailed numerical analysis of the impact of the various
MSSM sectors on the prediction of the EWPO. 
We find that the EWPO $\MW$, $\sweff$ and $\Ga_Z$
are sensitive to variations of
the $\Stop/\Sbot$~sector parameters
and to a lesser extent of the chargino/neutralino parameters.
The impact on the other $Z$~pole
observables was found to be much smaller and well within one
experimental standard deviation. 
The evaluation of $\MW$ and $\sweff$ incorporating higher-order
corrections to Higgs boson masses and couplings differs by about
$1\,\si$ from the predictions where the tree-level approximation for the
Higgs sector is used.

Particular emphasis in the analysis was put on the impact
of the complex phases at the one-loop level. The largest phase
dependence arises in general in the scalar top sector. Large effects are
also possible in the scalar bottom
sector if $\tb$ is large. Shifts induced by a phase variation exceeding
one experimental standard deviations can occur for $\MW$, $\sweff$ and
$\Ga_Z$, depending on the other MSSM parameters. 
It should be noted, however, that the dependence on the phases appearing
in the $\Stop/\Sbot$~sector at the one-loop level drop out of the explicit
one-loop terms for the effective electroweak couplings~$g^f_{\{V,A\},(1)}$
and thus only enter via their impact on the squark masses and mixing angles. 
Consequently, a measurement of the $\cp$-conserving EWPO
alone cannot serve to reveal the presence of the complex
phases. However, precise measurements of the EWPO in combination with
other experimental information will be very valuable for constraining
the SUSY parameters including the complex phases.

By comparing our MSSM result with the corresponding SM result 
we have illustrated the sensitivity of the EWPO to the virtual effects
of SUSY particles. We pointed out that the anticipated
experimental accuracy at future colliders, in particular at the ILC with
GigaZ option,
might resolve the virtual effects of SUSY particles, even in scenarios 
where the SUSY particles are so heavy that they escape direct detection 
at the LHC and the first phase of the ILC.

The EWPO have been analysed in the framework of several MSSM
scenarios. As typical examples we investigated three SPS-like
scenarios. Further studied scenarios were the ``focus point'' region, 
``split SUSY''
and the CPX scenario. In the ``focus point'' and ``split SUSY''
scenarios, both containing heavy
scalar fermions, only very small deviations from the corresponding SM result
could be observed. Within the CPX scenario we found that the 
current experimental precision of the EWPO is not yet sufficient to
probe the parameter
space at low Higgs boson masses that could not be covered by the LEP
Higgs searches.

Finally, we have analysed the theoretical uncertainty in the $\sweff$
prediction arising from the incomplete inclusion of complex phases
beyond the one-loop level. Our estimate yields that this (additional)
uncertainty can amount up to $2.6 \times 10^{-5}$, depending on the SUSY
parameters. The combination of this new uncertainty 
with the one estimated previously for the MSSM with real parameters
yields an estimate of the intrinsic uncertainty of up to 
$7 \times 10^{-5}$ for small SUSY masses. 

The results derived for the purpose of this paper have already found
application in various precision analysis 
projects~\cite{ehoww,AllanachFit,mastercode}, which are aimed 
at providing indirect constraints on the scale of supersymmetry
based on current constraints
from cosmology, $B$~physics, and electroweak precision data.


\subsection*{Acknowledgements}

We thank G.~Moortgat-Pick for interesting discussions concerning 
\refse{sec:ILCscen}. 
We thank J.~Erler and K.~M\"onig for discussions on the future
experimental uncertainties of the $Z$~pole observables. 
We are further grateful to P.~Bechtle for providing us with data to allow a
detailed analysis of the CPX benchmark scenario.
A.M.W.~would like to thank the Instituto de Fisica de Cantabria
(CSIC-UC), Santander for kind hospitality during part of this work.
Work supported in part by the European Community's Marie-Curie Research
Training Network under contract MRTN-CT-2006-035505 
`Tools and Precision Calculations for Physics Discoveries at Colliders'
(HEPTOOLS).


\newpage

\appendix

\section{Appendix}

Here we show generic Feynman-diagrams contributing to the $Z$~pole
observables in the MSSM.

\subsection{One-loop vertex graphs for the process $Z\to f \bar f$}

\begin{figure}[htb!]
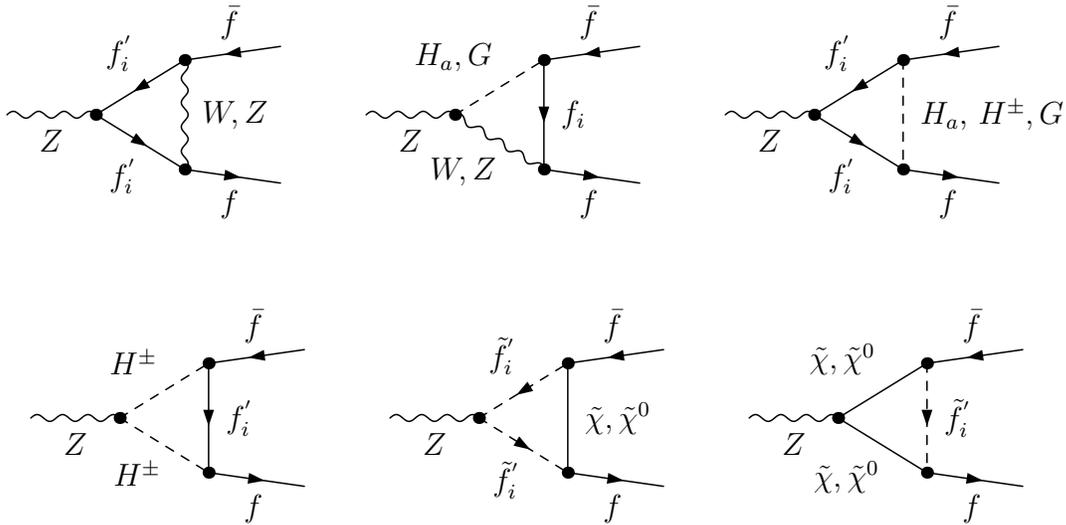

\centerline{
\unitlength=1.cm%
\begin{feynartspicture}(4,4)(1,1)
\FADiagram{}
\FAProp(0.,10.)(6.5,10.)(0.,){/Sine}{0}
\FALabel(3.25,8.93)[t]{$Z$}
\FAProp(20.,15.)(13.,14.)(0.,){/Straight}{1}
\FALabel(16.2808,15.5544)[b]{$\bar f$}
\FAProp(20.,5.)(13.,6.)(0.,){/Straight}{-1}
\FALabel(16.2808,4.44558)[t]{$f$}
\FAProp(6.5,10.)(13.,14.)(0.,){/Straight}{-1}
\FALabel(9.20801,13.1807)[br]{$f_{i}^{'}$}
\FAProp(6.5,10.)(13.,6.)(0.,){/Straight}{1}
\FALabel(9.5,6.81927)[tr]{$f_{i}^{'}$}
\FAProp(13.,14.)(13.,6.)(0.,){/Sine}{0}
\FALabel(14.274,10.)[l]{$W, Z$}
\FAVert(6.5,10.){0}
\FAVert(13.,14.){0}
\FAVert(13.,6.){0}
\end{feynartspicture}
\hspace{0.5cm}
\begin{feynartspicture}(4,4)(1,1)
\FADiagram{}
\FAProp(0.,10.)(6.5,10.)(0.,){/Sine}{0}
\FALabel(3.25,8.93)[t]{$Z$}
\FAProp(20.,15.)(13.,14.)(0.,){/Straight}{1}
\FALabel(16.2808,15.5544)[b]{$\bar f$}
\FAProp(20.,5.)(13.,6.)(0.,){/Straight}{-1}
\FALabel(16.2808,4.44558)[t]{$f$}
\FAProp(6.5,10.)(13.,14.)(0.,){/ScalarDash}{0}
\FALabel(9.20801,13.1807)[br]{$H_a, G$}
\FAProp(6.5,10.)(13.,6.)(0.,){/Sine}{0}
\FALabel(9.5,6.81927)[tr]{$W, Z$}
\FAProp(13.,14.)(13.,6.)(0.,){/Straight}{1}
\FALabel(14.274,10.)[l]{$f_i$}
\FAVert(6.5,10.){0}
\FAVert(13.,14.){0}
\FAVert(13.,6.){0}
\end{feynartspicture}
\hspace{0.5cm}
\begin{feynartspicture}(4,4)(1,1)
\FADiagram{}
\FAProp(0.,10.)(6.5,10.)(0.,){/Sine}{0}
\FALabel(3.25,8.93)[t]{$Z$}
\FAProp(20.,15.)(13.,14.)(0.,){/Straight}{1}
\FALabel(16.2808,15.5544)[b]{$\bar f$}
\FAProp(20.,5.)(13.,6.)(0.,){/Straight}{-1}
\FALabel(16.2808,4.44558)[t]{$f$}
\FAProp(6.5,10.)(13.,14.)(0.,){/Straight}{-1}
\FALabel(9.20801,13.1807)[br]{$f_i^{'}$}
\FAProp(6.5,10.)(13.,6.)(0.,){/Straight}{1}
\FALabel(9.5,6.81927)[tr]{$f_i^{'}$}
\FAProp(13.,14.)(13.,6.)(0.,){/ScalarDash}{0}
\FALabel(14.274,10.)[l]{$H_a$, $H^{\pm}, G$}
\FAVert(6.5,10.){0}
\FAVert(13.,14.){0}
\FAVert(13.,6.){0}
\end{feynartspicture}
\hspace{0.5cm}
}
\centerline{
\unitlength=1.cm%
\begin{feynartspicture}(4,4)(1,1)
\FADiagram{}
\FAProp(0.,10.)(6.5,10.)(0.,){/Sine}{0}
\FALabel(3.25,8.93)[t]{$Z$}
\FAProp(20.,15.)(13.,14.)(0.,){/Straight}{1}
\FALabel(16.2808,15.5544)[b]{$\bar f$}
\FAProp(20.,5.)(13.,6.)(0.,){/Straight}{-1}
\FALabel(16.2808,4.44558)[t]{$f$}
\FAProp(6.5,10.)(13.,14.)(0.,){/ScalarDash}{0}
\FALabel(9.20801,13.1807)[br]{$H^\pm$}
\FAProp(6.5,10.)(13.,6.)(0.,){/ScalarDash}{0}
\FALabel(9.5,6.81927)[tr]{$H^\pm$}
\FAProp(13.,14.)(13.,6.)(0.,){/Straight}{1}
\FALabel(14.274,10.)[l]{$f_i^{'}$}
\FAVert(6.5,10.){0}
\FAVert(13.,14.){0}
\FAVert(13.,6.){0}
\end{feynartspicture}
\hspace{0.5cm}
\begin{feynartspicture}(4,4)(1,1)
\FADiagram{}
\FAProp(0.,10.)(6.5,10.)(0.,){/Sine}{0}
\FALabel(3.25,8.93)[t]{$Z$}
\FAProp(20.,15.)(13.,14.)(0.,){/Straight}{1}
\FALabel(16.2808,15.5544)[b]{$\bar f$}
\FAProp(20.,5.)(13.,6.)(0.,){/Straight}{-1}
\FALabel(16.2808,4.44558)[t]{$f$}
\FAProp(6.5,10.)(13.,14.)(0.,){/ScalarDash}{-1}
\FALabel(9.20801,13.1807)[br]{$\tilde f_i^{'}$}
\FAProp(6.5,10.)(13.,6.)(0.,){/ScalarDash}{1}
\FALabel(9.5,6.81927)[tr]{$\tilde f_i^{'}$}
\FAProp(13.,14.)(13.,6.)(0.,){/Straight}{0}
\FALabel(14.274,10.)[l]{$\tilde \chi, \tilde \chi^0$}
\FAVert(6.5,10.){0}
\FAVert(13.,14.){0}
\FAVert(13.,6.){0}
\end{feynartspicture}
\hspace{0.5cm}
\begin{feynartspicture}(4,4)(1,1)
\FADiagram{}
\FAProp(0.,10.)(6.5,10.)(0.,){/Sine}{0}
\FALabel(3.25,8.93)[t]{$Z$}
\FAProp(20.,15.)(13.,14.)(0.,){/Straight}{1}
\FALabel(16.2808,15.5544)[b]{$\bar f$}
\FAProp(20.,5.)(13.,6.)(0.,){/Straight}{-1}
\FALabel(16.2808,4.44558)[t]{$f$}
\FAProp(6.5,10.)(13.,14.)(0.,){/Straight}{0}
\FALabel(9.20801,13.1807)[br]{$\tilde \chi, \tilde \chi^0$}
\FAProp(6.5,10.)(13.,6.)(0.,){/Straight}{0}
\FALabel(9.5,6.81927)[tr]{$\tilde \chi, \tilde \chi^0$}
\FAProp(13.,14.)(13.,6.)(0.,){/ScalarDash}{1}
\FALabel(14.274,10.)[l]{$\tilde f_i^{'}$}
\FAVert(6.5,10.){0}
\FAVert(13.,14.){0}
\FAVert(13.,6.){0}
\end{feynartspicture}
}
\caption{Generic vertex contributions to $Z\to f \bar f$.}
\label{fig:vertexoneloop}
\end{figure}

\subsection{SUSY QCD one-loop graphs for the    process $Z\to q\bar q$}
\begin{figure}[htb!]
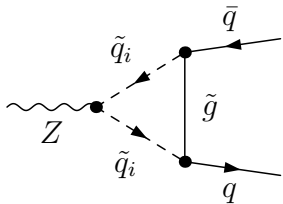

\centerline{
\unitlength=1.cm%
\begin{feynartspicture}(4,4)(1,1)
\FADiagram{}
\FAProp(0.,10.)(6.5,10.)(0.,){/Sine}{0}
\FALabel(3.25,8.93)[t]{$Z$}
\FAProp(20.,15.)(13.,14.)(0.,){/Straight}{1}
\FALabel(16.2808,15.5544)[b]{$\bar q$}
\FAProp(20.,5.)(13.,6.)(0.,){/Straight}{-1}
\FALabel(16.2808,4.44558)[t]{$q$}
\FAProp(6.5,10.)(13.,14.)(0.,){/ScalarDash}{-1}
\FALabel(9.20801,13.1807)[br]{$\tilde q_i$}
\FAProp(6.5,10.)(13.,6.)(0.,){/ScalarDash}{1}
\FALabel(9.5,6.81927)[tr]{$\tilde q_i$}
\FAProp(13.,14.)(13.,6.)(0.,){/Straight}{0}
\FALabel(14.274,10.)[l]{$\tilde g$}
\FAVert(6.5,10.){0}
\FAVert(13.,14.){0}
\FAVert(13.,6.){0}
\end{feynartspicture}
}
\caption{Generic SUSY QCD vertex contributions to $Z\to q \bar q$.}
\label{fig:vertexSUSYQCDoneloop}
\vspace{5em}
\end{figure}

\subsection{One-loop graphs for $Z \to \neu{1} \neu{1}$}

\begin{figure}[hb!]
\vspace{-2em}
\centerline{
\unitlength=1.cm%
\begin{feynartspicture}(4,4)(1,1)
\FADiagram{}
\FAProp(0.,10.)(6.5,10.)(0.,){/Sine}{0}
\FALabel(3.25,8.93)[t]{$Z$}
\FAProp(20.,15.)(13.,14.)(0.,){/Straight}{0}
\FALabel(16.2808,15.5544)[b]{$\tilde\chi^0_1$}
\FAProp(20.,5.)(13.,6.)(0.,){/Straight}{0}
\FALabel(16.2808,4.44558)[t]{$\tilde\chi^0_1$}
\FAProp(6.5,10.)(13.,14.)(0.,){/Straight}{0}
\FALabel(9.20801,13.1807)[br]{$\tilde\chi^0,\tilde\chi$}
\FAProp(6.5,10.)(13.,6.)(0.,){/Straight}{0}
\FALabel(9.5,6.81927)[tr]{$\tilde\chi^0,\tilde\chi$}
\FAProp(13.,14.)(13.,6.)(0.,){/ScalarDash}{0}
\FALabel(14.274,10.)[l]{$H_a, H^\pm, G$}
\FAVert(6.5,10.){0}
\FAVert(13.,14.){0}
\FAVert(13.,6.){0}
\end{feynartspicture}
\hspace{0.5cm}
\begin{feynartspicture}(4,4)(1,1)
\FADiagram{}
\FAProp(0.,10.)(6.5,10.)(0.,){/Sine}{0}
\FALabel(3.25,8.93)[t]{$Z$}
\FAProp(20.,15.)(13.,14.)(0.,){/Straight}{0}
\FALabel(16.2808,15.5544)[b]{$\tilde\chi^0_1$}
\FAProp(20.,5.)(13.,6.)(0.,){/Straight}{0}
\FALabel(16.2808,4.44558)[t]{$\tilde\chi^0_1$}
\FAProp(6.5,10.)(13.,14.)(0.,){/Straight}{-1}
\FALabel(9.20801,13.1807)[br]{$f_i, \nu_{f_i}$}
\FAProp(6.5,10.)(13.,6.)(0.,){/Straight}{1}
\FALabel(9.5,6.81927)[tr]{$f_i, \nu_{f_i}$}
\FAProp(13.,14.)(13.,6.)(0.,){/ScalarDash}{-1}
\FALabel(14.274,10.)[l]{$\tilde f_i, \tilde\nu_{f_i} $}
\FAVert(6.5,10.){0}
\FAVert(13.,14.){0}
\FAVert(13.,6.){0}
\end{feynartspicture}
\hspace{0.5cm}
\begin{feynartspicture}(4,4)(1,1)
\FADiagram{}
\FAProp(0.,10.)(6.5,10.)(0.,){/Sine}{0}
\FALabel(3.25,8.93)[t]{$Z$}
\FAProp(20.,15.)(13.,14.)(0.,){/Straight}{0}
\FALabel(16.2808,15.5544)[b]{$\tilde\chi^0_1$}
\FAProp(20.,5.)(13.,6.)(0.,){/Straight}{0}
\FALabel(16.2808,4.44558)[t]{$\tilde\chi^0_1$}
\FAProp(6.5,10.)(13.,14.)(0.,){/ScalarDash}{0}
\FALabel(9.20801,13.1807)[br]{$H_a, H^\pm, G$}
\FAProp(6.5,10.)(13.,6.)(0.,){/ScalarDash}{0}
\FALabel(9.5,6.81927)[tr]{$H_b, H^\pm, G$}
\FAProp(13.,14.)(13.,6.)(0.,){/Straight}{0}
\FALabel(14.274,10.)[l]{$\tilde \chi, \tilde \chi^0$}
\FAVert(6.5,10.){0}
\FAVert(13.,14.){0}
\FAVert(13.,6.){0}
\end{feynartspicture}
}
\vspace{-1em}
\centerline{
\unitlength=1.cm%
\begin{feynartspicture}(4,4)(1,1)
\FADiagram{}
\FAProp(0.,10.)(6.5,10.)(0.,){/Sine}{0}
\FALabel(3.25,8.93)[t]{$Z$}
\FAProp(20.,15.)(13.,14.)(0.,){/Straight}{0}
\FALabel(16.2808,15.5544)[b]{$\tilde\chi^0_1$}
\FAProp(20.,5.)(13.,6.)(0.,){/Straight}{0}
\FALabel(16.2808,4.44558)[t]{$\tilde\chi^0_1$}
\FAProp(6.5,10.)(13.,14.)(0.,){/ScalarDash}{1}
\FALabel(9.20801,13.1807)[br]{$\tilde f_i, \tilde \nu_{f_i}$}
\FAProp(6.5,10.)(13.,6.)(0.,){/ScalarDash}{-1}
\FALabel(9.5,6.81927)[tr]{$\tilde f_i, \tilde \nu_{f_i}$}
\FAProp(13.,14.)(13.,6.)(0.,){/Straight}{1}
\FALabel(14.274,10.)[l]{$f_i, \nu_{f_i}$}
\FAVert(6.5,10.){0}
\FAVert(13.,14.){0}
\FAVert(13.,6.){0}
\end{feynartspicture}
\hspace{0.5cm}
\begin{feynartspicture}(4,4)(1,1)
\FADiagram{}
\FAProp(0.,10.)(6.5,10.)(0.,){/Sine}{0}
\FALabel(3.25,8.93)[t]{$Z$}
\FAProp(20.,15.)(13.,14.)(0.,){/Straight}{0}
\FALabel(16.2808,15.5544)[b]{$\tilde\chi^0_1$}
\FAProp(20.,5.)(13.,6.)(0.,){/Straight}{0}
\FALabel(16.2808,4.44558)[t]{$\tilde\chi^0_1$}
\FAProp(6.5,10.)(13.,14.)(0.,){/Straight}{0}
\FALabel(9.20801,13.1807)[br]{$\tilde \chi^0, \tilde \chi$}
\FAProp(6.5,10.)(13.,6.)(0.,){/Straight}{0}
\FALabel(9.5,6.81927)[tr]{$\tilde \chi^0, \tilde \chi$}
\FAProp(13.,14.)(13.,6.)(0.,){/Sine}{0}
\FALabel(14.274,10.)[l]{$Z, W$}
\FAVert(6.5,10.){0}
\FAVert(13.,14.){0}
\FAVert(13.,6.){0}
\end{feynartspicture}
\hspace{0.5cm}
\begin{feynartspicture}(4,4)(1,1)
\FADiagram{}
\FAProp(0.,10.)(6.5,10.)(0.,){/Sine}{0}
\FALabel(3.25,8.93)[t]{$Z$}
\FAProp(20.,15.)(13.,14.)(0.,){/Straight}{0}
\FALabel(16.2808,15.5544)[b]{$\tilde\chi^0_1$}
\FAProp(20.,5.)(13.,6.)(0.,){/Straight}{0}
\FALabel(16.2808,4.44558)[t]{$\tilde\chi^0_1$}
\FAProp(6.5,10.)(13.,14.)(0.,){/ScalarDash}{0}
\FALabel(9.20801,13.1807)[br]{$H_a, H^\pm, G$}
\FAProp(6.5,10.)(13.,6.)(0.,){/Sine}{0}
\FALabel(9.5,6.81927)[tr]{$Z, W$}
\FAProp(13.,14.)(13.,6.)(0.,){/Straight}{0}
\FALabel(14.274,10.)[l]{$\tilde \chi^0, \tilde \chi$}
\FAVert(6.5,10.){0}
\FAVert(13.,14.){0}
\FAVert(13.,6.){0}
\end{feynartspicture}
}
\vspace{-1.5em}
\centerline{
\unitlength=1.cm%
\begin{feynartspicture}(4,4)(1,1)
\FADiagram{}
\FAProp(0.,10.)(6.5,10.)(0.,){/Sine}{0}
\FALabel(3.25,8.93)[t]{$Z$}
\FAProp(20.,15.)(13.,14.)(0.,){/Straight}{0}
\FALabel(16.2808,15.5544)[b]{$\tilde\chi^0_1$}
\FAProp(20.,5.)(13.,6.)(0.,){/Straight}{0}
\FALabel(16.2808,4.44558)[t]{$\tilde\chi^0_1$}
\FAProp(6.5,10.)(13.,14.)(0.,){/Sine}{1}
\FALabel(9.20801,13.1807)[br]{$W$}
\FAProp(6.5,10.)(13.,6.)(0.,){/Sine}{-1}
\FALabel(9.5,6.81927)[tr]{$W$}
\FAProp(13.,14.)(13.,6.)(0.,){/Straight}{1}
\FALabel(14.274,10.)[l]{$\tilde \chi$}
\FAVert(6.5,10.){0}
\FAVert(13.,14.){0}
\FAVert(13.,6.){0}
\end{feynartspicture}
}
\vspace{-1em}
\caption{Generic vertex contributions to  $Z\to \neu{1} \neu{1}$.}
\label{fig:vertexoneloopNeu1}
\end{figure}

\subsection{SUSY two-loop graphs contributing to $\De \rho$}

\begin{figure}[htb!]
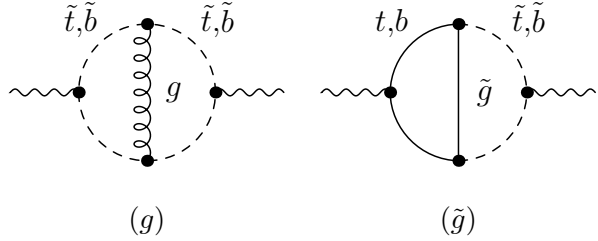

\vspace{-3em}
\centerline{
\unitlength=1.cm%
\begin{feynartspicture}(4,4)(1,1)
\FADiagram{$(g)$}
\FAVert(5,10){0}
\FAVert(15,10){0}
\FAProp(0,10)(5,10)(0.,){/Sine}{0}
\FAProp(15,10)(20,10)(0.,){/Sine}{0}
{
\FAProp(5,10)(15,10)(1.,){/ScalarDash}{0}
\FAProp(15,10)(5,10)(1.,){/ScalarDash}{0}
\FALabel(4,14)[b]{ ${\tilde{t}}$,${\tilde{b}}$}
\FALabel(14,14)[b]{ ${\tilde{t}}$,${\tilde{b}}$}}
\FAVert(10,15){0}
\FAVert(10,5){0}
\FAProp(10,5)(10,15)(0.,){/Cycles}{0}
\FALabel(11,10)[r]{$g$}
\end{feynartspicture}
\begin{feynartspicture}(4,4)(1,1)
\FADiagram{$(\tilde{g})$}
\FAVert(5,10){0}
\FAVert(15,10){0}
\FAProp(0,10)(5,10)(0.,){/Sine}{0}
\FAProp(15,10)(20,10)(0.,){/Sine}{0}
{
\FAProp(5,10)(10,15)(-.4,){/Straight}{0}
\FAProp(10,5)(5,10)(-.4,){/Straight}{0}
\FAProp(10,15)(15,10)(-.4,){/ScalarDash}{0}
\FAProp(10,5)(15,10)(.4,){/ScalarDash}{0}
\FALabel(4,14)[b]{ $t$,$b$}
\FALabel(14,14)[b]{ $\tilde{t}$,$\tilde{b}$}}
\FAVert(10,15){0}
\FAVert(10,5){0}
\FAProp(10,5)(10,15)(0.,){/Straight}{0}
\FALabel(11,10)[r]{$\tilde{g}$}
\end{feynartspicture}
}
\vspace{-1em}
\caption{Sample diagrams for the SUSY \order{\al \als} contributions to 
  $\De\rho$: $(g)$ squark loop with gluon
  exchange, $(\tilde{g})$ (s)quark loop with gluino exchange.}
\label{fig:samplediagramsQCD}
\end{figure}
%


\begin{figure}[htb!]
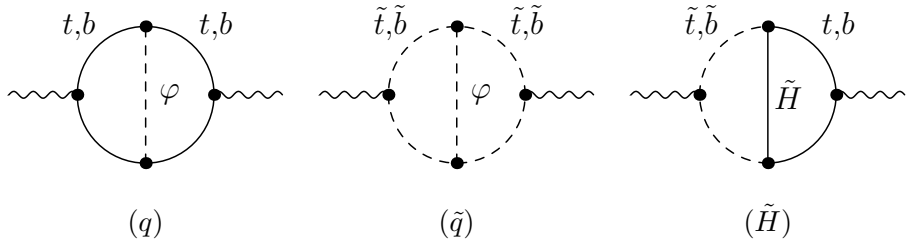

\vspace{-2em}
\centerline{
\unitlength=1.cm%
\begin{feynartspicture}(4,4)(1,1)
\FADiagram{$(q)$}
\FAVert(5,10){0}
\FAVert(15,10){0}
\FAProp(0,10)(5,10)(0.,){/Sine}{0}
\FAProp(15,10)(20,10)(0.,){/Sine}{0}
{
\FAProp(5,10)(15,10)(1.,){/Straight}{0}
\FAProp(15,10)(5,10)(1.,){/Straight}{0}
\FALabel(4,14)[b]{ ${t}$,${b}$}
\FALabel(14,14)[b]{ ${t}$,${b}$}}
\FAVert(10,15){0}
\FAVert(10,5){0}
\FAProp(10,5)(10,15)(0.,){/ScalarDash}{0}
\FALabel(11,10)[r]{$\varphi$}
\end{feynartspicture}
\unitlength=1.cm%
\begin{feynartspicture}(4,4)(1,1)
\FADiagram{$(\sq)$}
\FAVert(5,10){0}
\FAVert(15,10){0}
\FAProp(0,10)(5,10)(0.,){/Sine}{0}
\FAProp(15,10)(20,10)(0.,){/Sine}{0}
{
\FAProp(5,10)(15,10)(1.,){/ScalarDash}{0}
\FAProp(15,10)(5,10)(1.,){/ScalarDash}{0}
\FALabel(4,14)[b]{ $\tilde{t}$,$\tilde{b}$}
\FALabel(14,14)[b]{ $\tilde{t}$,$\tilde{b}$}}
\FAVert(10,15){0}
\FAVert(10,5){0}
\FAProp(10,5)(10,15)(0.,){/ScalarDash}{0}
\FALabel(11,10)[r]{$\varphi$}
\end{feynartspicture}
\unitlength=1.cm%
\begin{feynartspicture}(4,4)(1,1)
\FADiagram{$(\tilde{H})$}
\FAVert(5,10){0}
\FAVert(15,10){0}
\FAProp(0,10)(5,10)(0.,){/Sine}{0}
\FAProp(15,10)(20,10)(0.,){/Sine}{0}
{
\FAProp(5,10)(10,15)(-.4,){/ScalarDash}{0}
\FAProp(10,5)(5,10)(-.4,){/ScalarDash}{0}
\FAProp(10,15)(15,10)(-.4,){/Straight}{0}
\FAProp(10,5)(15,10)(.4,){/Straight}{0}
\FALabel(4,14)[b]{ $\tilde{t}$,$\tilde{b}$}
\FALabel(14,14)[b]{ ${t}$,${b}$}}
\FAVert(10,15){0}
\FAVert(10,5){0}
\FAProp(10,5)(10,15)(0.,){/Straight}{0}
\FALabel(11,10)[r]{$\tilde{H}$}
\end{feynartspicture}
}
\vspace{-1em}
\caption{Sample diagrams for the three classes of MSSM
  \order{\al_t^2}, \order{\al_b^2}, \order{\al_t \al_b} contributions
  to $\De\rho$: $(q)$ quark loop with Higgs
  exchange, $(\sq)$ squark loop with Higgs exchange, $(\tilde{H})$
  quark/squark loop with Higgsino exchange. $\varphi$ denotes Higgs and
  Goldstone boson exchange.}
\label{fig:samplediagramsYuk}
\vspace{-3em}
\end{figure}


\clearpage
\newpage
\pagebreak

\end{document}